\documentclass[11.5pt]{scrartcl}

\usepackage{xy-format}
\usepackage{xy-theorem}
\usepackage{fancyhdr}
\usepackage{amsmath}
\usepackage{longtable}
\title{From Baby Universes to Narain Moduli:\\Topological Boundary Averaging in SymTFTs}

\author[1]{Xingyang Yu}
\affil[1]{Department of Physics, Virginia Tech, Blacksburg, VA 24061, USA}

\date{}

\begin{document}
\maketitle

\begin{abstract}
Building on the viewpoint that ensemble averages in TQFT gravity can be organized by topological boundary data, we develop a SymTFT interpretation of ensemble averaging in low-dimensional holography.  The central operation is to keep fixed both the SymTFT and the physical boundary condition, while averaging over topological boundary conditions at the other end of the SymTFT slab.  Each such boundary condition gives an absolute completion of the same relative theory, so the ensemble is interpreted as an average over topological completions rather than over arbitrary local dynamics.  We formulate this construction in terms of cap functionals and their natural groupoid or Haar-type measures, and illustrate it in two examples.  In the closed-string sector of the Marolf--Maxfield model, topological boundary conditions are labelled by finite sets, and the groupoid sum reproduces the Poisson/Bell-polynomial moments.  In the Narain case, compact topological boundary conditions of an \(\mathbb{R}\)-valued BF SymTFT are identified with maximal isotropic subgroups, so that topological-boundary averaging reproduces the usual Narain moduli average with Zamolodchikov measure.  We also discuss possible extensions to JT gravity, random matrix theory, Virasoro T(Q)FT, and 3D gravity.
\end{abstract}

\newpage 
\tableofcontents
\section{Introduction}

A striking lesson from recent developments in low-dimensional holography is that a
gravitational path integral need not compute the partition function of a single boundary
theory.  When the bulk sum includes Euclidean wormholes connecting several asymptotic
boundaries, the result generally fails to factorize:
\begin{equation}
  \big\langle Z[J_1] Z[J_2] \big\rangle
  \neq
  \big\langle Z[J_1]\big\rangle
  \big\langle Z[J_2]\big\rangle.
\end{equation}
A natural interpretation is that the gravitational path integral computes moments in an
ensemble averaged boundary theories.  This idea appears concretely in JT gravity and its relation
to double-scaled random matrix theory, in baby-universe models of spacetime wormholes,
and in 3D examples where sums over bulk topologies reproduce averages
over families of 2D CFTs
\cite{1903.11115,1907.03363,2002.08950,2006.04855,2006.04839,2405.20366}.  These
examples suggest that ensemble averaging is not merely a technical device, but may be an
intrinsic feature of certain gravitational path integrals\footnote{See, e.g.,
\cite{2006.11317,2006.13414,2006.13971,2006.16289,
2103.16754,2104.01184,2105.02142,2105.08207,
2106.09048,2107.02178,2107.13130,Heckman:2021vzx,2111.07863,
2111.14856,2201.00903,2203.09537,2209.02131,
2404.10035, 2006.05499,2006.08648,2007.15653,2012.15830,
2102.03136,2102.12509,2103.15826,2104.10178,
2104.14710,2106.12760,2110.14649,2112.09143,
2203.06511,2208.14457,2304.13650,2306.07321,
2307.03707,2308.01787,2308.03829,2309.10846,
2310.06012,2310.13044,2311.00699,2311.08132,
2312.02276,2401.13900,2403.02976,2405.13111,
2407.02649,2503.00101,2504.08724,2506.19817,
2511.04311} and references therein for a partial list of references to recent literature on ensemble holography.}.

A second, largely independent, development is the modern understanding of generalized
global symmetries \cite{1412.5148}.  A D-dimensional QFT with generalized symmetry can often be
viewed as a boundary condition for a (D+1)-dimensional topological theory, the
Symmetry TFT or SymTFT \cite{1212.1692,2212.00195}\footnote{See, e.g., \cite{Reshetikhin:1991tc,Turaev:1992hq, Barrett:1993ab, hep-th/9812012, hep-th/0204148, Kirillov2010TuraevViroIA,
1012.0911, Kitaev:2011dxc, Fuchs:2012dt, 1212.1692, Kong:2014qka, Kong:2017hcw, Heckman:2017uxe, Freed:2018cec,
Gaiotto:2020iye, Kong:2020cie, Apruzzi:2021nmk, Freed:2022qnc, Kaidi:2022cpf, Antinucci:2022vyk, 2306.11783, Baume:2023kkf, Yu:2023nyn, 2401.06128,2401.10165, DelZotto:2024tae, Argurio:2024oym, Franco:2024mxa, Heckman:2024zdo, Gagliano:2024off, Cordova:2024iti, Cvetic:2024dzu, Bhardwaj:2024igy, Bonetti:2024cjk, Apruzzi:2024htg, 2411.14997, Jia:2025jmn,  Apruzzi:2025mdl, Heckman:2025lmw, Pace:2025hpb, Luo:2025phx, Apruzzi:2025hvs, 2510.06319, 2603.12323} and references therein for a partial list of references to foundational early work,
as well as more recent generalizations.}.  In this formulation the boundary QFT is a relative theory \cite{1212.1692}:
its partition function is not a number, but a vector in the Hilbert space of the SymTFT.
To obtain an ordinary absolute QFT, one must choose a topological boundary condition at
the other end of the SymTFT slab.  Different topological boundary conditions correspond
to different global forms, gaugings, or more generally different topological completions of
the same relative theory.

The purpose of this paper is to develop this connection in a form adapted to SymTFTs and
to two concrete examples.  The idea that ensemble averages can be organized as sums over
maximal gaugings, Lagrangian algebras, or topological boundary conditions of a fixed TQFT,
and related to sums over topologies, was put forward in
\cite{2310.13044,2405.20366}; see also the bordism-sum perspective of
\cite{2201.00903}.  Automorphism-weighted versions of this topological-boundary
average, including noncompact examples, were further discussed in
\cite{2511.04311,Dymarsky:2026asf}.  Building on this viewpoint, we study how, in a
class of low-dimensional ensemble holography examples, the ensemble average can be
understood as an average over topological boundary conditions of a fixed SymTFT.  The basic setup is
as follows.  We fix a SymTFT \(\mathfrak{T}_{\mathrm{sym}}\) and a physical boundary condition
\(B_{\mathrm{phys}}(J)\), depending on metric and source data \(J\).  This physical boundary
prepares a state
\begin{equation}
  \big|\Psi_{\mathrm{phys}}(M;J)\big\rangle
  \in
  \mathcal{H}_{\mathrm{sym}}(M)
\end{equation}
on a closed \(d\)-manifold \(M\).  A topological boundary condition \(L\) at the other end
of the slab defines a cap functional
\begin{equation}
  \big\langle L;M\big|
  \in
  \mathcal{H}_{\mathrm{sym}}(M)^* .
\end{equation}
The absolute partition function obtained by choosing the cap \(L\) is then
\begin{equation}
  Z_L(M;J)
  =
  \big\langle L;M \,\big|\,
  \Psi_{\mathrm{phys}}(M;J)
  \big\rangle .
\end{equation}
The ensemble average considered in this paper is obtained by varying \(L\), while keeping
both the SymTFT and the physical boundary fixed:
\begin{equation}
  \big\langle Z(M;J)\big\rangle_{\mathrm{top}}
  =
  \int_{\mathcal{L}_{\mathrm{top}}}
  d\mu(L)\,
  \big\langle L;M \,\big|\,
  \Psi_{\mathrm{phys}}(M;J)
  \big\rangle .
  \label{eq:intro-topological-average}
\end{equation}
For a finite family of topological boundaries, the integral is replaced by a groupoid sum,
with the usual factor \(1/|\operatorname{Aut}(L)|\).  For a continuous family, the measure is
a Haar-type measure on the corresponding space of Lagrangian boundary data.

This construction should be distinguished from averaging over arbitrary boundary
conditions.  A general boundary condition at the second end of the slab would usually carry
its own local degrees of freedom, its own dependence on metric or coupling data, and its
own Hamiltonian.  Such an average would be an average over local dynamics.  By contrast,
the topological boundary conditions considered here carry zero Hamiltonian.  They do not
change the physical time evolution supplied by \(B_{\mathrm{phys}}(J)\).  Instead, they specify
the Hilbert space, charge lattice, global form, or other topological completion on which the
fixed physical Hamiltonian acts.  This is the sense in which our ensemble average is an
average over topological completions of a fixed relative theory.

This viewpoint also gives a clean relation to summing over topologies.  In a functorial TFT, a
bordism \(Y:M\to \emptyset\) defines a linear functional on
\(\mathcal{H}_{\mathrm{sym}}(M)\).  Thus a sum over fillings of \(M\) becomes, after applying
the TFT functor, a sum over cap functionals.  A topological boundary condition also
defines such a cap.  Therefore, whenever the topology sum is only sensitive to the induced
topological cap data, it can be reorganized as a sum over topological boundary conditions.
This is the mechanism we isolate and study in this paper.

There are several operations in the literature that are all called ensemble averaging.
One may average over a moduli space of absolute theories, as in Narain averaging
\cite{2006.04855,2006.04839}; one may sum over modular images of a seed boundary
partition function, as in Poincar\'e-series constructions of 3D gravity
\cite{0712.0155,1407.6008}; or one may average over topological boundary conditions,
maximal gaugings, or Lagrangian algebras of a fixed TFT
\cite{2201.00903,2310.13044,2405.20366,2511.04311,Dymarsky:2026asf}.  These operations are not
identical in general.  In this paper we focus on the last type of operation.  In the examples
studied below, the SymTFT formulation identifies the relevant averaging variable with
topological boundary data.  The moduli being averaged over are reinterpreted as
topological completions, and the canonical measure on moduli space is reproduced as
the natural groupoid or Haar measure on the space of caps.

Our first example is the closed-string sector of the Marolf--Maxfield topological model \cite{2002.08950,2201.00903}.
The baby-universe analysis of this model gives moments of a Poisson random variable.  We
reproduce these moments from a fixed discrete SymTFT-like parent object
\(\mathfrak{S}_{\mathrm{Fin}}\), whose simple topological boundary conditions are labelled by
finite sets.  After choosing a finite set \(S\), the corresponding absolute closed 2D
TFT has Frobenius algebra
\begin{equation}
  A_S=\operatorname{Fun}(S,\mathbb{C}) .
\end{equation}
The physical boundary is represented by the unit \(1_S\), while the topological cap is
represented by the Frobenius counit.  Their pairing gives
\begin{equation}
  Z_S(S^1)=|S| .
\end{equation}
Averaging over finite sets up to bijection, with groupoid weight \(1/|\operatorname{Aut}(S)|\)
and fugacity \(\lambda^{|S|}\), gives
\begin{equation}
  \big\langle Z^n \big\rangle_{\mathrm{top}}
  =
  e^{-\lambda}
  \sum_{d=0}^{\infty}
  \frac{\lambda^d}{d!}\,d^n
  =
  B_n(\lambda),
\end{equation}
the Bell-polynomial moments of the Marolf--Maxfield ensemble.  At fixed \(S\) the theory
factorizes, while the non-factorization is produced entirely by the average over finite-set
topological caps.

Our second example is the Narain ensemble \cite{2006.04855,2006.04839}.  We describe the \(c\)-dimensional (target space dimension) compact
boson using a 3D \(\mathbb{R}\)-valued BF SymTFT.  The physical boundary
condition supplies the left- and right-moving current algebra, the oscillator descendants,
and the conformal Hamiltonian.  The compact topological boundary condition supplies the
even self-dual charge lattice, and hence the current-algebra primary spectrum.  For
\(c>1\), the compact component of the space of topological boundary conditions is
\begin{equation}
  \mathcal{L}_{\mathrm{Narain}}^{(c)}
  =
  O(c,c;\mathbb{Z})\backslash
  O(c,c;\mathbb{R})/
  \bigl(O(c)\times O(c)\bigr),
\end{equation}
which is precisely the Narain moduli space.  The Haar-induced measure on this space is
the usual Narain, or Zamolodchikov \cite{Zamolodchikov:1986gt}, measure.  For \(c>2\), the Siegel--Weil formula then
gives
\begin{equation}
  \big\langle Z(\tau)\big\rangle_{\mathrm{top}}^{(c)}
  =
  \frac{E_{c/2}(\tau)}
       {\tau_2^{c/2}|\eta(\tau)|^{2c}},
\end{equation}
reproducing the standard genus-one Narain average.  The \(c=1\) compact boson is useful
as a warm-up example: it makes the radius dictionary transparent\footnote{The interpretation of different topological boundary conditions of the \(U(1)\) SymTFT as labeling different radii of the \(c=1\) compact boson was also emphasized by H.~T.~Lam in talks \cite{Lam:KITP2025,Lam:Harvard2025}.} and also illustrates the
infinite-volume subtlety of rank-one Narain averaging.

The conceptual lesson of these examples is simple.  In the Marolf--Maxfield model, the
topological completion is a finite set.  In the Narain ensemble, it is an maximal isotropic subgroup of the defect group.  In both cases, the local physical boundary dynamics is held fixed, while the
topological boundary condition specifies the absolute theory.  The ensemble average is
therefore not an average over arbitrary dynamics, but an average over admissible
topological completions.

We also discuss two speculative directions..  JT gravity admits a BF-theory formulation \cite{1812.00918,1905.02726}, and
its relation to random matrix theory suggests that the random matrix ensemble may be
understood as an average over topological completions of a fixed relative one-dimensional
quantum mechanics.  Similarly, recent work on Virasoro TQFT \cite{2304.13650,2401.13900} suggests a possible
SymTFT framework for pure \(\mathrm{AdS}_3\) gravity, in which a doubled Virasoro TQFT
would play the role of the parent topological theory and the ensemble of 2D
CFT data would arise from averaging over suitable Virasoro topological boundary
conditions.  These directions are not developed into complete constructions here, but they
suggest that topological-boundary averaging may provide a useful organizing principle for
ensemble holography beyond the two examples explicitly analyzed in this paper.

Before presenting the organization of the paper, let us briefly mention some closely related literature.  SymTFTs and topological holography have
already appeared in discussions of gravitational path integrals.  In the
Liouville/Virasoro direction, higher-dimensional topological data and Virasoro
T(Q)FT have been used to reconstruct CFT path integrals and to formulate aspects of
non-perturbative \(\mathrm{AdS}_3/\mathrm{CFT}_2\) gravity
\cite{2210.12127,2311.18005,2403.03179,2412.11486,2412.12045,2504.21660}.
In particular, recent work relates fixed-topology \(\mathrm{AdS}_3\) gravity amplitudes
to Virasoro T(Q)FT amplitude-squared and to Conformal Turaev--Viro theory \cite{2507.12696}.
In a complementary TQFT-gravity direction, ensembles of boundary CFTs have been
related to sums over maximal gaugings, or equivalently to sums over topological
boundary conditions described by Lagrangian algebras
\cite{2310.13044,2405.20366,2511.04311}.  Our work is closer in spirit to this
second perspective, but provides a general prescription from a more TFT perspective which works nicely beyond finite RCFT cases. This distinction
is transparent from examples we considered: the Marolf--Maxfield model involves a countably
\emph{infinite} groupoid of finite-set caps, and the Narain example is governed by an
\(\mathbb{R}\)-valued BF SymTFT whose compact topological boundaries sweep out the
full Narain moduli space, including generic \emph{irrational} points.

The paper is organized as follows.  In Section~\ref{sec:topological-boundary-averaging}
we formulate the general construction of averaging over topological boundary conditions,
including the finite groupoid measure and its continuous Haar-measure analogue.  In
Section~\ref{sec:MM-from-SymTFT} we apply this framework to the closed-string sector of the
Marolf--Maxfield model and reproduce the Poisson/Bell-polynomial moments.  In
Section~\ref{sec:narain-symtft} we study the compact boson and higher-rank Narain
theories using \(\mathbb{R}\)-valued BF SymTFTs, and show that Narain moduli averaging is
topological-boundary averaging.  Section~\ref{sec:JT-Virasoro} discusses possible extensions
to JT gravity, random matrix theory, Virasoro TQFT, and 3D gravity.

\section{Averaging over Topological Boundary Conditions as Summing over Topologies}
\label{sec:topological-boundary-averaging}

In this section we formulate the basic operation that will be used throughout the
paper.  The point is simple.  In a functorial topological field theory, a bordism from a
manifold $M$ to the empty set defines a linear functional on the Hilbert space assigned
to $M$.  Thus, after applying the TFT functor, a sum over fillings of $M$ becomes a
sum over cap functionals.  In a SymTFT, there is a distinguished class of such caps:
those produced by topological boundary conditions.  We will use this observation to define
a SymTFT-controlled version of ensemble averaging.

The discussion in this section is not meant to define the most general holographic ensemble. Instead, we are  just 
isolating a more restrictive construction.  We keep fixed a relative theory, namely a
fixed SymTFT together with a fixed physical boundary condition, and we vary only the
topological boundary condition used to cap the SymTFT slab.  The resulting average is an
average over topological completions of a fixed relative theory.

\subsection{Caps and topological boundary conditions}
We start with a simple observation.  In a $(d+1)$-dimensional topological field theory,
a filling of a $d$-manifold $M$ is not merely seen as a geometry by itself, but as a linear
functional on the state space associated to $M$ \cite{Atiyah:1989vu}. Let $\mathfrak T$ be a $(d+1)$-dimensional TFT.  It assigns to a closed $d$-manifold
$M$ a vector space
\begin{equation}
    \mathcal H_{\mathfrak T}(M).
    \label{eq:tft-state-space}
\end{equation}
A bordism
\begin{equation}
    Y:M\longrightarrow \emptyset
    \label{eq:bordism-to-empty}
\end{equation}
defines a map
\begin{equation}
    Z_{\mathfrak T}(Y):
    \mathcal H_{\mathfrak T}(M)
    \longrightarrow
    \mathcal H_{\mathfrak T}(\emptyset)
    \cong
    \mathbb C .
    \label{eq:bordism-functional}
\end{equation}
Thus $Y$ defines an element of the dual space,
\begin{equation}
    Z_{\mathfrak T}(Y)\in \mathcal H_{\mathfrak T}(M)^* .
    \label{eq:bordism-as-cap}
\end{equation}
In this sense, a bordism from $M$ to the empty set is a cap.

A sum over bordisms with fixed boundary $M$ therefore gives a distinguished element of
$\mathcal H_{\mathfrak T}(M)^*$:
\begin{equation}
    \mathcal C_{\mathrm{bord}}(M)
    =
    \sum_{[Y:M\to\emptyset]}
    w(Y)\,
    Z_{\mathfrak T}(Y)
    \in
    \mathcal H_{\mathfrak T}(M)^* .
    \label{eq:bordism-cap-sum}
\end{equation}
The coefficient $w(Y)$ denotes whatever weight the theory assigns to the bordism $Y$.
At this stage we will not specify it.  In an ordinary gravitational path integral it would
come from the action and the path-integral measure.  In a purely topological or finite
groupoid version it includes the usual symmetry factors.  The point of
\eqref{eq:bordism-cap-sum} is not the precise choice of weight, but the fact that the
topology sum has become a sum of cap functionals.

This is a useful way to state the problem because a topological boundary condition also naturally 
defines a cap.  If $L$ is a topological boundary condition of $\mathfrak T$, then putting
$\mathfrak T$ on a collar ending on $L$ gives, for each $M$, a linear functional
\begin{equation}
    \langle L;M|
    \in
    \mathcal H_{\mathfrak T}(M)^* .
    \label{eq:topological-boundary-cap}
\end{equation}
This functional is not an arbitrary element of the dual vector space.  It is one that can be
realized by a boundary condition of the same topological theory.

The construction that we will use is that, in some cases, the bordism sum
\eqref{eq:bordism-cap-sum} can be reorganized in terms of these topological caps:
\begin{equation}
    \mathcal C_{\mathrm{bord}}(M)
    =
    \sum_{[L]}
    W_M(L)\,
    \langle L;M| .
    \label{eq:bordism-sum-as-boundary-sum}
\end{equation}
Here $W_M(L)$ is the total weight of all bordisms that give the same cap, or more
generally the same cap data, in the TFT.  Equivalently, the statement is that the image of
the bordism sum in $\mathcal H_{\mathfrak T}(M)^*$ factors through the set of topological
boundary conditions.

This should not be interpreted as saying that a geometric filling $Y$ is literally the same
thing as a topological boundary condition $L$.  The statement is weaker, and more natural
from the TFT point of view.  After applying the functor $Z_{\mathfrak T}$, the filling $Y$
is remembered only through the functional, or physically speaking the ``wave function'' $Z_{\mathfrak T}(Y)$.  Distinct fillings may
therefore define the same cap.  If the caps that occur in the sum are naturally labelled by
topological boundary conditions, then the topology sum descends to a sum over such
boundary conditions.

This is the sense in which we will relate topology sums to boundary condition averaging.  The
geometric objects may be  handlebodies or more exotic fillings.  The TFT sees
only the corresponding elements of the dual Hilbert space.  In favorable examples, these
elements are classified by Lagrangian data.  This is what happens in the Narain example \cite{2006.04855}:
for connected $\Sigma$, a handlebody determines a Lagrangian sublattice of
$H_1(\Sigma,\mathbb Z)$; after rewriting the formula in these terms, the same expression
continues to make sense also when $\Sigma$ is disconnected, even though there is no longer
a preferred ordinary handlebody associated to each Lagrangian sublattice.  This suggests
that the more intrinsic object is the Lagrangian cap data, not the ordinary smooth manifold
itself. 

We now specialize this observation to SymTFTs. Let $\mathfrak T_{\mathrm{sym}}$ be the $(d+1)$-dimensional SymTFT associated
with a $d$-dimensional QFT.  The SymTFT is placed
on a slab
\begin{equation}
    M\times[0,1].
    \label{eq:symtft-slab}
\end{equation}
One end of the slab is coupled to the physical $d$-dimensional relative theory \cite{1212.1692,2212.00195}.  This boundary in general is
not topological.  It depends on the metric and possible source data on $M$, which we denote
collectively by $J$.  We then write the physical boundary condition as
\begin{equation}
    B_{\mathrm{phys}}(J).
    \label{eq:physical-boundary}
\end{equation}
This physical boundary provides a state in the Hilbert space on $M$:
\begin{equation}
    |\Psi_{\mathrm{phys}}(M;J)\rangle
    \in
    \mathcal H_{\mathrm{sym}}(M),
    \label{eq:physical-state}
\end{equation}
which is the partition vector, instead of a number, of the relative physical boundary theory. 

To obtain an absolute theory, one chooses a topological boundary condition $L$ at the
other end of the slab.  $L$ specifies which bulk topological operators can end on the
topological boundary.  The remaining
bulk operators, modulo those trivialized by $L$, become the symmetry operators of the
absolute theory.
In finite semisimple SymTFTs,  $L$ is often described as a Lagrangian algebra
of the category of bulk topological operators \cite{hep-th/0204148, 1008.0654,1012.0911}.  In the continuous abelian examples considered in the following sections, the
analogous object will be a maximal isotropic subset in the continuous family of topological operators in the SymTFT \cite{hep-th/9812012, 2010.15890, 2203.09537, 2306.11783}.

For each $M$, $L$ defines a cap functional
\begin{equation}
    \langle L;M|
    \in
    \mathcal H_{\mathrm{sym}}(M)^* .
    \label{eq:symtft-cap}
\end{equation}
The partition function of the absolute theory specified by $L$ is
\begin{equation}
    Z_L(M;J)
    =
    \langle L;M|\Psi_{\mathrm{phys}}(M;J)\rangle .
    \label{eq:absolute-partition-function}
\end{equation}

We here emphasize an important restriction in this construction.  We do not average over
arbitrary boundary conditions at the second end of the slab.  A general boundary condition
$B$ would also define a linear functional
\begin{equation}
    \langle B;M,J_B|:
    \mathcal H_{\mathrm{sym}}(M)
    \longrightarrow
    \mathbb C,
    \label{eq:general-boundary-functional}
\end{equation}
which can be regarded as another physical boundary with reversed orientation of $M$. 
However, such a boundary condition would usually carry its own local degrees of freedom and
its own dependence on metric and coupling data. In Hamiltonian language, such a boundary would come with its own nontrivial boundary
Hamiltonian.  Summing over general boundary conditions would therefore be a sum over
local dynamics.  This is too large an operation for our purposes.  In the ensemble examples
we have in mind, the average is over a specified family of theories, such as a fixed theory with random couplings drawn from a prescribed ensemble with measure.  An arbitrary sum over boundary conditions
would not single out such an ensemble, and would not be determined by the SymTFT alone.

The SymTFT construction uses a smaller class of caps.  The boundary $L$ is topological:
it carries zero Hamiltonian at the end of the slab. In this sense, the Hamiltonian of the boundary absolute QFTs solely
come from $B_{\mathrm{phys}}(J)$, while $L$ supplies which Hilbert space the Hamiltonian acting upon.

We may now construct the object whose measure will be discussed in the next subsection.
Let $\mathcal L_{\mathrm{top}}$ denote the collection of admissible  topological boundary
conditions\footnote{More precisely, simple topological boundary conditions  are those cannot be reduced to simpler topological boundaries. Since here we are only interested in semisimple TFTs, simple topological boundaries are equivalently those which are indecomposible into sums of other topological boundaries.} of the SymTFT.  Before specifying a measure, the possible averaged cap
functionals lie in the span
\begin{equation}
    \operatorname{Span}
    \left\{
        \langle L;M|:
        L\in\mathcal L_{\mathrm{top}}
    \right\}
    \subset
    \mathcal H_{\mathrm{sym}}(M)^* .
    \label{eq:span-of-topological-caps}
\end{equation}
An averaged partition function will be obtained by choosing a linear combination, or in
continuous examples an integral, of these cap functionals and applying it to
$|\Psi_{\mathrm{phys}}(M;J)\rangle$.

The analogy with summing over topologies is now direct.  A bordism $Y:M\to\emptyset$
defines a cap in $\mathcal H_{\mathrm{sym}}(M)^*$.  A topological boundary condition
$L$ also defines a cap in $\mathcal H_{\mathrm{sym}}(M)^*$.  If, after applying the SymTFT functor, the bordism sum only depends on the filling
through its induced topological cap, then the sum over topologies can be reorganized as
a sum over topological boundary conditions. Schematically,
\begin{equation}
    \text{bordisms }Y:M\to\emptyset
    \xrightarrow{\;Z_{\mathrm{sym}}\;}
    \text{cap functionals in }\mathcal H_{\mathrm{sym}}(M)^*
    \leadsto
    \text{topological boundary conditions }L .
    \label{eq:bordisms-to-caps-to-boundaries}
\end{equation}

Note that we by no means claim that every ensemble in holography arises this way.  We also do not claim that
every geometric filling is literally a topological boundary condition.  The claim is that, for
a fixed SymTFT, the natural topological caps are supplied by its topological boundary
conditions, and in examples where the bulk topology sum is only sensitive to this cap data,
the topology sum can be rewritten as an average over such boundary conditions.

It remains to specify how the different caps should be weighted.  For a finite SymTFT,
the natural object is not a set but a groupoid: topological boundary conditions can have
automorphisms, and equivalent boundary conditions should not be counted independently.
The corresponding average is therefore a groupoid sum over isomorphism classes of
topological boundary conditions, in the same spirit as the groupoid-cardinality factors that
appear in sums over fields or bordisms.  The same principle applies to discrete, possibly
infinite, families whenever the sum is well-defined.  For a continuous SymTFT, this counting
problem is replaced by a measure problem.  The space of Lagrangian boundary data should be
equipped with a measure invariant under the natural duality action of the SymTFT.  We now
turn to this question.

\subsection{Weights and measures on the space of topological boundary conditions}
\label{subsec:weights-and-measures}

Let us specify weights in the sum over topological caps. If the weights depended on the spacetime $M$, or on the sources $J$,
then the resulting object would not be an ensemble average in the usual sense.  It would
instead be a separate prescription for each observable.  An ensemble measure should be
chosen once and for all on the space of caps, and the same measure should then be used
for all boundary manifolds and all insertions.

Let
\begin{equation}
    F_{M,J}(L)
    =
    Z_L(M;J)
    =
    \langle L;M|\Psi_{\mathrm{phys}}(M;J)\rangle
    \label{eq:function-on-cap-space}
\end{equation}
be the function on the space of topological boundary conditions obtained by evaluating
the physical boundary state against the cap, i.e. topological boundary $L$.  The problem is to define an integral
of this function over the allowed topological boundary conditions.

We first consider the case where the collection of  topological boundary conditions is finite.  The correct
notion is generally not a merely set but a groupoid\footnote{For the physics reader comfortable with categories, a groupoid can be regarded as a special category with all objects and morphisms invertible.}. The associated physical facts include that two topological boundary conditions may be
equivalent, and/or a given topological boundary condition may have automorphisms.  Therefore
one should not simply sum over all topological boundaries with the same weight.  For a finite groupoid $\mathcal G$,
the natural integral of a function $F$ on its objects is the groupoid cardinality integral
\begin{equation}
    \int_{\mathcal G} F
    :=
    \sum_{[x]\in \pi_0(\mathcal G)}
    \frac{1}{|\operatorname{Aut}_{\mathcal G}(x)|}\,
    F(x).
    \label{eq:groupoid-cardinality-integral}
\end{equation}
Here the sum is over isomorphism classes of objects, and
$\operatorname{Aut}_{\mathcal G}(x)$ is the automorphism group of the object $x$.
This is the same counting rule that appears in finite group gauge theory (see, e.g., \cite{1511.00295}), in summed-bordism
constructions (see, e.g., \cite{2201.00903}), and in automorphism-weighted topological-boundary
ensembles \cite{2511.04311,Dymarsky:2026asf}: configurations with nontrivial automorphisms are weighted by the inverse
order of their automorphism group.

Applying this rule to the groupoid $\mathcal L_{\mathrm{top}}$ of topological boundary
conditions gives
\begin{equation}
\boxed{
    \left\langle Z(M;J)\right\rangle_{\mathrm{top}}
    =
    \sum_{[L]}
    \frac{1}{|\operatorname{Aut}(L)|}\,
    \langle L;M|\Psi_{\mathrm{phys}}(M;J)\rangle .
}
    \label{eq:finite-groupoid-cap-average}
\end{equation}
This formula is the finite version of the topological-boundary ensemble.  The automorphism
group in \eqref{eq:finite-groupoid-cap-average} is the automorphism group of the cap as an
object of the groupoid of topological boundary conditions.  It is not an automorphism group
of the spacetime $M$.  The spacetime data have already entered through the function
$F_{M,J}(L)=\langle L;M|\Psi_{\mathrm{phys}}(M;J)\rangle$.

The same principle applies to discrete but possibly infinite families of caps.  If the set of
isomorphism classes is countable and the automorphism groups are finite, the ensemble averaging is again \eqref{eq:finite-groupoid-cap-average}. Now, however, convergence is part of the problem.  The counting prescription tells us how
each cap should be weighted, but it does not by itself guarantee that the total sum defines
a finite partition function.

For continuous SymTFTs, the groupoid sum is replaced by the integral with a measure.  The space
of topological boundary conditions can have continuous components, and the sum over
$[L]$ is promoted to an integral
\begin{equation}
\boxed{
    \left\langle Z(M;J)\right\rangle_{\mathrm{top}}
    =
    \int_{\mathcal L_{\mathrm{top}}}
    d\mu(L)\,
    \langle L;M|\Psi_{\mathrm{phys}}(M;J)\rangle.
}
    \label{eq:continuous-cap-average}
\end{equation}
The measure $d\mu(L)$ should be intrinsic to the SymTFT.  In the examples of interest, 
there is a natural duality group acting on the space of topological boundary, and the
measure is fixed by invariance under this action.

More explicitly, suppose that a generic component of the space of topological boundary conditions reads
\begin{equation}
    \mathcal L_{\mathrm{gen}}
    \simeq
    \Gamma\backslash G/H .
    \label{eq:lagrangian-homogeneous-space}
\end{equation}
Here $G$ is the continuous group of automorphisms of the topological defect data preserving the bulk TFT structure, e.g., braiding or quadratic form, $H$ is the stabilizer of a reference topological boundary condition, and
$\Gamma$ is the discrete duality group by which physically equivalent data are identified.
The measure on $G/H$ is induced from Haar measure \cite{9006cc9e-2dcc-3fd8-aada-e4af19b6e225} on $G$, and it descends to a
measure on $\Gamma\backslash G/H$.  Equivalently, it is characterized by
\begin{equation}
    d\mu(gL)=d\mu(L),
    \qquad
    g\in G,
    \label{eq:invariant-measure-condition}
\end{equation}
together with the quotient by $\Gamma$.

If the quotient has finite volume, one can either use the unnormalized measure or normalize
it to unit total volume.  Thus one may define
\begin{equation}
    \left\langle Z(M;J)\right\rangle_{\mathrm{top}}^{\mathrm{norm}}
    =
    \frac{1}{\operatorname{Vol}_{\mu}(\mathcal L_{\mathrm{top}})}
    \int_{\mathcal L_{\mathrm{top}}}
    d\mu(L)\,
    Z_L(M;J),
    \label{eq:normalized-continuous-cap-average}
\end{equation}
provided
\begin{equation}
    \operatorname{Vol}_{\mu}(\mathcal L_{\mathrm{top}})
    =
    \int_{\mathcal L_{\mathrm{top}}}d\mu(L)
    <
    \infty .
    \label{eq:finite-volume-condition}
\end{equation}
In many cases of interest, the overall normalization is conventional, so we will not specify it. 

The finite and continuous prescriptions are compatible.  If the quotient
$\Gamma\backslash G/H$ has orbifold points, then the invariant measure should be
understood as an orbifold measure.  Locally, if a point has a finite stabilizer group
$G_L$, integration over a small neighborhood $U/G_L$ is defined by
\begin{equation}
    \int_{U/G_L} f
    =
    \frac{1}{|G_L|}
    \int_U f .
    \label{eq:orbifold-measure-local}
\end{equation}
Thus the familiar factor $1/|\operatorname{Aut}(L)|$ is the discrete version of the same
principle.

\section{A SymTFT Realization of the Marolf--Maxfield Topological Ensemble}
\label{sec:MM-from-SymTFT}

In this section we explain, in a deliberately minimal setting, how the closed-string sector 
ensemble of Marolf--Maxfield \cite{2002.08950} can be reproduced as a sum over topological boundary
conditions of a fixed SymTFT.  By the closed-string sector we mean the part of the model
without end-of-the-world branes.  We use the corresponding topological-boundary-condition formalism and
show that the sum over topological boundary conditions reproduces the normalized Bell-polynomial
moments of the Marolf--Maxfield model.  We leave the open-string sector, including
end-of-the-world branes, for future work.

\subsection{The closed-string sector of Marolf--Maxfield}

We begin by recalling the general baby-universe interpretation of
ensemble averaging, and then specialize it to the closed-string sector of
the Marolf--Maxfield topological model \cite{2002.08950}. This review is slightly longer
than what is strictly needed for the computation, but it will be useful
for explaining what the SymTFT construction in the next subsection is
supposed to reproduce.

Consider a gravitational path integral with asymptotic boundary
conditions $J_i$ imposed on $n$ disconnected boundary components.
Schematically one writes
\begin{equation}
  \big\langle Z[J_1]\cdots Z[J_n]\big\rangle_{\mathrm{grav}}
  =
  \int_{\Phi\sim \{J_i\}} \mathcal D\Phi\, e^{-S[\Phi]} .
  \label{eq:general-grav-moment}
\end{equation}
If the bulk path integral includes connected geometries whose conformal
boundary has several connected components, then the answer need not
factorize. For example,
\begin{equation}
  \big\langle Z[J_1] Z[J_2]\big\rangle_{\mathrm{grav}}
  \neq
  \big\langle Z[J_1]\big\rangle_{\mathrm{grav}}
  \big\langle Z[J_2]\big\rangle_{\mathrm{grav}} .
  \label{eq:wormhole-nonfactorization}
\end{equation}
The connected contribution is interpreted as a spacetime wormhole between
the two asymptotic boundaries. This is the basic origin of the ensemble
interpretation.

A convenient way to organize this non-factorization is to cut open the
gravitational path integral along an intermediate slice. The slice may
contain components that do not reach any asymptotic boundary. These
closed spatial components are called baby universes. The states obtained
in this way span the baby-universe Hilbert space \cite{Coleman:1988cy, Giddings:1988cx, Giddings:1988wv}, denoted
$\mathcal H_{\mathrm{BU}}$. Given $n$ boundary insertions, there is a
corresponding state
\begin{equation}
  \big| Z[J_1]\cdots Z[J_n]\big\rangle
  \in \mathcal H_{\mathrm{BU}} .
  \label{eq:BU-state-boundaries}
\end{equation}
The state with no asymptotic boundary is the Hartle--Hawking state,
\begin{equation}
  |\mathrm{HH}\rangle \in \mathcal H_{\mathrm{BU}},
  \qquad
  \mathcal N := \langle \mathrm{HH}|\mathrm{HH}\rangle
  =
  \langle 1\rangle_{\mathrm{grav}} .
  \label{eq:HH-state-norm}
\end{equation}
Here $\mathcal N$ is the no-boundary amplitude, or cosmological
partition function. In a completely precise construction one should also
quotient by null states in the inner product defined by the gravitational
path integral. We will not need this refinement explicitly in the
closed-string sector example below.

For each asymptotic boundary condition $J$, one defines an operator
$\widehat Z[J]$ on $\mathcal H_{\mathrm{BU}}$ by adding one more
boundary component:
\begin{equation}
  \widehat Z[J]\,
  \big|Z[J_1]\cdots Z[J_n]\big\rangle
  =
  \big|Z[J]Z[J_1]\cdots Z[J_n]\big\rangle .
  \label{eq:Zhat-creation}
\end{equation}
Since the order of the boundary components in the gravitational path
integral is irrelevant, these operators commute:
\begin{equation}
  [\widehat Z[J_1],\widehat Z[J_2]]=0 .
  \label{eq:Zhat-commuting}
\end{equation}
Thus they can be diagonalized simultaneously. We denote their common
eigenstates by $\alpha$ :
\begin{equation}
  \widehat Z[J]|\alpha\rangle
  =
  Z_\alpha[J]\,|\alpha\rangle ,
  \qquad \forall\,J .
  \label{eq:alpha-state-general}
\end{equation}
Inserting a resolution of the identity in the $\alpha$-basis gives
\begin{equation}
  \frac{
  \big\langle Z[J_1]\cdots Z[J_n]\big\rangle_{\mathrm{grav}}
  }{
  \langle 1\rangle_{\mathrm{grav}}
  }
  =
  \sum_\alpha
  p_\alpha\,
  Z_\alpha[J_1]\cdots Z_\alpha[J_n],
  \qquad
  p_\alpha
  =
  \frac{|\langle \mathrm{HH}|\alpha\rangle|^2}
  {\langle \mathrm{HH}|\mathrm{HH}\rangle}.
  \label{eq:alpha-ensemble-general}
\end{equation}
This is the ensemble interpretation. A fixed $\alpha$-sector is a
factorizing theory, while the Hartle--Hawking state prepares a probability
distribution over such sectors. If the $\alpha$-spectrum is continuous,
the sum in \eqref{eq:alpha-ensemble-general} is replaced by an integral
with the corresponding probability measure.

We now specialize this discussion to the closed-string sector of the
Marolf--Maxfield topological model. By ``closed-string sector'' we mean the sector
with closed circular asymptotic boundaries and no end-of-the-world branes.
There is only one type of closed asymptotic boundary $S^1$, so there is a single
boundary observable, which we denote by $Z$. Equivalently, the general
operator $\widehat Z[J]$ above reduces to one baby-universe operator,
which we denote by a reduced symbol $\widehat{Z}$. The gravitational path integral with $n$ labelled circular boundaries
computes the moment $\big\langle Z^n\big\rangle_{\mathrm{MM}}$.
The boundaries are held fixed and labelled. Thus automorphisms of a bulk
surface are required to act trivially on the boundary components.

The combinatorics of the closed sector is simple. A connected component
of the bulk can end on any nonempty subset of the $n$ boundary circles.
After summing over the genus of that connected component, and after
absorbing the boundary weight into the normalization of $Z$, every
connected component with at least one boundary contributes the same
effective factor. We denote this factor by $\lambda$. Equivalently, this is to say that all connected closed-boundary amplitudes are normalized to
the same $\lambda$:
\begin{equation}
  \frac{
  \big\langle Z^m\big\rangle_{\mathrm{conn}}
  }{
  \langle 1\rangle_{\mathrm{MM}}
  }
  =
  \lambda ,
  \qquad m\geq 1 .
  \label{eq:all-connected-MM}
\end{equation}
The detailed dependence of $\lambda$ on the topological action is not
important for us. It can be regarded as the effective fugacity for one connected bulk
component that touches the asymptotic boundary.

For $n$ labelled boundaries, a bulk configuration is therefore specified
by a partition $\pi$ of the set $[n]=\{1,\ldots,n\}$. Each block of
$\pi$ records the set of boundary circles that lie on one connected
component of the bulk. Since each block contributes a factor $\lambda$,
we obtain
\begin{equation}
  \frac{\big\langle Z^n\big\rangle_{\mathrm{MM}}}
  {\langle 1\rangle_{\mathrm{MM}}}
  =
  \sum_{\pi\in \operatorname{Part}([n])}
  \lambda^{|\pi|}
  =
  B_n(\lambda)\equiv
  e^{-\lambda}
  \sum_{d=0}^{\infty}
  \frac{\lambda^d}{d!}\, d^n .
  \label{eq:MM-Bell}
\end{equation}
where $B_n(\lambda)$ is the Bell (or Touchard) polynomial. 
The normalized generating function thus is
\begin{equation}
  \frac{
  \big\langle e^{uZ}\big\rangle_{\mathrm{MM}}
  }{
  \langle 1\rangle_{\mathrm{MM}}
  }
  =
  \exp\!\left[\lambda(e^u-1)\right].
  \label{eq:MM-generating-function}
\end{equation}

Equations 
\eqref{eq:MM-generating-function} show that the closed-string sector
Marolf--Maxfield moments are precisely the moments of a Poisson random
variable of mean $\lambda$. Thus the $\alpha$-sectors can be labelled
by a non-negative integer $d$, and the baby-universe operator has
spectrum
\begin{equation}
  \widehat{Z}|d\rangle
  =
  d\,|d\rangle,
  \qquad d\in \mathbb Z_{\geq 0}.
  \label{eq:MM-Zb-spectrum}
\end{equation}
The Hartle--Hawking state prepares the probability distribution
\begin{equation}
  p_d
  =
  \frac{|\langle \mathrm{HH}|d\rangle|^2}
  {\langle \mathrm{HH}|\mathrm{HH}\rangle}
  =
  e^{-\lambda}\frac{\lambda^d}{d!}.
  \label{eq:MM-Poisson-prob}
\end{equation}
In a fixed $d$-sector the theory factorizes:
\begin{equation}
  \big\langle Z^n\big\rangle_d
  =
  d^n .
  \label{eq:fixed-d-factorization}
\end{equation}
The non-factorizing Marolf--Maxfield answer \eqref{eq:MM-Bell} is obtained only after
averaging over the $d$-sectors with the Poisson weights
\eqref{eq:MM-Poisson-prob}.

\subsection{SymTFT approach to Marolf--Maxfield ensemble}
\label{subsec:finite-set-symtft}

We now give a SymTFT realization of the closed-string sector ensemble averaging problem reviewed above.
The model does not enjoy a nice Lagrangian description such as BF or Chern--Simons theory.  The fixed parent object we use is instead a discrete, stacky, groupoid-completed 2D SymTFT analogue, which we denote by
\(\mathfrak S_{\mathrm{Fin}}\).  It is not the ordinary TFT associated with one chosen finite set.  Rather, it is a single parent object whose admissible topological boundary conditions, to be summed over below, are labelled by finite sets.  Since the closed-string sector of Marolf--Maxfield has only one asymptotic boundary observable $Z$, the topological data needed to reproduce their result is very small and explicit.

A mathematical fact we will use is that oriented 2D TFTs can be constructed via Frobenius algebras \cite{Abrams:1996ty, Kock_2003, 2206.12448}\footnote{We refer the reader to \cite{2311.16230} for a physics friendly introduction to Frobenius algebra.}.  We will use this fact only after a topological cap has been chosen.  Let $S$ be a finite set.  The cap labelled by $S$ produces an ordinary absolute closed 2D TFT whose Frobenius algebra is
\begin{equation}
   \mathcal{A}_S
    =
    \operatorname{Fun}(S,\mathbb C)
    =
    \bigoplus_{s\in S}\mathbb C e_s ,
    \label{eq:finite-set-frobenius-algebra-MM}
\end{equation}
where $e_s$ is the delta-function supported at $s$. The categorical data of this algebra is labeled by the Frobenius object $( \mathcal{A}_S,\mu, \eta, \delta, \varepsilon)$:
\begin{equation}
    \begin{split}
        \text{multiplication } \mu&:  \mathcal{A}_S \otimes  \mathcal{A}_S \rightarrow  \mathcal{A}_S,\\
        \text{unit } \eta&: I \rightarrow  \mathcal{A}_S,\\
        \text{comultiplication } \delta&:  \mathcal{A}_S \rightarrow  \mathcal{A}_S\otimes  \mathcal{A}_S,\\
        \text{counit } \varepsilon&:  \mathcal{A}_S \rightarrow I.
    \end{split}
\end{equation}
which respectively read
\begin{equation}
   e_s e_{s'}
    =
    \delta_{s,s'} e_s,
    \qquad
    1_S
    =
    \sum_{s\in S} e_s, 
    \qquad 
    \Delta_S(e_s)
    =e_s\otimes e_s,
    \qquad
      \varepsilon_S(e_s)=1 .
    \label{eq:finite-set-frobenius-structure-MM}
\end{equation}
The Frobenius pairing is 
\begin{equation}
    \langle e_s,e_{s'}\rangle_S
    =
    \varepsilon_S(e_s e_{s'})
    =
    \delta_{s,s'} .
    \label{eq:finite-set-frobenius-pairing-MM}
\end{equation}

By the mathematical construction of oriented 2D TFTs via Frobenius algebras, our
algebra $ \mathcal{A}_S$ defines an oriented closed worldsheet TFT, which we denote by $\mathcal{T}_S$.  Its
state space on a circle is
\begin{equation}
    \mathcal{H}_{\mathcal{T}_S}(S^1)
    =
     \mathcal{A}_S .
    \label{eq:TL-circle-state-space-MM}
\end{equation}
The case $S=\varnothing$ will also be included.  We regard it as the formal zero-dimensional
sector
\begin{equation}
     \mathcal{A}_\varnothing=0 .
    \label{eq:empty-set-zero-sector-MM}
\end{equation}
For $d>0$, we choose the
representative
\begin{equation}
        [d]=\{1,\ldots,d\},
    \qquad
    \mathcal A_{[d]}\cong \mathbb C^d .
    \label{eq:finite-set-representative-Ld-MM}
\end{equation}
We should not regard the theories $\mathcal T_S$ as different parent SymTFTs that are subsequently summed over.  Rather, the fixed parent object is a discrete analogue
\begin{equation}
    \mathfrak S_{\mathrm{Fin}},
\end{equation}
whose simple topological boundary conditions form the groupoid
\begin{equation}
    \operatorname{TopBdy}(\mathfrak S_{\mathrm{Fin}})
    \simeq
    \mathsf{FinSet}^{\simeq}.
    \label{eq:topbdy-SFin-FinSet-MM}
\end{equation}
Since finite sets can have arbitrary cardinality, $\mathfrak S_{\mathrm{Fin}}$ is not a finite SymTFT in the usual finite-semisimple sense; it is a countably semisimple, discrete non-compact SymTFT.  A choice of topological boundary condition $\mathsf B_S$ labelled by a finite set $S$ produces the absolute closed TFT $\mathcal T_S$ described above.  Thus
\begin{equation}
    \mathcal T_S(S^1)
    =
    \mathcal A_S
    =
    \operatorname{Fun}(S,\mathbb C)
\end{equation}
is the closed algebra of the fixed-cap absolute theory, not the parent SymTFT itself.

This is analogous to a finite discrete gauge-theory SymTFT.  For example, in 2D $\mathbb Z_N$ BF theory, with schematic action
\begin{equation}
    \frac{iN}{2\pi}\int a_0\, d b_1,
\end{equation}
one fixed SymTFT controls a family of absolute theories labelled by a discrete parameter in $\mathbb Z_N$ .  A topological boundary condition selects one member of that family.  In the present finite-set model, the discrete label $r\in\mathbb Z_N$ is replaced by a finite set $S$, and the selected absolute theory is $\mathcal T_S$.  

\paragraph{The physical boundary and the topological boundary conditions.}
\label{par:physical-boundary-and-finite-set-caps-MM}

We next specify the physical boundary condition.  In the closed-string sector of the
Marolf--Maxfield model there is a single asymptotic boundary condition, corresponding to
one circular boundary and one observable $Z$.  The physical boundary should not be understood as a boundary condition of one already chosen absolute TFT $\mathcal T_S$.  It is instead a universal identity section over the groupoid of topological caps.  After the topological cap $\mathsf B_S$ is chosen, this same physical boundary is represented in the fixed-cap theory $\mathcal T_S$ by the unit state:
\begin{equation}
    B_{\mathrm{phys}}(S^1):
    \qquad
    S \longmapsto
    |\Psi_{\mathrm{phys}}(S^1)\rangle_S
    =
    1_S
    \in \mathcal A_S .
    \label{eq:physical-boundary-prepares-unit-MM}
\end{equation}
Here the subscript $S$ on the state is only a reminder that this is the representation of the same physical boundary after the topological cap $\mathsf B_S$ has been chosen.  Thus $B_{\mathrm{phys}}$ itself is not $S$-dependent; only its fixed-cap description is.

The topological boundary condition determined by the finite set $S$, which we denote by $\mathsf B_S$, is the cap of the parent theory $\mathfrak S_{\mathrm{Fin}}$.  In the fixed-cap absolute description, this cap is represented by the Frobenius counit $\varepsilon_S:
    \mathcal A_S \longrightarrow \mathbb C$ in \eqref{eq:finite-set-frobenius-structure-MM}:
\begin{equation}
   B_{\mathrm{top}}(S^1): \langle \mathsf B_S;S^1|
    =
    \varepsilon_S.
    \label{eq:topological-cap-is-counit-MM}
\end{equation}
The resulting partition function by pairing the physical boundary \eqref{eq:physical-boundary-prepares-unit-MM} and topological boundary \eqref{eq:topological-cap-is-counit-MM} is
\begin{equation}
    Z_S(S^1)
    =
    \langle \mathsf B_S;S^1|\Psi_{\mathrm{phys}}(S^1)\rangle_S
    =
    \varepsilon_S(1_S)
    =
    \sum_{s\in S}\varepsilon_S(e_s)
    =
    |S| .
    \label{eq:one-boundary-partition-function-S-MM}
\end{equation}
For $S=\varnothing$, this formula gives $Z_\varnothing(S^1)=0$, which aligns with our convention that $\mathcal A_\varnothing=0$ is the formal zero-dimensional sector.

It is now straightforward to see that a fixed topological boundary condition $\mathsf B_S$, labelled by the finite set $S$, realizes an $\alpha$-sector in which the closed-boundary
baby-universe operator has eigenvalue
\begin{equation}
    \widehat Z(S^1)\,|S\rangle
    =
   Z_S(S^1) \,|S\rangle= |S|\,|S\rangle .
    \label{eq:Zhat-S-eigenvalue-MM}
\end{equation}
For the representative $[d]=\{1,\ldots,d\}, d\geq 0$, we write $|d\rangle:=|[d]\rangle$, which allows us to  reproduce \eqref{eq:MM-Zb-spectrum}
\begin{equation}
    \widehat Z(S^1)\,|d\rangle
    =
    d\,|d\rangle .
    \label{eq:Zhat-d-eigenvalue-MM}
\end{equation}

\paragraph{The groupoid of topological boundary conditions.}
\label{par:finite-set-cap-groupoid-MM}

By definition of the parent finite-set SymTFT analogue $\mathfrak S_{\mathrm{Fin}}$, a simple topological boundary condition is labelled by a finite set $S$.  We now discuss how to sum over all such topological boundary conditions to obtain an ensemble average.  As discussed in general strategy in Section~\ref{subsec:weights-and-measures}, we must specify what it means to sum over such topological boundary conditions, which amounts to determine the weight/measure structure over the space of topological boundary conditions.  

The point is
that the elements of \(S\) do not enjoy a specific way of labeling.  Relabelling the points of \(S\) gives an
isomorphic Frobenius algebra and hence the same topological boundary condition.  Therefore the space
of topological boundary conditions is not merely a set, but a groupoid, whose objects and morphisms are
\begin{equation}
    \begin{split}
         \text{objects (0-morphisms):}& \qquad \text{finite sets}\\
        \text{1-morphisms:}& \qquad \text{bijections between sets}
    \end{split}
\end{equation}

In the Frobenius-algebra language, a bijection \(f:S\to S'\) induces an isomorphism
\begin{equation}
    f^*:
    \mathcal A_{S'}
    \longrightarrow
    \mathcal A_S,
    \qquad
    e_{s'}
    \longmapsto
    e_{f^{-1}(s')} .
    \label{eq:bijection-induces-frobenius-isomorphism-MM}
\end{equation}
Conversely, every algebra automorphism of \(\mathcal A_S=\operatorname{Fun}(S,\mathbb C)\)
permutes the primitive idempotents \(e_s\).\footnote{
An idempotent is an element \(p\) satisfying \(p^2=p\).  It is called primitive if it is
nonzero and cannot be decomposed as \(p=p_1+p_2\), where \(p_1,p_2\) are nonzero
orthogonal idempotents, \(p_i^2=p_i\) and \(p_1p_2=0\).  In
\(\mathcal A_S=\operatorname{Fun}(S,\mathbb C)\), the idempotents are characteristic
functions of subsets of \(S\), and the primitive idempotents are precisely the delta
functions \(e_s\) supported at single points \(s\in S\).
}
Since all primitive idempotents have the same Frobenius trace,
\begin{equation}
    \varepsilon_S(e_s)=1 ,
    \label{eq:equal-trace-primitive-idempotents-MM}
\end{equation}
every permutation preserves the Frobenius counit, and it also preserves the
comultiplication \(\Delta_S(e_s)=e_s\otimes e_s\).  Hence
\begin{equation}
    \operatorname{Aut}_{\mathrm{Frob}}(\mathcal A_S)
    \cong
    \operatorname{Aut}(S)
    \cong
    \mathfrak S_{|S|}.
    \label{eq:frobenius-automorphism-symmetric-group-MM}
\end{equation}

Thus the groupoid of topological boundary conditions of $\mathfrak S_{\mathrm{Fin}}$ is the groupoid of finite sets,
\begin{equation}
    \mathfrak B_{\mathrm{top}}
    \simeq
    \mathsf{FinSet}^{\simeq}.
    \label{eq:cap-groupoid-is-FinSet-MM}
\end{equation}
After choosing the representative \([d]=\{1,\ldots,d\}\) in each
isomorphism class, this groupoid decomposes as
\begin{equation}
    \mathfrak B_{\mathrm{top}}
    \simeq
    \mathsf{FinSet}^{\simeq}
    \simeq
    \coprod_{d\geq 0} B\mathfrak S_d .
    \label{eq:cap-groupoid-FinSet-MM}
\end{equation}
Here \(B\mathfrak S_d\) denotes the one-object groupoid whose automorphism group is
\(\mathfrak S_d =\mathfrak S_{|S|}    \cong \operatorname{Aut}(S)\).  In other words, there is one finite-set topological boundary condition of each cardinality \(d\), but the topological boundary condition
with \(d\) points has an automorphism group \(\mathfrak S_d\) coming from relabellings of
the points.  The Frobenius algebras $\mathcal A_S$ are the absolute closed TFTs obtained after choosing these caps.

This is the origin of the weighted factor in the topological-boundary sum.  For any function \(F\) of the topological-boundary-condition
data, the groupoid-counting prescription gives
\begin{equation}
    \int_{\mathfrak B_{\mathrm{top}}} F
    :=
    \sum_{[S]\in \pi_0(\mathfrak B_{\mathrm{top}})}
    \frac{1}{|\operatorname{Aut}(S)|}\,F(S)
    =
    \sum_{d=0}^{\infty}
    \frac{1}{|\mathfrak S_d|}\,F([d])
    =
    \sum_{d=0}^{\infty}
    \frac{1}{d!}\,F([d]).
    \label{eq:finite-set-groupoid-integral-MM}
\end{equation}

\paragraph{The groupoid average and (non-)factorization.}
\label{par:finite-set-groupoid-average-MM}

We are now ready to reproduce the Marolf--Maxfield closed-string sector
ensemble.  In addition to the groupoid-counting factor, we assign a fugacity
\(\lambda\) to each point of the finite set \(S\), or equivalently to each primitive
idempotent of the fixed-cap algebra \(\mathcal A_S\).  The factor $1/|\operatorname{Aut}(S)|$ is the intrinsic groupoid-counting factor, while $\lambda^{|S|}$ is the fugacity, equivalently the choice of Hartle--Hawking/ensemble state in this discrete model.  Thus the unnormalized boundary-condition measure is
\begin{equation}
    d\mu_\lambda(S)
    =
    \frac{\lambda^{|S|}}{|\operatorname{Aut}(S)|}.
    \label{eq:finite-set-cap-measure-MM}
\end{equation}
Here \(\operatorname{Aut}(S)\) is the group of bijections from \(S\) to itself.  The
normalization is the weighted groupoid cardinality
\begin{equation}
    \mathcal N_\lambda
    =
    \sum_{[S]\in \pi_0(\mathfrak B_{\mathrm{top}})}
    \frac{\lambda^{|S|}}{|\operatorname{Aut}(S)|}
    =
    \sum_{d=0}^{\infty}
    \frac{\lambda^d}{|\mathfrak S_d|}
    =
    \sum_{d=0}^{\infty}
    \frac{\lambda^d}{d!}
    =
    e^\lambda .
    \label{eq:finite-set-normalization-MM}
\end{equation}

The normalized one-boundary average is then
\begin{equation}
    \bigl\langle Z\bigr\rangle_{\mathrm{top}}
    =
    \frac{1}{\mathcal N_\lambda}
    \sum_{[S]\in\pi_0(\mathfrak B_{\mathrm{top}})}
    \frac{\lambda^{|S|}}{|\operatorname{Aut}(S)|}
    Z_S(S^1).
    \label{eq:one-boundary-cap-average-raw-MM}
\end{equation}
Using \(Z_S(S^1)=\langle \mathsf B_S;S^1|\Psi_{\mathrm{phys}}(S^1)\rangle_S=|S|\), this becomes
\begin{equation}
    \bigl\langle Z\bigr\rangle_{\mathrm{top}}
    =
    e^{-\lambda}
    \sum_{d=0}^{\infty}
    \frac{\lambda^d}{d!}\,d
    =
    \lambda .
    \label{eq:one-boundary-cap-average-MM}
\end{equation}
This agrees with the
Marolf--Maxfield closed-string sector result
\begin{equation}
    \frac{\left\langle Z\right\rangle_{\mathrm{MM}}}
    {\left\langle 1\right\rangle_{\mathrm{MM}}}
    =
    \lambda .
    \label{eq:one-boundary-MM-match-MM}
\end{equation}
The integer \(d\) in the
Marolf--Maxfield \(\alpha\)-sector is thus interpreted as the cardinality of the finite-set topological boundary condition, i.e., $d=|S|$, while 
the Poisson probability is the normalized groupoid measure on finite-set topological boundary conditions with fugacity
\(\lambda\):
\begin{equation}
    p_d
    =
    \frac{1}{\mathcal N_\lambda}
    \frac{\lambda^d}{|\operatorname{Aut}([d])|}
    =
    \frac{1}{\mathcal N_\lambda}
    \frac{\lambda^d}{|\mathfrak S_d|}
    =
    e^{-\lambda}
    \frac{\lambda^d}{d!}.
    \label{eq:Poisson-from-groupoid-measure-MM}
\end{equation}

It is also straightforward to generalize to the case with multiple boundary components. Concretely, we consider \(n\) labelled closed boundaries as 
\begin{equation}
    M_n
    =
    \bigsqcup_{i=1}^n S^1_i ,
    \label{eq:n-labelled-circles-MM}
\end{equation}
where $S_i^1$ labels the $i$-th $S^1$ component of the boundary manifold $M_n$. After capping by $\mathsf B_S$, the state space on this boundary manifold is a tensor product Hilbert space 
\begin{equation}
    \mathcal{H}_{\mathcal{T}_S}(M_n)=\mathcal{A}_S^{\otimes n}
\end{equation}
The same universal physical boundary and the topological cap are then represented, in the fixed-cap theory, by
\begin{equation}
    |\Psi_{\mathrm{phys}}(M_n)\rangle_S
    =
    1_S^{\otimes n}
    \in
    \mathcal A_S^{\otimes n},
    \qquad
    \langle \mathsf B_S;M_n|
    =
    \varepsilon_S^{\otimes n}.
    \label{eq:fixed-S-n-boundary-state-and-cap-MM}
\end{equation}
Therefore, the partition function for a given topological boundary condition $\langle \mathsf B_S; M_n|$ reads
\begin{equation}
    Z_S(M_n)
    =
    \langle \mathsf B_S;M_n|\Psi_{\mathrm{phys}}(M_n)\rangle_S
    =
    \prod_{i=1}^n Z_S(S^1_i)
    =
    |S|^n .
    \label{eq:fixed-S-n-boundary-factorization-MM}
\end{equation}
This corresponds to the \(\alpha\)-sector represented by \([d]\) with \(d=|S|\),
\begin{equation}
    \langle d|\widehat Z(S^1)^n|d\rangle
    =
    d^n .
    \label{eq:fixed-d-nth-moment-MM}
\end{equation}

The normalized average over the space of topological boundary conditions is thus
\begin{equation}
    \bigl\langle Z^n\bigr\rangle_{\mathrm{top}}
    =
    \frac{1}{\mathcal N_\lambda}
    \sum_{[S]\in\pi_0(\mathfrak B_{\mathrm{top}})}
    \frac{\lambda^{|S|}}{|\operatorname{Aut}(S)|}
    |S|^n .
    \label{eq:n-boundary-cap-average-raw-MM}
\end{equation}
Under the representatives \([d]\), this becomes
\begin{equation}
    \bigl\langle Z^n\bigr\rangle_{\mathrm{top}}
    =
    e^{-\lambda}
    \sum_{d=0}^{\infty}
    \frac{\lambda^d}{d!}\,d^n ,
    \label{eq:n-boundary-cap-average-MM}
\end{equation}
which is exactly the \(n\)-th moment of a Poisson random variable of mean \(\lambda\).  Hence, we obtain 
\begin{equation}
    \bigl\langle Z^n\bigr\rangle_{\mathrm{top}}
=B_n(\lambda)=\frac{\left\langle Z^n\right\rangle_{\mathrm{MM}}}
    {\left\langle 1\right\rangle_{\mathrm{MM}}},
    \label{eq:cap-average-Bell-polynomial-MM}
\end{equation}
reproducing the
Marolf--Maxfield moment formula as the Bell polynomial. 
The generating function is likewise
\begin{equation}
    \bigl\langle e^{uZ}\bigr\rangle_{\mathrm{top}}
    =
    e^{-\lambda}
    \sum_{d=0}^{\infty}
    \frac{\lambda^d}{d!}\,e^{ud}
    =
    \exp\!\left[\lambda(e^u-1)\right],
    \label{eq:cap-generating-function-MM}
\end{equation}
again matching the normalized Marolf--Maxfield result.

This also provides a SymTFT perspective for the (non-)factorization.  Without loss of generality, consider two $S^1$ components for the boundary. At a fixed topological boundary condition $\mathsf B_S$ labelled by \(S\), or equivalently in the fixed-cap absolute theory with algebra $\mathcal{A}_S$, we have
\begin{equation}
    Z_S(S^1\sqcup S^1)
    =
    Z_S(S^1)^2
    =
    |S|^2 .
    \label{eq:fixed-S-two-boundary-factorization-MM}
\end{equation}
But after averaging over topological boundary conditions, i.e. over the caps $\mathsf B_S$ or equivalently over finite sets $S$ up to bijection, we have
\begin{equation}
    \bigl\langle Z^2\bigr\rangle_{\mathrm{top}}
    =
    e^{-\lambda}
    \sum_{d=0}^{\infty}
    \frac{\lambda^d}{d!}\,d^2
    =
    \lambda^2+\lambda ,
    \label{eq:cap-second-moment-MM}
\end{equation}
whereas, \eqref{eq:one-boundary-cap-average-MM} gives rise to
\begin{equation}
    \bigl\langle Z\bigr\rangle_{\mathrm{top}}^2
    =
    \lambda^2 .
    \label{eq:cap-first-moment-squared-MM}
\end{equation}
Therefore, the connected two-boundary correlator is
\begin{equation}
    \bigl\langle Z^2\bigr\rangle_{\mathrm{top}}
    -
    \bigl\langle Z\bigr\rangle_{\mathrm{top}}^2
    =
    \lambda \neq 0.
    \label{eq:cap-connected-two-point-MM}
\end{equation}
This is the wormhole contribution in the Marolf--Maxfield model,
reproduced in the SymTFT language as the variance of the finite-set cardinality \(|S|\) over the space of topological boundary conditions\footnote{This is reminiscent of the entangled state in SymTFT with multiple physical boundaries and its resulting non-factorization in \cite{2510.06319}.}.

We close this section by pointing out that a careful treatment of Marolf--Maxfield model via summing over TFT bordisms is discussed in \cite{2201.00903}. It would be interesting to build an explicit correspondence between that treatment with the SymTFT perspective in this section, which we leave it as a future work.

\section{A SymTFT Realization of Averaging over Narain Moduli}
\label{sec:narain-symtft}

We now turn from the discrete example of the previous section to a genuinely
continuous example. The guiding principle is the same as before. We keep fixed a
relative theory, or equivalently a fixed SymTFT together with a fixed physical boundary
condition, and we vary the topological boundary condition used to cap the SymTFT
slab. The difference is that the relevant topological boundary conditions now form a
continuous space rather than a finite groupoid. Thus the groupoid sum of Section
\ref{sec:MM-from-SymTFT} is replaced by an integral over a space of topological boundary conditions.

In this section we use the central charge \(c=c_L=c_R\) to denote the rank of the Narain lattice, or equivalently the
dimension of the target space torus $T^c$ in the boundary 2D compact bosons. The SymTFT we use is the 3D
abelian BF theory with $\mathbb{R}$-valued gauge fields \cite{2401.10165,2401.06128}
\begin{equation}
  S_{\mathrm{BF}}
  =
  {1\over 2\pi}
  \int_{X_3}
  \sum_{i=1}^c a_i \wedge d b_i ,
  \qquad
  a_i,b_i\in \Omega^1(X_3;\mathbb R).
  \label{eq:RBF-action-d}
\end{equation}
For \(c=1\) this is the \(\mathbb{R}\)-valued SymTFT of the compact boson. For general \(c\),
it is the natural continuous analogue of the finite abelian SymTFTs discussed in the
Chern--Simons/code-CFT literature \cite{2310.06012,2310.13044,Barbar:2025krh}.
A closely related recent discussion of flat gauging in compact boson theories and of the
non-compact BF SymTFT description of Narain moduli and \(O(c,c;\mathbb Z)\) duality
appears in \cite{Jia:2026tfh}.

The theory has topological line operators
\begin{equation}
  U_{\vec{\alpha}}[\gamma]
  =
  \exp\left(i\alpha_i\oint_\gamma  a_i\right),
  \qquad
  V_{\vec{\beta}}[\gamma]
  =
  \exp\left(i\beta_i\oint_\gamma  b_i\right),
  \qquad
  \vec{\alpha},\vec{\beta} \in \mathbb{R}^c .
  \label{eq:RBF-lines-d}
\end{equation}
Since the gauge group is \(\mathbb R\), there are no large gauge transformations, and
\(\alpha,\beta\) are real rather than integral labels. The braiding of two such lines is given by
\begin{equation}
   \left\langle
  U_\alpha[\gamma_1] V_\beta[\gamma_2]
  \right\rangle
  =
  \exp\left(
    2\pi i\,\vec{\alpha}\cdot \vec{\beta}\,\mathrm{Link}(\gamma_1,\gamma_2)
  \right).
  \label{eq:RBF-braiding-general}
\end{equation}

The defect group is
\begin{equation}
  \mathbb{D}_c = \mathbb{R}^{c}\oplus \mathbb{R}^{c}
\end{equation}
equipped with the Dirac pairing
\begin{equation}
  \big\langle (\alpha,\beta),(\alpha',\beta')\big\rangle
  =
  \alpha\cdot \beta' + \alpha'\cdot \beta
  \quad \mathrm{mod}\ \mathbb Z .
  \label{eq:RBF-defect-pairing}
\end{equation}
The topological boundary conditions, in this continuous setting, correspond to maximal isotropic subgroups (or Lagrangian subgroups,
analogue of Lagrangian algebras for finite semisimple TFTs) \cite{hep-th/9812012, 2010.15890, 2203.09537, 2306.11783}
\begin{equation}
  L \subset \mathbb{D}_c
\end{equation}
with respect to \eqref{eq:RBF-defect-pairing}. Physically, a topological boundary condition is a Dirichlet boundary condition trivializing a maximal subset of lines operators $U_{\vec{\alpha}}V_{\vec{\beta}}$ with trivial braiding.  We will continue to occasionally call such an
object a Lagrangian boundary condition, although one should keep in mind that this
is a continuous, non-finite version of the usual finite semisimple terminology.

Let \(\Sigma\) be the physical 2D boundary, equipped with the moduli data (e.g., $\tau$ for a $T^2$) collectively denoted by \(\Omega\). It gives rise to a conformal boundary condition for the 3D BF theory, equipped with local Hamiltonian of $c$ compact bosons as well as their conformal blocks. From the BF theory quantization perspective, the physical boundary condition prepares a state in the state space on $\Sigma$
\begin{equation}
  \left|\Psi_{\mathrm{phys}}(\Sigma;\Omega)\right\rangle
  \in
  \mathcal H_{\mathrm{BF}}(\Sigma),
\end{equation}
and a topological boundary condition \(L\) defines a cap functional
\begin{equation}
  \langle L;\Sigma|
  \in
  \mathcal H_{\mathrm{BF}}(\Sigma)^* .
\end{equation}
The absolute theory obtained by capping the slab with \(L\) has partition function
\begin{equation}
  Z_L(\Sigma;\Omega)
  =
  \langle L;\Sigma|\Psi_{\mathrm{phys}}(\Sigma;\Omega)\rangle ,
  \label{eq:ZL-pairing}
\end{equation}
which is the partition function of the Narain CFT on 2-manifold $\Sigma$ with modulus parameter $\Omega$.

In this section, we will show from the compact Narain perspective, the possible \(L\)'s are precisely the data that
determine points on the moduli space of Narain compactification. More precisely, naive space of topological boundary conditions will be covering space for the moduli space, up to certain limit points. After taking into account the symmetry/reparameterization of the SymTFT, the genuine space of topological boundary conditions to be averaged over is
\begin{equation}
  \mathcal L_{\mathrm{Narain}}^{(c)}
  \simeq
  O(c,c;\mathbb Z)\backslash O(c,c;\mathbb R)/
  \bigl(O(c)\times O(c)\bigr).
  \label{eq:narain-lagrangian-space}
\end{equation}
This is exactly the usual Narain moduli space \cite{Narain:1985jj,Narain:1986am}. The continuous integral measure for ensemble averaging discussed
in Section \ref{subsec:weights-and-measures} is the invariant \emph{Haar measure} \cite{9006cc9e-2dcc-3fd8-aada-e4af19b6e225} on this quotient, which
is the same measure that obtained from the Zamolodchikov metric in \cite{2006.04855}. 

The resulting topological-boundary average takes the form
\begin{equation}
\begin{split}
    \big\langle Z_\Sigma(\Omega)\big\rangle_{\mathrm{top}}
  &=
  {1\over \mathrm{Vol}(\mathcal L_{\mathrm{Narain}}^{(d)})}
  \int_{\mathcal L_{\mathrm{Narain}}^{(d)}} d\mu(L)\,
  Z_L(\Sigma;\Omega), \\
  &=
  {1\over \mathrm{Vol}(\mathcal L_{\mathrm{Narain}}^{(d)})}
  \int_{\mathcal L_{\mathrm{Narain}}^{(d)}} d\mu(L)\,
  \langle L;\Sigma|\Psi_{\mathrm{phys}}(\Sigma;\Omega)\rangle .
\end{split}
  \label{eq:narain-topological-average}
\end{equation}
This is the continuous SymTFT analogue of the discrete SymTFT average in Section
\ref{sec:MM-from-SymTFT}. The ensemble is not an average over arbitrary states in
\(\mathcal H_{\mathrm{BF}}(\Sigma)\), nor over arbitrary boundary dynamics. It is an
average over topological completions of one fixed relative theory.

There are also boundary/infinity points of the space of Lagrangian subgroups. Physically these
correspond to partial decompactification limits, i.e., noncompact bosons and their noncompact winding-mode
duals. The compact Narain average is obtained by integrating over the compact
Narain space of Lagrangian boundaries \eqref{eq:narain-lagrangian-space}. The role of the noncompact point is
particularly visible for \(c=1\), where the formal average diverges. We now discuss
this rank-one case in detail.

\subsection{\texorpdfstring{\(c=1\)}{c=1}}
\label{subsec:narain-d1}

For \(c=1\), the SymTFT action \eqref{eq:RBF-action-d} reduces to the
\(\mathbb R\)-valued BF theory \cite{2401.10165}
\begin{equation}
  S_{\mathrm{BF}}
  =
  {1\over 2\pi}
  \int_{X_3} a\wedge db,
  \qquad
  a,b\in \Omega^1(X_3;\mathbb R).
  \label{eq:RBF-action-c1}
\end{equation}
The topological line
operators are
\begin{equation}
  U_\alpha[\gamma]
  =
  \exp\left(i\alpha\oint_\gamma a\right),
  \qquad
  V_\beta[\gamma]
  =
  \exp\left(i\beta\oint_\gamma b\right),
  \qquad
  \alpha,\beta\in\mathbb R .
  \label{eq:RBF-lines-c1}
\end{equation}
Their braiding is
\begin{equation}
  \left\langle
  U_\alpha[\gamma_1] V_\beta[\gamma_2]
  \right\rangle
  =
  \exp\left(
    2\pi i\,\alpha\beta\,\mathrm{Link}(\gamma_1,\gamma_2)
  \right).
  \label{eq:RBF-braiding-c1}
\end{equation}
Equivalently, the defect group is
\begin{equation}
  \mathbb{D}_1=\mathbb R\oplus \mathbb R
\end{equation}
with Dirac pairing
\begin{equation}
  \big\langle(\alpha,\beta),(\alpha',\beta')\big\rangle
  =
  \alpha\beta' + \alpha'\beta
  \quad \mathrm{mod}\ \mathbb Z .
  \label{eq:defect-pairing-c1}
\end{equation}

\paragraph{Conformal boundary condition as the physical boundary.} Let us first specify the physical boundary 
condition that prepares the state
\begin{equation}
  |\Psi_{\mathrm{phys}}(\Sigma;\Omega)\rangle
  \in
  \mathcal H_{\mathrm{BF}}(\Sigma).
  \label{eq:c1-physical-state}
\end{equation}
This is the fixed conformal boundary condition at the physical end of the SymTFT slab.
Introduce the linear combinations
\begin{equation}
  A_{\mathrm L}=a+b,
  \qquad
  A_{\mathrm R}=a-b .
  \label{eq:AL-AR-definition-c1}
\end{equation}
On the physical boundary \(\Sigma\), with local complex coordinates \(z,\bar z\), the
conformal boundary condition is
\begin{equation}
  (A_{\mathrm L})_{\bar z}\big|_\Sigma = 0,
  \qquad
  (A_{\mathrm R})_{z}\big|_\Sigma = 0.
  \label{eq:c1-conformal-boundary-condition}
\end{equation}
Equivalently,
\begin{equation}
  (a+b)_{\bar z}\big|_\Sigma=0,
  \qquad
  (a-b)_z\big|_\Sigma=0 .
  \label{eq:c1-conformal-boundary-condition-ab}
\end{equation}
Thus \(A_{\mathrm L}\) supplies the left-moving current algebra, while
\(A_{\mathrm R}\) supplies the right-moving current algebra. If one works on the
Lorentzian boundary cylinder \(\Sigma=S^1_\varphi\times\mathbb R_t\), this same
condition becomes
\begin{equation}
  (a_t+b_t)\big|_\Sigma=(a_\varphi+b_\varphi)\big|_\Sigma,
  \qquad
  (a_t-b_t)\big|_\Sigma=-(a_\varphi-b_\varphi)\big|_\Sigma,
  \label{eq:c1-lorentzian-conformal-bc}
\end{equation}
or, equivalently,
\begin{equation}
  a_t\big|_\Sigma=b_\varphi\big|_\Sigma,
  \qquad
  b_t\big|_\Sigma=a_\varphi\big|_\Sigma .
  \label{eq:c1-lorentzian-conformal-bc-ab}
\end{equation}

The corresponding physical boundary Hamiltonian is fixed once and for all by this
conformal boundary condition. In the normalization of \eqref{eq:RBF-action-c1}, it is
\begin{equation}
  H_{\mathrm{phys}}
  =
  {1\over 4\pi}
  \int_{S^1} d\varphi\,
  :\!\left(a_\varphi^2+b_\varphi^2\right)\!: ,
  \label{eq:c1-boundary-hamiltonian}
\end{equation}
whose corresponding spatial momentum reads
\begin{equation}
  P_{\mathrm{phys}}
  =
  {1\over 2\pi}
  \int_{S^1} d\varphi\,
  :\!a_\varphi b_\varphi\!: .
  \label{eq:c1-boundary-momentum}
\end{equation}
Let \(\mathsf P_{\mathrm L}\) and \(\mathsf P_{\mathrm R}\) denote the zero-mode
charge operators of the left- and right-moving current algebras supplied by
\(A_{\mathrm L}\) and \(A_{\mathrm R}\). A bulk line labelled by
\((\alpha,\beta)\in\mathbb{D}_1\) specifies a charge sector of the physical boundary
theory. Here \(\alpha\) and \(\beta\) are real charge labels, not dynamical fields. In the
charge sector labelled by \((\alpha,\beta)\), the zero-mode operators
\(\mathsf P_{\mathrm L}\) and \(\mathsf P_{\mathrm R}\) have eigenvalues
\begin{equation}
  p_{\mathrm L}(\alpha,\beta)=\alpha+\beta,
  \qquad
  p_{\mathrm R}(\alpha,\beta)=\alpha-\beta .
  \label{eq:c1-left-right-momenta}
\end{equation}
Equivalently, the Virasoro zero-mode operators are
\begin{equation}
  L_0
  =
  {1\over 4}\mathsf P_{\mathrm L}^2+N_{\mathrm L},
  \qquad
  \bar L_0
  =
  {1\over 4}\mathsf P_{\mathrm R}^2+N_{\mathrm R}.
  \label{eq:c1-L0-L0bar-operator}
\end{equation}
Therefore, in the charge sector \((\alpha,\beta)\), their eigenvalues are
\begin{equation}
  h_{\alpha,\beta;N_{\mathrm L}}
  =
  {1\over 4}\bigl(\alpha+\beta\bigr)^2+N_{\mathrm L},
  \qquad
  \bar h_{\alpha,\beta;N_{\mathrm R}}
  =
  {1\over 4}\bigl(\alpha-\beta\bigr)^2+N_{\mathrm R}.
  \label{eq:c1-L0-L0bar-sector}
\end{equation}
We then have the familiar relation
\begin{equation}
  H_{\mathrm{phys}}
  =
  L_0+\bar L_0-{1\over 12},
  \qquad
  P_{\mathrm{phys}}
  =
  L_0-\bar L_0 .
  \label{eq:c1-H-P-sector}
\end{equation}
The radius \(R\) has not appeared
in the local physical boundary dynamics. It will enter only through the topological boundary condition,
which selects the allowed charge sectors.

\paragraph{Continuous compact topological boundary conditions.}
We now turn to the topological boundary conditions. As discussed in \cite{2401.10165}, there are boundary conditions realizing $U(1)\times U(1)$ symmetry in 2D QFTs, as well as two boundary conditions realizing $\mathbb{R}$ symmetries. In the former case, the topological
boundary conditions are associated to the maximal isotropic subgroups of the defect group $\mathbb{D}_1$:
\begin{equation}
  L_R
  =
  \left\{
    \left({n\over R},wR\right)
    \;:\;
    n,w\in\mathbb Z
  \right\},
  \qquad
  R\in\mathbb R_{>0}.
  \label{eq:LR-definition}
\end{equation}
Here \(R\) will play the role as the compactification radius of the target $S^1$, in the convention where the self-dual radius is
\(R=1\). The identification of the one-parameter family of compact Lagrangians \(L_R\) with the radius modulus of the \(c=1\) compact boson was also emphasized by H.~T.~Lam in talks \cite{Lam:KITP2025,Lam:Harvard2025}.

Let us check that \(L_R\) is Lagrangian. For two elements of \(L_R\),
\begin{equation}
  x_1=\left({n_1\over R},w_1R\right),
  \qquad
  x_2=\left({n_2\over R},w_2R\right),
\end{equation}
the pairing is
\begin{equation}
  \langle x_1,x_2\rangle
  =
  n_1w_2+n_2w_1
  \in \mathbb Z ,
\end{equation}
so \(L_R\) is isotropic. Conversely, suppose that
\((\alpha,\beta)\in\mathbb R^2\) pairs trivially with every element of \(L_R\), i.e.,
\begin{equation}
  \alpha\,wR + {n\over R}\,\beta \in \mathbb Z
  \qquad
  \forall\, n,w\in\mathbb Z .
\end{equation}
Setting \(n=0\) gives \(\alpha R\in\mathbb Z\), and setting \(w=0\) gives
\(\beta/R\in\mathbb Z\). Therefore
\begin{equation}
  \alpha\in R^{-1}\mathbb Z,
  \qquad
  \beta\in R\mathbb Z,
\end{equation}
which exactly means \((\alpha,\beta)\in L_R\). Hence \(L_R\) is also maximal.

The boundary condition \(L_R\) trivializes the lines
\begin{equation}
  U_{n/R},
  \qquad
  V_{wR},
  \qquad n,w\in\mathbb Z .
\end{equation}
More precisely, for each element
\begin{equation}
  \left({n\over R},wR\right)\in L_R ,
\end{equation}
the composite bulk line
\begin{equation}
  \mathcal U_{n,w}
  :=
  U_{n/R}V_{wR}
  \label{eq:c1-condensed-line-primary}
\end{equation}
can end on the topological boundary. Take into account also the other endpoint on the
physical boundary, it creates a local operator $\mathcal O_{n,w}$ of the absolute 2D compact boson CFT.
These are precisely the primary fields of the compact boson at radius
\(R\). In the convention used above, their left- and right-moving zero-mode eigenvalues are
\begin{equation}
  p_{\mathrm L}={n\over R}+wR,
  \qquad
  p_{\mathrm R}={n\over R}-wR,
  \label{eq:c1-primary-left-right-momenta}
\end{equation}
and hence
\begin{equation}
  h_{n,w}
  =
  {1\over 4}\left({n\over R}+wR\right)^2,
  \qquad
  \bar h_{n,w}
  =
  {1\over 4}\left({n\over R}-wR\right)^2,
  \label{eq:c1-primary-dimensions}
\end{equation}
before adding oscillator excitations. 

The surviving topological line operators on the boundary are therefore labelled by
\begin{equation}
  \alpha\in \mathbb R/R^{-1}\mathbb Z \cong U(1),
  \qquad 
  \beta\in \mathbb R/R\mathbb Z \cong U(1) .
  \label{eq:surviving-lines-c1}
\end{equation}
These two \(U(1)\)'s are the momentum and winding symmetries of the compact boson.
Their action on the local primary \(\mathcal O_{n,w}\) is determined by the braiding in
the \(\mathbb R\)-valued BF theory. A surviving line \(U_\alpha\) linked with the
endpoint of \(\mathcal U_{n,w}\) gives
\begin{equation}
  U_\alpha:\quad
  \mathcal O_{n,w}
  \longmapsto
  \exp\!\left(2\pi i\,\alpha\,wR\right)\mathcal O_{n,w},
  \label{eq:c1-winding-charge-action}
\end{equation}
while a surviving line \(V_\beta\) gives
\begin{equation}
  V_\beta:\quad
  \mathcal O_{n,w}
  \longmapsto
  \exp\!\left(2\pi i\,{\beta n\over R}\right)\mathcal O_{n,w}.
  \label{eq:c1-momentum-charge-action}
\end{equation}
Equivalently, writing
\begin{equation}
  \alpha={\vartheta_{\mathrm w}\over R},
  \qquad
  \beta=\vartheta_{\mathrm m} R,
  \qquad
  \vartheta_{\mathrm w},\vartheta_{\mathrm m}\in \mathbb R/\mathbb Z,
\end{equation}
the two symmetry actions become
\begin{equation}
  U_{\vartheta_{\mathrm w}/R}:\quad
  \mathcal O_{n,w}\longmapsto
  e^{2\pi i\,\vartheta_{\mathrm w} w}\mathcal O_{n,w},
  \qquad
  V_{\vartheta_{\mathrm m}R}:\quad
  \mathcal O_{n,w}\longmapsto
  e^{2\pi i\,\vartheta_{\mathrm m} n}\mathcal O_{n,w}.
  \label{eq:c1-U1mw-charge-action}
\end{equation}
Thus \(\mathcal O_{n,w}\) carries
\begin{equation}
  (Q_{\mathrm m},Q_{\mathrm w})=(n,w)
  \label{eq:c1-primary-U1mw-charges}
\end{equation}
under \(U(1)_{\mathrm m}\times U(1)_{\mathrm w}\), where \(U(1)_{\mathrm m}\) is generated by
the \(V\)-type surviving lines and \(U(1)_{\mathrm w}\) is generated by the \(U\)-type
surviving lines.

\paragraph{The two noncompact topological boundary conditions.}
The above compact boundary conditions \(L_R\) have two degenerate limits \cite{2401.10165},
\begin{equation}
  L_{\infty}
  =
  \mathbb R\oplus 0,
  \qquad
  L_{0}
  =
  0\oplus \mathbb R .
  \label{eq:c1-degenerate-R-boundaries}
\end{equation}
They are again maximal isotropic subgroups of \(\mathbb D_1=\mathbb R\oplus\mathbb R\).
Indeed, \(L_{\infty}\) has trivial self-pairing, and an element
\((\alpha,\beta)\in\mathbb D_1\) pairs trivially with every \((x,0)\in L_{\infty}\)
only if
\begin{equation}
  x\beta\in\mathbb Z
  \qquad
  \forall\,x\in\mathbb R,
\end{equation}
which implies \(\beta=0\).

Let us first consider \(L_{\infty}\). Since \(L_{\infty}\) contains all \(U_\alpha\) lines, they can end on the topological boundary. This leads to 2D local operators $\mathcal O^{(\infty)}_\alpha, \alpha\in\mathbb R$.
They lie in the charge sector
\begin{equation}
  (\alpha,\beta)=(\alpha,0),
\end{equation}
and therefore have left- and right-moving zero-mode eigenvalues
\begin{equation}
  p_{\mathrm L}=\alpha,
  \qquad
  p_{\mathrm R}=\alpha .
  \label{eq:c1-Linfty-momenta}
\end{equation}
Before adding oscillator descendants, their conformal weights are
\begin{equation}
  h_\alpha^{(\infty)}
  =
  {1\over 4}\alpha^2,
  \qquad
  \bar h_\alpha^{(\infty)}
  =
  {1\over 4}\alpha^2 ,
  \label{eq:c1-Linfty-dimensions}
\end{equation}
which precisely match the primaries of the noncompact boson. Equivalently, one may think of them
as the \(R\to\infty\) limit of the compact-boson primaries with \(w=0\), where
\(n/R\) becomes a continuous momentum.

The topological lines that survive on the \(L_\infty\) boundary are the quotient
\begin{equation}
  \mathbb D_1/L_\infty
  \simeq
  0\oplus\mathbb R,
\end{equation}
represented by the \(V_\beta\) lines. Thus the noncompact boson has an
\(\mathbb R\)-valued topological symmetry generated by
\begin{equation}
  V_\beta,
  \qquad
  \beta\in\mathbb R .
\end{equation}
Its action on the local primary \(\mathcal O^{(\infty)}_\alpha\) is determined by the
BF braiding:
\begin{equation}
  V_\beta:\quad
  \mathcal O^{(\infty)}_\alpha
  \longmapsto
  \exp\!\left(2\pi i\,\beta\alpha\right)
  \mathcal O^{(\infty)}_\alpha .
  \label{eq:c1-Linfty-R-symmetry-action}
\end{equation}
Thus \(\mathcal O^{(\infty)}_\alpha\) carries continuous \(\mathbb R\)-charge \(\alpha\).
This is the \(\mathbb R\) symmetry of the noncompact boson.

The other degenerate boundary condition is \(L_0=0\oplus\mathbb R\). Now all
\(V_\beta\) lines can end, and they give rise to local operators $\widetilde{\mathcal O}^{(0)}_\beta, \beta\in\mathbb R$. 
They lie in the charge sector $(\alpha,\beta)=(0,\beta)$
with conformal weights are
\begin{equation}
  h_\beta^{(0)}
  =
  {1\over 4}\beta^2,
  \qquad
  \bar h_\beta^{(0)}
  =
  {1\over 4}\beta^2 .
  \label{eq:c1-L0-dimensions}
\end{equation}
These are the primaries of the noncompact dual boson, or equivalently the \(R\to 0\)
limit of the compact boson. The surviving topological lines on the \(L_0\) boundary are
\begin{equation}
  \mathbb D_1/L_0
  \simeq
  \mathbb R\oplus 0,
\end{equation}
represented by \(U_\alpha\), \(\alpha\in\mathbb R\). Their action on the local primary
\(\widetilde{\mathcal O}^{(0)}_\beta\) is
\begin{equation}
  U_\alpha:\quad
  \widetilde{\mathcal O}^{(0)}_\beta
  \longmapsto
  \exp\!\left(2\pi i\,\alpha\beta\right)
  \widetilde{\mathcal O}^{(0)}_\beta .
  \label{eq:c1-L0-R-symmetry-action}
\end{equation}
Thus \(\widetilde{\mathcal O}^{(0)}_\beta\) carries continuous \(\mathbb R\)-charge
\(\beta\) under the dual \(\mathbb R\) symmetry.

Alternatively, these two degenerate boundary conditions also have a simple interpretation as
topological gaugings of the compact-boson symmetries. 
Flat gauging a finite subgroup of, e.g., \(\mathbb Z_N\subset U(1)_w\) sends
\begin{equation}
  L_R
  \longmapsto
  L_{NR}
  =
  \left\{
    \left({n\over NR},wNR\right)
    :
    n,w\in\mathbb Z
  \right\}.
  \label{eq:c1-ZN-gauging-U}
\end{equation}
Taking \(N\to\infty\), this sequence approaches
\begin{equation}
  L_{NR}
  \longrightarrow
  L_\infty
  =
  \mathbb R\oplus 0 .
  \label{eq:c1-flat-gauging-U-to-Linfty}
\end{equation}
This aligns with the fact that flat gauging the full winding symmetry $U(1)_w$ produces the noncompact boson. Similarly, flat gauging a finite subgroup \(\mathbb Z_N\subset U(1)_m\) sends
\begin{equation}
  L_R
  \longmapsto
  L_{R/N}
  =
  \left\{
    \left({Nn\over R},{wR\over N}\right)
    :
    n,w\in\mathbb Z
  \right\}.
  \label{eq:c1-ZN-gauging-V}
\end{equation}
Taking \(N\to\infty\), this approaches
\begin{equation}
  L_{R/N}
  \longrightarrow
  L_0
  =
  0\oplus\mathbb R,
  \label{eq:c1-flat-gauging-V-to-L0}
\end{equation}
producing the noncompact dual boson by flat gauging the full \(U(1)_m\) symmetry. 
In this sense the two boundary conditions \(L_\infty\) and \(L_0\) are precisely the
SymTFT realizations of the \(R=\infty\) and \(R=0\) endpoints of the
compact-boson radius line.

\paragraph{Duality and integral measure.}
The compact locus of topological boundary conditions contains a discrete redundancy
coming from an automorphism of the \(\mathbb R\)-valued BF SymTFT. Namely, the bulk
topological theory has an electric-magnetic duality as an invertible automorphism 
\begin{equation}
  \mathsf T:\qquad
  a\longleftrightarrow b ,
  \label{eq:c1-T-duality-bulk-automorphism}
\end{equation}
which exchanges the two factors of the defect group
\(\mathbb D_1=\mathbb R\oplus\mathbb R\). On line labels this acts as
\begin{equation}
  \mathsf T:\qquad
  (\alpha,\beta)\longmapsto(\beta,\alpha),
  \qquad
  U_\alpha\longleftrightarrow V_\alpha .
  \label{eq:c1-T-duality-line-action}
\end{equation}
This should be regarded as a duality automorphism of the SymTFT, or equivalently as an
invertible 2D topological interface.

The reason it becomes the usual \(T\)-duality of the absolute 2D theory is
that it acts compatibly with the fixed physical boundary condition. Indeed, using
\begin{equation}
  A_{\mathrm L}=a+b,
  \qquad
  A_{\mathrm R}=a-b ,
\end{equation}
the automorphism \(\mathsf T\) sends
\begin{equation}
  A_{\mathrm L}\longmapsto A_{\mathrm L},
  \qquad
  A_{\mathrm R}\longmapsto -A_{\mathrm R}.
\end{equation}
The second transformation is just charge conjugation of the right-moving current algebra.
Thus, with our fixed physical boundary condition, \(\mathsf T\) maps the absolute theory
obtained from one topological cap to an isomorphic absolute theory obtained from the
transformed cap.

On the compact-boson Lagrangian boundary condition,
\begin{equation}
  L_R
  =
  \left\{
    \left({n\over R},wR\right)
    :
    n,w\in\mathbb Z
  \right\},
\end{equation}
this gives
\begin{equation}
  \mathsf T(L_R)=L_{1/R}.
  \label{eq:c1-T-duality-LR}
\end{equation}
Therefore \(R\) and \(1/R\) label the same absolute compact-boson CFT. In the language of
the space of topological boundary conditions, the \(\mathbb Z_2\) generated by
\(\mathsf T\) is a gauge redundancy of the moduli problem, and we should quotient by it:
\begin{equation}
  \mathcal L_{\mathrm{comp}}^{(1)}
  =
  \mathbb R_{>0}/(R\sim R^{-1})
  \simeq
  [1,\infty).
  \label{eq:c1-lagrangian-moduli}
\end{equation}
At the self-dual point \(R=1\), the \(\mathbb Z_2\) automorphism has a fixed point. Thus the
quotient should be understood as an orbifold\footnote{More precisely as a quotient stack.}, although
this fixed point has measure zero in the continuous integral below.

The two degenerate maximal isotropic subgroups, $L_\infty$ and $L_0$ are exchanged by \(\mathsf T\), aligning with the fact that they correspond to the two endpoints of the radius line. In what follows we first integrate over the compact locus
\eqref{eq:c1-lagrangian-moduli}; the role of the degenerate loci will be visible as a
divergence at the end of the integral.

In addition to this discrete duality, the SymTFT also has a continuous automorphism, which is a reparameterization
\begin{equation}
  a\longmapsto \lambda^{-1}a,
  \qquad
  b\longmapsto \lambda b,
  \qquad
  \lambda\in\mathbb R_{>0}.
  \label{eq:BF-rescaling-c1}
\end{equation}
At the level of labels for topological line operators, this acts as
\begin{equation}
  (\alpha,\beta)
  \longmapsto
  (\lambda^{-1}\alpha,\lambda\beta),
\end{equation}
and therefore scaling the $R$ for topological boundary conditions:
\begin{equation}
  L_R\longmapsto L_{\lambda R}.
  \label{eq:c1-rescaling-LR}
\end{equation}
Unlike the discrete \(T\)-duality above, this continuous automorphism is not quotienting out
the radius modulus of the absolute compact-boson CFT. With the fixed physical boundary condition, it moves us along the family of topological caps and hence along the compact-boson
radius line. Its role is instead to determine the invariant measure on this continuous family of
topological boundary conditions.

Before imposing the \(T\)-duality quotient, the generic compact locus is a free and
transitive \(\mathbb R_{>0}\)-orbit. Therefore, the most natural measure, which is invariant under the continuous automorphism 
\eqref{eq:BF-rescaling-c1} is the Haar measure on \(\mathbb R_{>0}\),
\begin{equation}
  d\mu_C(R)
  =
  C\,{dR\over R},
  \qquad C>0 .
  \label{eq:c1-haar-measure-with-C}
\end{equation}
The constant \(C\) is not fixed by this automorphism. In fact, Haar measure is unique only up to an
overall multiplicative constant.\footnote{
In the \(c=1\) warm-up of Maloney--Witten \cite{2006.04855}, the Zamolodchikov metric is written as
\(ds^2=4\,dR^2/R^2\), while the measure used in the formal average is
\(dR/(2R)\). If one instead quotes the raw Riemannian volume form associated to that metric,
one obtains \(2\,dR/R\). These differ only by an overall constant. Since the \(c=1\) average
diverges, and since finite-volume Narain averages are usually normalized separately, this constant
is a convention rather than physical data.}
After quotienting by \(R\sim 1/R\), the same measure can be written on the fundamental
domain as
\begin{equation}
  d\mu_C(R)=C\,{dR\over R},
  \qquad
  R\in[1,\infty).
  \label{eq:c1-haar-measure-fundamental-domain}
\end{equation}
For a \(T\)-duality-invariant integrand this is equivalently obtained from
\(\frac12\int_0^\infty C\,dR/R\), with the factor of \(\frac12\) implementing the
\(\mathbb Z_2\) quotient. The self-dual fixed point should be treated with the usual orbifold
stabilizer factor, but it does not affect the continuous integral.

\paragraph{Averaging over topological boundary conditions.}
Pairing the physical boundary state with \(L_R\) gives the usual compact-boson
partition function
\begin{equation}
\begin{split}
    Z_R(\tau)
  =
  Z_{L_R}(T^2;\tau)
  &=
  \langle L_R;T^2|\Psi_{\mathrm{phys}}(T^2;\tau)\rangle\\
  &=
  {1\over |\eta(\tau)|^2}
  \sum_{n,w\in\mathbb Z}
  q^{{1\over 4}(n/R+wR)^2}
  \bar q^{{1\over 4}(n/R-wR)^2},~~
  q=e^{2\pi i\tau}.
\end{split}
  \label{eq:compact-boson-partition}
\end{equation}
The factor \(1/|\eta(\tau)|^2\) comes from the oscillator part of the fixed physical
boundary Hamiltonian \eqref{eq:c1-boundary-hamiltonian}. The discrete sum is supplied
by the topological boundary condition \(L_R\), which restricts the allowed charge sectors to
\begin{equation}
  (\alpha,\beta)
  =
  \left({n\over R},wR\right),
  \qquad n,w\in\mathbb Z.
  \label{eq:c1-allowed-charge-sectors-LR}
\end{equation}
Substituting these charge labels into the zero-mode eigenvalue formula
\eqref{eq:c1-left-right-momenta} gives
\begin{equation}
  p_{\mathrm L}
  =
  {n\over R}+wR,
  \qquad
  p_{\mathrm R}
  =
  {n\over R}-wR.
  \label{eq:c1-pL-pR-LR}
\end{equation}
This matches with the fact that the exponents in \eqref{eq:compact-boson-partition} are precisely the
eigenvalues of the current-algebra zero-mode contribution to
\(L_0\) and \(\bar L_0\).
Equivalently, the lattice sum in \eqref{eq:compact-boson-partition} is the sum over the
local primaries \(\mathcal O_{n,w}\) selected by the topological boundary condition \(L_R\).

Combining the previous discussion of duality quotient and the integral measure, the topological-boundary average over the compact-boson locus is thus
\begin{equation}
  \big\langle Z(\tau)\big\rangle_{\mathrm{top},C}^{(c=1)}
  =
  \int_{\mathcal L_{\mathrm{comp}}^{(1)}} d\mu_C(L)\,
  Z_L(T^2;\tau)
  =
  C\int_1^\infty {dR\over R}\,Z_R(\tau).
  \label{eq:c1-formal-average}
\end{equation}
In the rest of
this subsection we set \(C=1\) unless otherwise stated. The integral in
\eqref{eq:c1-formal-average} is over the compact-boson locus only. The degenerate
topological boundary conditions \(L_\infty\) and \(L_0\) are not integrated over as ordinary
points of this Haar orbit; rather, they appear as endpoints in a partial compactification of
the radius line.

The expression \eqref{eq:c1-formal-average} is the rank-one analogue of the Narain
average in \cite{2006.04855}. Note that it is not a finitely normalized average. Indeed, for \(R\to\infty\), the
winding sectors with \(w\neq 0\) are exponentially suppressed, and the \(w=0\) sector gives
\begin{equation}
  Z_R(\tau)
  \sim
  {1\over |\eta(\tau)|^2}
  \sum_{n\in\mathbb Z}
  \exp\left(-{\pi\tau_2 n^2\over R^2}\right).
  \label{eq:c1-large-R-w0-sector}
\end{equation}
By Poisson resummation,
\begin{equation}
  \sum_{n\in\mathbb Z}
  \exp\left(-{\pi\tau_2 n^2\over R^2}\right)
  \sim
  {R\over \sqrt{\tau_2}},
  \qquad R\to\infty.
  \label{eq:c1-poisson-large-R}
\end{equation}
Thus
\begin{equation}
  Z_R(\tau)
  \sim
  {R\over \sqrt{\tau_2}\,|\eta(\tau)|^2},
  \qquad R\to\infty,
  \label{eq:c1-large-R-asymptotic}
\end{equation}
and hence
\begin{equation}
  \int^\infty {dR\over R}\, Z_R(\tau)
  \sim
  {1\over \sqrt{\tau_2}\,|\eta(\tau)|^2}
  \int^\infty dR
  =
  \infty .
  \label{eq:c1-divergence}
\end{equation}
The rank-one average therefore has no normalized probabilistic interpretation. From the
present SymTFT viewpoint, this divergence has a simple structural origin: the  space of topological boundary conditions as the
radius line is noncompact, and its large-radius end approaches the degenerate topological
boundary condition \(L_\infty=\mathbb R\oplus 0\), which describes the noncompact boson.
The dual endpoint \(L_0=0\oplus\mathbb R\) is obtained by applying \(T\)-duality. This reintreprets the \(c=1\) pathology of Narain averaging as an endpoint effect in the
space of topological boundary conditions.

This example is nevertheless useful because it makes the general mechanism explicit. The
compact-boson radius is not inserted as a coupling in the local physical boundary condition,
nor in the boundary Hamiltonian \eqref{eq:c1-boundary-hamiltonian}. It is supplied by the
choice of a topological boundary condition \(L_R\) of the fixed \(\mathbb R\)-valued BF
SymTFT. Averaging over radii is therefore identified with integrating over topological boundary conditions
of the SymTFT, where the duality quotient and the integral measure both enjoy a SymTFT interpretation. For higher Narain rank, the same construction leads to the Narain
moduli space reproduced by the space topological boundary conditions, and the invariant integral measure becomes the Haar-induced measure on the
Narain quotient by dualities.

\subsection{\texorpdfstring{\(c>1\)}{c > 1}}
\label{subsec:narain-cgt1}

We now pass from the rank-one compact boson to higher-rank Narain theories \cite{2006.04855,2006.04839}. Throughout
this subsection, \(c\) denotes the rank of the Narain lattice, equivalently the dimension of
the target torus \(T^c\), i.e., $c_{\mathrm L}=c_{\mathrm R}=c$.
We use the \(\mathbb R\)-valued BF SymTFT introduced at the beginning of this section,
with action 
\begin{equation}
  S_{\mathrm{BF}}
  =
  {1\over 2\pi}
  \int_{X_3}
  \sum_{i=1}^c a_i \wedge d b_i ,
  \qquad
  a_i,b_i\in \Omega^1(X_3;\mathbb R), 
\end{equation}
and line operators
\begin{equation}
  U_{\vec{\alpha}}[\gamma]
  =
  \exp\left(i\alpha_i\oint_\gamma  a_i\right),
  \qquad
  V_{\vec{\beta}}[\gamma]
  =
  \exp\left(i\beta_i\oint_\gamma  b_i\right),
  \qquad
  \vec{\alpha},\vec{\beta} \in \mathbb{R}^c .
  \label{eq:RBF-lines-d}
\end{equation}
The defect
group reads
\[
  \mathbb D_c=\mathbb R^c\oplus\mathbb R^c,
\]
with the Dirac pairing 
\begin{equation}
  \big\langle
  (\vec\alpha,\vec\beta),(\vec\alpha',\vec\beta')
  \big\rangle
  =
  \vec\alpha\cdot\vec\beta'
  +
  \vec\alpha'\cdot\vec\beta
  \quad \mathrm{mod}\ \mathbb Z .
  \label{eq:cgt1-defect-pairing}
\end{equation}
We also use the quadratic refinement
\begin{equation}
  q(\vec\alpha,\vec\beta)
  =
  \vec\alpha\cdot\vec\beta
  \quad \mathrm{mod}\ \mathbb Z ,
  \label{eq:cgt1-quadratic-refinement}
\end{equation}
which controls the spin of the corresponding line.

\paragraph{Physical boundary.}
The physical boundary condition is the fixed conformal boundary condition at the physical end of
the BF slab. It prepares the state
\begin{equation}
  |\Psi_{\mathrm{phys}}(\Sigma;\Omega)\rangle
  \in
  \mathcal H_{\mathrm{BF}}(\Sigma).
  \label{eq:cgt1-physical-state}
\end{equation}
Introduce the left- and right-moving combinations
\begin{equation}
  A_{\mathrm L,i}=a_i+b_i,
  \qquad
  A_{\mathrm R,i}=a_i-b_i,
  \qquad
  i=1,\ldots,c .
  \label{eq:cgt1-AL-AR-definition}
\end{equation}
On the physical boundary \(\Sigma\), with local complex coordinates \(z,\bar z\), the
conformal boundary condition is
\begin{equation}
  (A_{\mathrm L,i})_{\bar z}\big|_\Sigma=0,
  \qquad
  (A_{\mathrm R,i})_{z}\big|_\Sigma=0,
  \qquad
  i=1,\ldots,c .
  \label{eq:cgt1-conformal-boundary-condition}
\end{equation}
Equivalently,
\begin{equation}
  (a_i+b_i)_{\bar z}\big|_\Sigma=0,
  \qquad
  (a_i-b_i)_z\big|_\Sigma=0 .
  \label{eq:cgt1-conformal-boundary-condition-ab}
\end{equation}
Thus the physical boundary supplies \(c\) left-moving and \(c\) right-moving abelian current
algebras.

On the Lorentzian boundary cylinder
\(\Sigma=S^1_\varphi\times\mathbb R_t\), the same boundary condition can be written as
\begin{equation}
  (a_i)_t\big|_\Sigma=(b_i)_\varphi\big|_\Sigma,
  \qquad
  (b_i)_t\big|_\Sigma=(a_i)_\varphi\big|_\Sigma .
  \label{eq:cgt1-lorentzian-bc}
\end{equation}
The corresponding physical boundary Hamiltonian is fixed once and for all by this conformal
boundary condition:
\begin{equation}
  H_{\mathrm{phys}}
  =
  {1\over 4\pi}
  \int_{S^1}d\varphi\,
  :\!\sum_{i=1}^c
  \left(
    (a_i)_\varphi^2+(b_i)_\varphi^2
  \right)\!: ,
  \label{eq:cgt1-boundary-hamiltonian}
\end{equation}
and the spatial momentum is
\begin{equation}
  P_{\mathrm{phys}}
  =
  {1\over 2\pi}
  \int_{S^1}d\varphi\,
  :\!\sum_{i=1}^c
  (a_i)_\varphi(b_i)_\varphi\!: .
  \label{eq:cgt1-boundary-momentum}
\end{equation}

Let \(\mathsf P_{\mathrm L,i}\) and \(\mathsf P_{\mathrm R,i}\) denote the zero-mode charge
operators of the left- and right-moving current algebras. A bulk line labelled by
\((\vec\alpha,\vec\beta)\in\mathbb D_c\) specifies a charge sector of the physical boundary
theory. Here \(\vec\alpha\) and \(\vec\beta\) are real charge labels, not dynamical fields. In the
charge sector labelled by \((\vec\alpha,\vec\beta)\), the zero-mode operators have eigenvalues
\begin{equation}
  \vec p_{\mathrm L}(\vec\alpha,\vec\beta)
  =
  \vec\alpha+\vec\beta,
  \qquad
  \vec p_{\mathrm R}(\vec\alpha,\vec\beta)
  =
  \vec\alpha-\vec\beta .
  \label{eq:cgt1-left-right-momenta}
\end{equation}
The Virasoro zero-mode operators are
\begin{equation}
  L_0
  =
  {1\over 4}\left|\mathsf P_{\mathrm L}\right|^2
  +
  N_{\mathrm L},
  \qquad
  \bar L_0
  =
  {1\over 4}\left|\mathsf P_{\mathrm R}\right|^2
  +
  N_{\mathrm R}.
  \label{eq:cgt1-L0-L0bar-operator}
\end{equation}
Therefore, in the charge sector \((\vec\alpha,\vec\beta)\), their eigenvalues are
\begin{equation}
  h_{\vec\alpha,\vec\beta;N_{\mathrm L}}
  =
  {1\over 4}
  \left|\vec\alpha+\vec\beta\right|^2
  +
  N_{\mathrm L},
  \qquad
  \bar h_{\vec\alpha,\vec\beta;N_{\mathrm R}}
  =
  {1\over 4}
  \left|\vec\alpha-\vec\beta\right|^2
  +
  N_{\mathrm R}.
  \label{eq:cgt1-L0-L0bar-sector}
\end{equation}
Thus
\begin{equation}
  H_{\mathrm{phys}}
  =
  L_0+\bar L_0-{c\over 12},
  \qquad
  P_{\mathrm{phys}}
  =
  L_0-\bar L_0 .
  \label{eq:cgt1-H-P-sector}
\end{equation}
The Narain moduli do not appear in this local physical boundary Hamiltonian. They enter only
through the choice of topological cap.

\paragraph{Continuous compact topological boundary conditions.}
A compact topological boundary condition is a discrete Lagrangian subgroup
\begin{equation}
  L\subset\mathbb D_c .
\end{equation}
More explicitly, \(L\) is associated with a rank-\(2c\) lattice satisfying
\begin{equation}
  q(\ell)\in\mathbb Z,
  \qquad
  \big\langle \ell,\ell'\big\rangle\in\mathbb Z,
  \qquad
  \ell,\ell'\in L,
  \label{eq:cgt1-integral-isotropic-condition}
\end{equation}
together with the maximality condition
\begin{equation}
  L\cong L^\perp,
  \qquad
  L^\perp
  :=
  \left\{
    x\in\mathbb D_c:
    \big\langle x,\ell\big\rangle\in\mathbb Z
    \ \mathrm{for\ all}\ \ell\in L
  \right\}.
  \label{eq:cgt1-self-dual-lattice-condition}
\end{equation}
The condition \(q(\ell)\in\mathbb Z\) says that the lines ending on the topological boundary
are bosonic\footnote{Since we want the resulting absolute theories being bosonic CFTs instead of their fermionization.}. The condition \(\langle\ell,\ell'\rangle\in\mathbb Z\) says that all lines within $L$ have mutually trivial braiding. The maximality condition says that every line trivially-braiding with those in \(L\) is already in
\(L\). In other words, this maximal isotropic condition tells us the lattice associated to \(L\) is even self-dual in \(\mathbb R^{c,c}\), but in the SymTFT language written in the
\((\vec\alpha,\vec\beta)\) polarization determined by the BF line operators.

We now parameterize these compact topological boundary conditions explicitly from the
SymTFT. Start from a compact boundary as the reference boundary condition
\begin{equation}
  L_\circ
  =
  \left\{
    (\vec n,\vec w):
    \vec n,\vec w\in\mathbb Z^c
  \right\}
  \subset
  \mathbb D_c=\mathbb R^c\oplus\mathbb R^c .
  \label{eq:cgt1-reference-lattice}
\end{equation}
The pairing on \(L_\circ\) is
\begin{equation}
  \big\langle
  (\vec n,\vec w),(\vec n',\vec w')
  \big\rangle
  =
  \vec n\cdot\vec w'
  +
  \vec n'\cdot\vec w
  \in\mathbb Z ,
  \label{eq:cgt1-reference-pairing}
\end{equation}
and the quadratic refinement is
\begin{equation}
  q(\vec n,\vec w)=\vec n\cdot\vec w\in\mathbb Z.
  \label{eq:cgt1-reference-quadratic}
\end{equation}
Conversely, if \(x=(\vec\alpha,\vec\beta)\in\mathbb D_c\) pairs integrally with every
\((\vec n,\vec w)\in L_\circ\), then
\begin{equation}
  \vec\alpha\cdot\vec w+\vec\beta\cdot\vec n\in\mathbb Z
  \qquad
  \forall\,\vec n,\vec w\in\mathbb Z^c .
\end{equation}
Taking \(\vec w=0\) implies \(\vec\beta\in\mathbb Z^c\), and taking \(\vec n=0\) implies
\(\vec\alpha\in\mathbb Z^c\). Hence \(x\in L_\circ\), so
\begin{equation}
  L_\circ=L_\circ^\perp .
\end{equation}
Therefore, \(L_\circ\) is a compact Lagrangian boundary condition of the BF SymTFT. In fact, it is the
higher-rank analogue of the self-dual-radius boundary at $R=1$ in the \(c=1\) example.

In order to find other topological boundary conditions from this reference one, we determine the relevant automorphism group of the BF theory.
Introduce the \(2c\)-component vector of gauge fields
\begin{equation}
  \mathcal A
  =
  \begin{pmatrix}
    a_1\\
    \vdots\\
    a_c\\
    b_1\\
    \vdots\\
    b_c
  \end{pmatrix},
  \qquad
  \eta
  =
  \begin{pmatrix}
    0 & \mathbf 1_c\\
    \mathbf 1_c & 0
  \end{pmatrix}.
  \label{eq:cgt1-A-vector-eta}
\end{equation}
Up to a total derivative, the BF action can be written in the following symmetric form
\begin{equation}
  S_{\mathrm{BF}}
  =
  {1\over 4\pi}
  \int_{X_3}
  \mathcal A^T\eta\, d\mathcal A .
  \label{eq:cgt1-BF-symmetric-form}
\end{equation}
Indeed,
\begin{equation}
  {1\over 4\pi}
  \int
  \mathcal A^T\eta\,d\mathcal A
  =
  {1\over 4\pi}
  \int
  \sum_{i=1}^c
  \left(
    a_i\wedge db_i+b_i\wedge da_i
  \right)
  =
  {1\over 2\pi}
  \int
  \sum_{i=1}^c a_i\wedge db_i
\end{equation}
on a closed three-manifold, or modulo the boundary polarization on a slab.

Now consider a constant linear transformation of variables
\begin{equation}
  \mathcal A\longmapsto N\mathcal A,
  \qquad
  N\in GL(2c,\mathbb R).
\end{equation}
The action becomes
\begin{equation}
  S_{\mathrm{BF}}
  \longmapsto
  {1\over 4\pi}
  \int
  \mathcal A^T N^T\eta N\,d\mathcal A .
\end{equation}
This field redefinition is an automorphism of the BF SymTFT precisely when
\begin{equation}
  N^T\eta N=\eta .
  \label{eq:cgt1-Occ-condition-field}
\end{equation}
Therefore, the bulk SymTFT heory has a continuous group of linear automorphisms
\begin{equation}
  O(c,c;\mathbb R)
  =
  \left\{
    N\in GL(2c,\mathbb R):
    N^T\eta N=\eta
  \right\}.
  \label{eq:cgt1-Occ-field-automorphism}
\end{equation}
This is a statement about the automorphisms of the \(\mathbb R\)-valued BF theory, or
equivalently about invertible topological interfaces (i.e. 0-form symmetry operators) of the SymTFT. It is not yet a quotient
by physical equivalences of the absolute boundary CFT.

Let us now see the same group directly from the line operators. For a line labelled by
\begin{equation}
  x=(\vec\alpha,\vec\beta)\in\mathbb D_c=\mathbb R^c\oplus\mathbb R^c,
\end{equation}
write
\begin{equation}
  \mathcal W_x[\gamma]
  :=
  U_{\vec\alpha}[\gamma]V_{\vec\beta}[\gamma]
  =
  \exp\left(
    i\oint_\gamma x^T\mathcal A
  \right).
  \label{eq:cgt1-line-Wx}
\end{equation}
The Dirac pairing can be written as
\begin{equation}
  \langle x,y\rangle
  =
  x^T\eta y
  \quad \mathrm{mod}\ \mathbb Z .
  \label{eq:cgt1-Dirac-pairing-matrix}
\end{equation}
Equivalently, the braiding phase is
\begin{equation}
  \mathcal W_x[\gamma_1]\mathcal W_y[\gamma_2]
  =
  \exp\left(
    2\pi i\,x^T\eta y\,
    \mathrm{Link}(\gamma_1,\gamma_2)
  \right)
  \mathcal W_y[\gamma_2]\mathcal W_x[\gamma_1].
  \label{eq:cgt1-braiding-matrix}
\end{equation}
A general field transformation \(\mathcal A\mapsto N\mathcal A\) sends the line label to
\begin{equation}
  x\longmapsto N^T x .
\end{equation}
Given \(N\in O(c,c;\mathbb R)\), this preserves the Dirac pairing / line braiding:
\begin{equation}
  \langle N^Tx,N^Ty\rangle
  =
  x^T N\eta N^T y
  =
  x^T\eta y
  =
  \langle x,y\rangle .
\end{equation}
Equivalently, we may describe the induced automorphism directly on the defect group by
\begin{equation}
  x\longmapsto g x,
  \qquad
  g\in O(c,c;\mathbb R),
  \qquad
  g^T\eta g=\eta .
  \label{eq:cgt1-Occ-defect-automorphism}
\end{equation}
Note that we haven not used the Narain CFT as an input to obtain this \(O(c,c;\mathbb R)\); it is the group of
linear automorphisms of the SymTFT defect data preserving the braiding.

Now choose the reference compact topological boundary condition
\begin{equation}
  L_\circ
  =
  \left\{
    (\vec n,\vec w):
    \vec n,\vec w\in\mathbb Z^c
  \right\}
  \subset \mathbb D_c .
  \label{eq:cgt1-reference-lattice}
\end{equation}
For any $g\in O(c,c;\mathbb R)$
we can define
\begin{equation}
  L_g
  =
  gL_\circ .
  \label{eq:cgt1-Lg-definition}
\end{equation}
This is again a compact Lagrangian topological boundary condition. To see this, let
\(\ell_1,\ell_2\in L_\circ\), then the braiding-preserving tells us
\begin{equation}
  \langle g\ell_1,g\ell_2\rangle
  =
  \langle \ell_1,\ell_2\rangle
  \in\mathbb Z,
\end{equation}
and similarly
\begin{equation}
  q(g\ell)=q(\ell)\in\mathbb Z .
\end{equation}
Moreover,
\begin{equation}
  (gL_\circ)^\perp
  =
  g(L_\circ^\perp)
  =
  gL_\circ .
  \label{eq:cgt1-Lg-self-dual}
\end{equation}
Thus acting with the SymTFT automorphism group on the reference Lagrangian boundary
produces another allowed topological cap. From a defect perspective, this amounts to pushing invertible topological surfaces labeled by $O(c,c;\mathbb R)$ elements onto the reference topological boundary $L_\circ$ and transform it to $gL_\circ$. This is the intrinsic SymTFT origin of the
continuous family of compact topological boundary conditions, which correspond to Narain moduli before the quotient.

As in $c=1$, where the modulus $R$ labeling different CFTs can enter the parameterization of topological boundary conditions explicitly, we now present how the usual torus sigma-model moduli, metric and $B$-field, can enter the space of topological boundary conditions. Without any input from Narain CFTs, we set
\begin{equation}
  G=G^T>0,
  \qquad
  B^T=-B ,
  \label{eq:cgt1-GB-data}
\end{equation}
where \(G\) is a positive definite symmetric matrix and \(B\) is an antisymmetric matrix.
We choose the positive square root of \(G\),
\begin{equation}
  e=G^{1/2},
  \qquad
  e^T=e,
  \qquad
  e^2=G,
  \label{eq:cgt1-positive-vielbein}
\end{equation}
and then define a linear map on the defect group
\(\mathbb D_c=\mathbb R^c\oplus\mathbb R^c\) by
\begin{equation}
  g_{G,B}
  =
  \begin{pmatrix}
    e^{-T} & e^{-T}B \\
    0      & e
  \end{pmatrix}.
  \label{eq:cgt1-gGB-definition}
\end{equation}
More explicitly, on a line defect labeled by \(x=(\vec\alpha,\vec\beta)\), \(g_{G,B}\) acts as
\begin{equation}
  g_{G,B}:
  \begin{pmatrix}
    \vec\alpha \\
    \vec\beta
  \end{pmatrix}
  \longmapsto
  \begin{pmatrix}
    e^{-T}(\vec\alpha+B\vec\beta) \\
    e\vec\beta
  \end{pmatrix}.
  \label{eq:cgt1-gGB-action}
\end{equation}
It is easy to check $g_{G,B}$ satisfies
\begin{equation}
\begin{split}
  g_{G,B}^T\eta g_{G,B}
  &=
  \begin{pmatrix}
    e^{-1} & 0 \\
    B^T e^{-1} & e^T
  \end{pmatrix}
  \begin{pmatrix}
    0 & e \\
    e^{-T} & e^{-T}B
  \end{pmatrix}
  \\
  &=
  \begin{pmatrix}
    0 & \mathbf 1_c \\
    \mathbf 1_c & B^T+B
  \end{pmatrix}
  =
  \begin{pmatrix}
    0 & \mathbf 1_c \\
    \mathbf 1_c & 0
  \end{pmatrix}=\eta,
\end{split}
\end{equation}
where we used \(B^T=-B\). Therefore \(g_{G,B} \) parametrizes the \(O(c,c;\mathbb R)\) automorphism  of the SymTFT defect data.

Applying this defect automorphism to the reference boundary \(L_\circ\) gives
an explicit expression of the topological boundary condition
\begin{equation}
  L_{G,B}
  =
  \left\{
    \left(
      e^{-T}(\vec n+B\vec w),
      e\vec w
    \right)
    :
    \vec n,\vec w\in\mathbb Z^c
  \right\}
  \subset \mathbb D_c .
  \label{eq:cgt1-LGB-definition}
\end{equation}
That is to say, \(G\) and \(B\) appear as coordinates on the SymTFT family of compact
topological boundary conditions. Note that the above derivation requires no input from the physical boundary Hamiltonian. Rather, we are just characterizing which SymTFT line defects 
\((\vec\alpha,\vec\beta)\) are allowed to end on the topological boundary, and the characterization happens to match the Narain moduli $G$ and $B$.

Although the above discussion is sufficient to see $L_{G,B}$ generated by $g_{G,B}$ is topological boundary condition, let us still check the Lagrangian condition explicitly. 
Write an element in the defect group as 
\begin{equation}
  \ell_{G,B}(\vec n,\vec w)
  =
  \left(
    \vec\alpha_{G,B}(\vec n,\vec w),
    \vec\beta_{G,B}(\vec n,\vec w)
  \right)
  =
  \left(
    e^{-T}(\vec n+B\vec w),
    e\vec w
  \right).
  \label{eq:cgt1-ell-GB}
\end{equation}
For two such elements,
\[
  \ell=\ell_{G,B}(\vec n,\vec w),
  \qquad
  \ell'=\ell_{G,B}(\vec n',\vec w'),
\]
the Dirac pairing is
\begin{equation}
\begin{split}
  \langle \ell,\ell'\rangle
  &=
  \vec\alpha_{G,B}(\vec n,\vec w)\cdot
  \vec\beta_{G,B}(\vec n',\vec w')
  +
  \vec\alpha_{G,B}(\vec n',\vec w')\cdot
  \vec\beta_{G,B}(\vec n,\vec w)
  \\
  &=
  (\vec n+B\vec w)\cdot\vec w'
  +
  (\vec n'+B\vec w')\cdot\vec w
  \\
  &=
  \vec n\cdot\vec w'
  +
  \vec n'\cdot\vec w
  \in\mathbb Z .
\end{split}
  \label{eq:cgt1-LGB-pairing}
\end{equation}
The \(B\)-dependent terms cancel because \(B\) is antisymmetric:
\[
  (B\vec w)\cdot\vec w'
  +
  (B\vec w')\cdot\vec w
  =
  \vec w^T B^T\vec w'
  +
  \vec w'^T B^T\vec w
  =
  -\vec w^T B\vec w'
  -
  \vec w'^T B\vec w
  =
  0 .
\]
Similarly,
\begin{equation}
  q(\ell)
  =
  \vec\alpha_{G,B}(\vec n,\vec w)\cdot
  \vec\beta_{G,B}(\vec n,\vec w)
  =
  (\vec n+B\vec w)\cdot \vec w
  =
  \vec n\cdot\vec w
  \in\mathbb Z,
  \label{eq:cgt1-LGB-evenness}
\end{equation}
because \(\vec w^T B\vec w=0\). Thus the lines in \(L_{G,B}\) are bosonic and have mutually
trivial braiding. Finally, since \(g_{G,B}\) preserves the pairing,
\begin{equation}
  L_{G,B}^{\perp}
  =
  (g_{G,B}L_\circ)^\perp
  =
  g_{G,B}(L_\circ^\perp)
  =
  g_{G,B}L_\circ
  =
  L_{G,B}.
  \label{eq:cgt1-LGB-self-dual}
\end{equation}
Therefore \(L_{G,B}\) is a compact Lagrangian topological boundary condition.

Another quick consistency check is to consider \(c=1\), where we have
\begin{equation}
  G=R^2,
  \qquad
  B=0,
  \qquad
  e=R,
\end{equation}
and thus
\begin{equation}
  L_{G,B}
  =
  \left\{
    \left({n\over R},wR\right):
    n,w\in\mathbb Z
  \right\}
  =
  L_R,
  \label{eq:cgt1-LGB-reduces-to-LR}
\end{equation}
which exactly reduces to the $R$-labeled topological boundary conditions.

\paragraph{Local primaries and surviving symmetry lines.}
We now explain the operator content of the absolute 2D theory obtained via a topological boundary condition \(L\). The elements of
\(L\) are precisely the bulk line labels that become trivial on the topological boundary. Thus, for each $\ell=(\vec\alpha,\vec\beta)\in L$,
the composite bulk line
\begin{equation}
  \mathcal W_\ell[\gamma]
  :=
  U_{\vec\alpha}[\gamma]V_{\vec\beta}[\gamma]
  \label{eq:cgt1-condensed-line-Well}
\end{equation}
can end on the topological boundary. If we bring the other endpoint of this line to the
physical boundary, it creates a local operator of the absolute 2D theory. We
denote this local operator by $\mathcal O_\ell$ .
In this sense, the Lagrangian subgroup \(L\) gives the current-algebra primary spectrum of the
absolute CFT.

The conformal weights of \(\mathcal O_\ell\) are read using the fixed physical boundary
condition. Namely, the zero-mode eigenvalues in the charge sector
\(\ell=(\vec\alpha,\vec\beta)\) are
\begin{equation}
  \vec p_{\mathrm L}(\ell)=\vec\alpha+\vec\beta,
  \qquad
  \vec p_{\mathrm R}(\ell)=\vec\alpha-\vec\beta .
  \label{eq:cgt1-primary-pLR-general}
\end{equation}
Therefore, before adding oscillator descendants,
\begin{equation}
  h_\ell
  =
  {1\over 4}
  \left|\vec p_{\mathrm L}(\ell)\right|^2,
  \qquad
  \bar h_\ell
  =
  {1\over 4}
  \left|\vec p_{\mathrm R}(\ell)\right|^2 .
  \label{eq:cgt1-primary-dimensions-general}
\end{equation}
Moreover,
\begin{equation}
  h_\ell-\bar h_\ell
  =
  \vec\alpha\cdot\vec\beta
  =
  q(\ell)
  \in\mathbb Z ,
  \label{eq:cgt1-locality-spin-condition}
\end{equation}
so the endpoint operator has integer spin in the bosonic 2D theory. This is the
operator-theoretic meaning of the condition \(q(\ell)\in\mathbb Z\).

For the boundary condition \(L_{G,B}\), parameterized by $G$ and $B$ , the condensed lines are labelled by
  $\vec n,\vec w\in\mathbb Z^c$
through
\begin{equation}
  \ell_{G,B}(\vec n,\vec w)
  =
  \left(
    e^{-T}(\vec n+B\vec w),
    e\vec w
  \right)
  \in L_{G,B}.
  \label{eq:cgt1-ellGB-primary-label}
\end{equation}
The corresponding local primary is denoted by
  $\mathcal O_{\vec n,\vec w}^{G,B}$.
Its left- and right-moving
momenta are
\begin{equation}
  \vec p_{\mathrm L}(\vec n,\vec w;G,B)
  =
  e^{-T}
  \bigl(
    \vec n+(B+G)\vec w
  \bigr),
  \qquad
  \vec p_{\mathrm R}(\vec n,\vec w;G,B)
  =
  e^{-T}
  \bigl(
    \vec n+(B-G)\vec w
  \bigr).
  \label{eq:cgt1-primary-pLR-GB}
\end{equation}
and the corresponding conformal weights are
\begin{equation}
  h_{\vec n,\vec w}^{G,B}
  =
  {1\over 4}
  \left|
  e^{-T}
  \bigl(
    \vec n+(B+G)\vec w
  \bigr)
  \right|^2,
  \qquad
  \bar h_{\vec n,\vec w}^{G,B}
  =
  {1\over 4}
  \left|
  e^{-T}
  \bigl(
    \vec n+(B-G)\vec w
  \bigr)
  \right|^2,
  \label{eq:cgt1-primary-dimensions-GB}
\end{equation}
That is to say, the familiar Narain primary labeled by momentum and winding quantum numbers
\((\vec n,\vec w)\) is now realized from the SymTFT line
\(\mathcal W_{\ell_{G,B}(\vec n,\vec w)}\).

We now discuss the topological symmetry lines of the absolute theory. A general bulk line
labelled by
\begin{equation}
  \gamma=(\vec\rho,\vec\sigma)\in\mathbb D_c
\end{equation}
can be laid on the topological boundary. However, two such lines define the same boundary
topological defect if they differ by a line that is trivialized by the boundary condition \(L\).
Therefore the surviving topological line operators are labelled by 
  $\mathbb D_c/L$ \cite{2401.10165}.
Their action on the local primaries is determined by the braiding of bulk lines. Namely,
\begin{equation}
  \gamma:
  \quad
  \mathcal O_\ell
  \longmapsto
  \exp\!\left(
    2\pi i\,
    \big\langle \gamma,\ell\big\rangle
  \right)
  \mathcal O_\ell .
  \label{eq:cgt1-surviving-line-action-general}
\end{equation}
This action is well-defined on the quotient \(\mathbb D_c/L\), because shifting
\(\gamma\) by an element of \(L\) changes the phase by an integer multiple of \(2\pi i\).

For \(L=L_{G,B}\), the quotient \(\mathbb D_c/L_{G,B}\) is a \(2c\)-torus. A convenient
representative of its elements is
\begin{equation}
  \gamma_{\vec\vartheta_{\mathrm m},\vec\vartheta_{\mathrm w}}
  =
  \left(
    e^{-T}
    \bigl(
      \vec\vartheta_{\mathrm w}
      +
      B\vec\vartheta_{\mathrm m}
    \bigr),
    e\vec\vartheta_{\mathrm m}
  \right),
  \qquad
  \vec\vartheta_{\mathrm m},\vec\vartheta_{\mathrm w}
  \in
  \mathbb R^c/\mathbb Z^c .
  \label{eq:cgt1-U1mw-representative}
\end{equation}
The periodicity follows because shifting
\[
  \vec\vartheta_{\mathrm m}\mapsto \vec\vartheta_{\mathrm m}+\vec r,
  \qquad
  \vec\vartheta_{\mathrm w}\mapsto \vec\vartheta_{\mathrm w}+\vec s,
  \qquad
  \vec r,\vec s\in\mathbb Z^c,
\]
changes \(\gamma_{\vec\vartheta_{\mathrm m},\vec\vartheta_{\mathrm w}}\) by
\[
  \left(
    e^{-T}(\vec s+B\vec r),
    e\vec r
  \right)
  =
  \ell_{G,B}(\vec s,\vec r)
  \in L_{G,B}.
\]
Therefore \(\vec\vartheta_{\mathrm m}\) and \(\vec\vartheta_{\mathrm w}\) indeed parametrize
two \(c\)-dimensional compact symmetry groups.

The action of this surviving line on the primary
\(\mathcal O_{\vec n,\vec w}^{G,B}\) is
\begin{equation}
\begin{split}
  \big\langle
  \gamma_{\vec\vartheta_{\mathrm m},\vec\vartheta_{\mathrm w}},
  \ell_{G,B}(\vec n,\vec w)
  \big\rangle
  &=
  \bigl(
    \vec\vartheta_{\mathrm w}
    +
    B\vec\vartheta_{\mathrm m}
  \bigr)\cdot \vec w
  +
  \bigl(
    \vec n+B\vec w
  \bigr)\cdot \vec\vartheta_{\mathrm m}
  \\
  &=
  \vec\vartheta_{\mathrm w}\cdot\vec w
  +
  \vec\vartheta_{\mathrm m}\cdot\vec n
  \quad \mathrm{mod}\ \mathbb Z ,
\end{split}
  \label{eq:cgt1-U1mw-pairing}
\end{equation}
where the \(B\)-dependent terms cancel because \(B^T=-B\). Hence
\begin{equation}
  \gamma_{\vec\vartheta_{\mathrm m},\vec\vartheta_{\mathrm w}}:
  \quad
  \mathcal O_{\vec n,\vec w}^{G,B}
  \longmapsto
  \exp\!\left[
    2\pi i
    \left(
      \vec\vartheta_{\mathrm m}\cdot\vec n
      +
      \vec\vartheta_{\mathrm w}\cdot\vec w
    \right)
  \right]
  \mathcal O_{\vec n,\vec w}^{G,B}.
  \label{eq:cgt1-U1mw-action}
\end{equation}
Thus \(\mathcal O_{\vec n,\vec w}^{G,B}\) carries charges
\begin{equation}
  (Q_{\mathrm m},Q_{\mathrm w})=(\vec n,\vec w)
  \label{eq:cgt1-primary-U1mw-charges}
\end{equation}
under
\begin{equation}
  U(1)_{\mathrm m}^{\,c}\times U(1)_{\mathrm w}^{\,c}.
\end{equation}
Here \(U(1)_{\mathrm m}^{\,c}\) is generated by the parameter
\(\vec\vartheta_{\mathrm m}\), and measures momentum charge \(\vec n\), while
\(U(1)_{\mathrm w}^{\,c}\) is generated by \(\vec\vartheta_{\mathrm w}\), and measures
winding charge \(\vec w\).

\paragraph{Noncompact topological boundary conditions.}
The compact Narain boundary conditions \(L_{G,B}\) are not the only maximal isotropic
subgroups of \(\mathbb D_c\). They are the ones relevant for compact torus CFTs, since
they associate with discrete rank-\(2c\) lattices. The enlarged space of topological boundary conditions
also contains degenerate maximal isotropic subgroups which are  continuous. The
simplest examples are
\begin{equation}
  L_{\infty}^{(c)}
  =
  \mathbb R^c\oplus 0,
  \qquad
  L_{0}^{(c)}
  =
  0\oplus \mathbb R^c .
  \label{eq:cgt1-full-degenerate-boundaries}
\end{equation}
Similarly to $c=1$ case, it is straightforward to check they are Lagrangian. Physically, \(L_{\infty}^{(c)}\) describes \(c\) noncompact bosons, while
\(L_{0}^{(c)}\) describes their dual noncompact bosons.  They are the higher-rank analogues of the
\(R=\infty\) and \(R=0\) endpoints in the \(c=1\) discussion.

More generally, one may decompactify only part of the compact Lagrangian boundary condition. After choosing a splitting
\begin{equation}
  \mathbb D_c
  =
  \mathbb D_r\oplus\mathbb D_{c-r},
  \qquad
  \mathbb D_r=\mathbb R^r\oplus\mathbb R^r ,
  \label{eq:cgt1-defect-splitting}
\end{equation}
one can consider boundary conditions of the form
\begin{equation}
  L_{\infty,r;G_\perp,B_\perp}
  =
  \bigl(\mathbb R^r\oplus0\bigr)
  \oplus
  L_{G_\perp,B_\perp},
  \label{eq:cgt1-partial-Linfty}
\end{equation}
or
\begin{equation}
  L_{0,r;G_\perp,B_\perp}
  =
  \bigl(0\oplus\mathbb R^r\bigr)
  \oplus
  L_{G_\perp,B_\perp}.
  \label{eq:cgt1-partial-L0}
\end{equation}
Here \(L_{G_\perp,B_\perp}\subset\mathbb D_{c-r}\) is an ordinary compact Narain
boundary condition for the remaining \(c-r\) compact directions. Thus
\(L_{\infty,r;G_\perp,B_\perp}\) describes \(r\) noncompact bosons together with a
compact Narain theory of rank \(c-r\), while \(L_{0,r;G_\perp,B_\perp}\) describes the
corresponding dual noncompact directions.

These noncompact boundary conditions arise as degeneration of compact Narain topological boundaries. For example,
take \(B=0\) and
\begin{equation}
  G=
  \mathrm{diag}
  \bigl(
    R_1^2,\ldots,R_r^2,G_\perp
  \bigr).
\end{equation}
Then the compact boundary condition contains charge labels
\begin{equation}
  \left(
    {n_1\over R_1},\ldots,{n_r\over R_r};
    w_1R_1,\ldots,w_rR_r
  \right)
\end{equation}
in the first \(r\) directions. Sending \(R_1,\ldots,R_r\to\infty\), the momentum labels
\(n_i/R_i\) become continuous while the winding sectors with \(w_i\neq0\) are pushed to
infinite energy in the physical boundary Hamiltonian. The limiting topological boundary
condition is thus of the form
\eqref{eq:cgt1-partial-Linfty}. Sending \(R_i\to0\) instead gives the dual boundary
condition \eqref{eq:cgt1-partial-L0}.

The compact Narain average considered below integrates only over the compact Narain
locus of discrete lattices \(L_{G,B}\). The degenerate boundary conditions
\eqref{eq:cgt1-full-degenerate-boundaries}--\eqref{eq:cgt1-partial-L0} are not ordinary
points of the space of Lagrangian boundaries whose integral measure for averaging will be shown to be Haar measure as in $c=1$ case. Rather, they should be regarded as boundary or cusp loci in a
partial compactification of the topological-boundary space. Their role is nevertheless
important for understanding convergence: divergences of Narain averages arise precisely
from regions of the compact Narain locus that approach such noncompact limits.

\paragraph{Dualities and integral measure.}
Having determined the parametrization of the (compact) topological boundary conditions is given by  $O(c,c;\mathbb R)$, the next natural question to ask is what is the genuine space we want to average over, and whether that matches the Narain moduli space. This requires us to identify redundancies in this parametrization. First, the reference Lagrangian lattice
\(L_\circ\simeq \mathbb Z^c\oplus\mathbb Z^c\) has an automorphism group,
\begin{equation}
  O(c,c;\mathbb Z)
  =
  \mathrm{Aut}(L_\circ,\langle\, ,\,\rangle,q).
  \label{eq:cgt1-integral-t-duality-group}
\end{equation}
Acting on the integer labels \((\vec n,\vec w)\), this group preserves the integral Dirac
pairing and the quadratic refinement associated to spins. We thus have
\begin{equation}
  g\gamma L_\circ = gL_\circ,
  \qquad
  \gamma\in O(c,c;\mathbb Z),
  \label{eq:cgt1-right-integral-redundancy}
\end{equation}
which means \(O(c,c;\mathbb Z)\) is a redundancy in the parametrization of the same compact
topological boundary condition. In the absolute 2D CFT, this is the familiar integral
T-duality group. It includes changes of integral torus basis, integral \(B\)-field shifts, and
factorized dualities.

There is also a continuous frame redundancy associated with the fixed physical boundary
condition\footnote{Note that this redundancy is due to this specific physical boundary, but not SymTFT itself. That is to say, given a SymTFT, different physical boundaries may require different redundancy of space of topological boundary conditions.}. Recall that the  Hamiltonian supported on the physical boundary only depends on the Euclidean norms
\begin{equation}
  |\vec p_{\mathrm L}|^2,
  \qquad
  |\vec p_{\mathrm R}|^2 .
\end{equation}
Therefore independent orthogonal rotations
\begin{equation}
  \vec p_{\mathrm L}\longmapsto O_{\mathrm L}\vec p_{\mathrm L},
  \qquad
  \vec p_{\mathrm R}\longmapsto O_{\mathrm R}\vec p_{\mathrm R},
  \qquad
  O_{\mathrm L},O_{\mathrm R}\in O(c),
  \label{eq:cgt1-left-right-frame-redundancy}
\end{equation}
only change the orthonormal frame of the left- and right-moving current algebras. They do
not change the absolute CFT. Thus, after fixing the physical boundary condition, the compact
component of the space of topological boundary conditions is
\begin{equation}
  \mathcal L_{\mathrm{Narain}}^{(c)}
  =
  O(c,c;\mathbb Z)\backslash O(c,c;\mathbb R)/
  \bigl(O(c)\times O(c)\bigr).
  \label{eq:cgt1-narain-lagrangian-space}
\end{equation}
Equivalently, with the convention \(L_g=gL_\circ\), one may first quotient on the right by
the automorphisms of \(L_\circ\) and then quotient by the left/right frame rotations; the two
descriptions are related by \(g\mapsto g^{-1}\). This is exactly the usual Narain moduli
space, now obtained as the space of compact topological boundary conditions of the fixed
\(\mathbb R\)-valued BF SymTFT.

Before performing the averaging, let us decide the measure from the same SymTFT automorphism principle. Since
\(O(c,c;\mathbb R)\) is the continuous automorphism group preserving the BF braiding data,
the natural measure on the orbit of compact Lagrangian boundaries is the measure induced from Haar measure on
\(O(c,c;\mathbb R)\). Descending this measure through the discrete quotient by
\(O(c,c;\mathbb Z)\) and the frame quotient by \(O(c)\times O(c)\) gives an invariant measure
on \(\mathcal L_{\mathrm{Narain}}^{(c)}\). We denote it by
\begin{equation}
  d\mu(L).
  \label{eq:cgt1-narain-haar-measure}
\end{equation}
This is the continuous version of the groupoid measure discussed in
Section \ref{subsec:weights-and-measures}: the finite groupoid factor \(1/|\mathrm{Aut}(L)|\) is replaced by
the Haar measure on the quotient of the continuous automorphism orbit.

The overall normalization of \(d\mu(L)\) is conventional. In a \(G,B\) coordinate patch, the
Haar-induced measure can be written, up to an overall constant, as
\begin{equation}
  d\mu(G,B)
  =
  C_{\mathrm{Nar}}\,
  { \displaystyle
    \prod_{1\leq i\leq j\leq c} dG_{ij}
    \prod_{1\leq i<j\leq c} dB_{ij}
  \over
    (\det G)^c
  } .
  \label{eq:cgt1-GB-haar-measure}
\end{equation}
For \(c=1\), this reduces to \(dG/G=2\,dR/R\), which agrees with the
radius Haar measure of the previous subsection up to the expected overall normalization.

This Haar-induced measure is the same measure that appears in the standard Narain
average. From the 2D CFT point of view, the Narain moduli space carries the
Zamolodchikov metric, and its associated volume form agrees with the locally homogeneous
measure on
\[
  O(c,c;\mathbb Z)\backslash O(c,c;\mathbb R)/(O(c)\times O(c)).
\]
From the SymTFT viewpoint, the same measure is obtained because the space of compact
topological caps is an \(O(c,c;\mathbb R)\)-orbit and the averaging measure must be invariant
under this defect-data automorphism group. Thus the SymTFT construction reproduces both
the Narain moduli space \cite{Narain:1985jj,Narain:1986am} and the natural Narain averaging measure \cite{2006.04839, 2006.04855}.

For \(c>1\), the Haar volume of the compact Narain quotient is finite. Hence one may use the
normalized measure
\begin{equation}
  d\widehat\mu(L)
  =
  {d\mu(L)\over
  \mathrm{Vol}\!\left(\mathcal L_{\mathrm{Narain}}^{(c)}\right)} .
  \label{eq:cgt1-normalized-narain-measure}
\end{equation}
The compact Narain average below is taken with respect to this measure, whenever the
observable being averaged is integrable.

\paragraph{Averaging over topological-boundary conditions.}
Pairing the fixed physical boundary state with a compact topological boundary condition \(L\) gives
the Narain partition function
\begin{equation}
  Z_L(T^2;\tau)
  =
  \langle L;T^2|\Psi_{\mathrm{phys}}(T^2;\tau)\rangle .
  \label{eq:cgt1-ZL-pairing}
\end{equation}
Using the Lagrangian lattice \(L\subset \mathbb D_c\), this is
\begin{equation}
  Z_L(\tau)
  =
  {1\over |\eta(\tau)|^{2c}}
  \Theta_L(\tau),
  \label{eq:cgt1-ZL-theta-factorization}
\end{equation}
where
\begin{equation}
  \Theta_L(\tau)
  =
  \sum_{\ell=(\vec\alpha,\vec\beta)\in L}
  q^{{1\over 4}
  \left|\vec\alpha+\vec\beta\right|^2}
  \bar q^{{1\over 4}
  \left|\vec\alpha-\vec\beta\right|^2},
  \qquad
  q=e^{2\pi i\tau}.
  \label{eq:cgt1-siegel-narain-theta-L}
\end{equation}
The factor \(1/|\eta(\tau)|^{2c}\) comes from the oscillator part of the fixed physical
boundary Hamiltonian \eqref{eq:cgt1-boundary-hamiltonian}. The theta function
\(\Theta_L(\tau)\) is supplied entirely by the topological boundary condition \(L\). This aligns with our previous discussion that 
changing \(L\) changes the allowed current-algebra primary sectors, while the local physical
boundary dynamics is held fixed.

For the explicit compact topological boundary condition \(L_{G,B}\), this becomes
\begin{equation}
  Z_{G,B}(\tau)
  =
  {1\over |\eta(\tau)|^{2c}}
  \sum_{\vec n,\vec w\in\mathbb Z^c}
  q^{{1\over 4}
  \left|
  e^{-T}
  \bigl(
    \vec n+(B+G)\vec w
  \bigr)
  \right|^2}
  \bar q^{{1\over 4}
  \left|
  e^{-T}
  \bigl(
    \vec n+(B-G)\vec w
  \bigr)
  \right|^2}.
  \label{eq:cgt1-GB-narain-partition-function}
\end{equation}
This is the standard genus-one partition function of the toroidal Narain compactification
with metric \(G\) and \(B\)-field \(B\) \cite{2006.04855}. In the present SymTFT interpretation, however,
\((G,B)\) are not couplings in the physical boundary Hamiltonian. They are coordinates on
the space of topological boundaries \(L_{G,B}\), specifying which bulk line-defect labels are allowed to end
on the topological boundary.

The topological-boundary average over compact Narain caps is therefore
\begin{equation}
  \big\langle Z(\tau)\big\rangle_{\mathrm{top}}^{(c)}
  =
  \int_{\mathcal L_{\mathrm{Narain}}^{(c)}}
  d\widehat\mu(L)\,
  Z_L(T^2;\tau),
  \label{eq:cgt1-narain-average}
\end{equation}
whenever the integral converges as an ordinary integral. Equivalently, using the pairing
notation of \eqref{eq:cgt1-ZL-pairing},
\begin{equation}
  \big\langle Z(\tau)\big\rangle_{\mathrm{top}}^{(c)}
  =
  \int_{\mathcal L_{\mathrm{Narain}}^{(c)}}
  d\widehat\mu(L)\,
  \langle L;T^2|\Psi_{\mathrm{phys}}(T^2;\tau)\rangle .
  \label{eq:cgt1-narain-average-pairing}
\end{equation}
Here \(d\widehat\mu(L)\) is the normalized Haar-induced measure introduced in \eqref{eq:cgt1-normalized-narain-measure}. If one
instead works with the unnormalized Haar measure \(d\mu(L)\), then
\eqref{eq:cgt1-narain-average} should be replaced by
\[
  {1\over \mathrm{Vol}\!\left(\mathcal L_{\mathrm{Narain}}^{(c)}\right)}
  \int_{\mathcal L_{\mathrm{Narain}}^{(c)}}d\mu(L)\,
  Z_L(T^2;\tau).
\]
The compact Narain average integrates only over the compact locus of discrete
rank-\(2c\) lattices \(L\). The degenerate noncompact endpoints discussed above
are not included as ordinary points of the Haar integral; they appear as cusp or boundary
loci approached by sequences of compact caps.

Since the oscillator factor is independent of \(L\), the average reduces to the Haar average
of the Siegel--Narain theta function:
\begin{equation}
  \big\langle Z(\tau)\big\rangle_{\mathrm{top}}^{(c)}
  =
  {1\over |\eta(\tau)|^{2c}}
  \int_{\mathcal L_{\mathrm{Narain}}^{(c)}}
  d\widehat\mu(L)\,
  \Theta_L(\tau).
  \label{eq:cgt1-average-reduces-to-theta-average}
\end{equation}
This is the higher-rank version of the \(c=1\) radius average. 

There is an important distinction between the finite volume of the compact Narain quotient
and the convergence of the partition-function integral. For \(c>1\), the Haar volume of
\(\mathcal L_{\mathrm{Narain}}^{(c)}\) is finite, so the normalized measure
\(d\widehat\mu(L)\) exists. However, the genus-one theta integral
\[
  \int_{\mathcal L_{\mathrm{Narain}}^{(c)}}
  d\widehat\mu(L)\,
  \Theta_L(\tau)
\]
is finite only for \(c>2\) \cite{2006.04855}. At \(c=2\), the moduli-space volume is finite, but the theta
integral still diverges at the cusps, for example in large-volume and large-complex-structure
limits. Thus \(c=2\) is closer to the \(c=1\) example from the viewpoint of the unregularized
genus-one partition-function average, even though the quotient itself has finite Haar volume.

For \(c>2\), the normalized genus-one average is finite. In the convention where
\(d\widehat\mu(L)\) has unit total volume, the Siegel--Weil formula gives \cite{2006.04855}
\begin{equation}
  \int_{\mathcal L_{\mathrm{Narain}}^{(c)}}
  d\widehat\mu(L)\,
  \Theta_L(\tau)
  =
  {E_{c/2}(\tau)\over \tau_2^{c/2}},
  \qquad
  c>2,
  \label{eq:cgt1-siegel-weil-theta-average}
\end{equation}
up to the convention used for the normalization of the real-analytic Eisenstein series
\(E_{c/2}(\tau)\). Therefore
\begin{equation}
  \big\langle Z(\tau)\big\rangle_{\mathrm{top}}^{(c)}
  =
  {E_{c/2}(\tau)\over
  \tau_2^{c/2}|\eta(\tau)|^{2c}},
  \qquad
  c>2.
  \label{eq:cgt1-siegel-weil-partition-average}
\end{equation}
The measure in this statement is the Haar-induced measure on the Narain quotient, or
equivalently the measure obtained from the Zamolodchikov metric on the Narain moduli
space. The Siegel--Weil formula is the identity that evaluates the corresponding Haar
average of the Siegel--Narain theta function; it is not a separate choice of measure.

This completes the higher-rank version of the \(c=1\) dictionary. The fixed physical
boundary condition supplies the current algebra, oscillator descendants, and Hamiltonian.
The compact topological boundary condition \(L\) supplies the even self-dual charge lattice,
hence the current-algebra primary spectrum. Averaging over Narain CFTs is therefore
realized as averaging over compact topological boundary conditions of a fixed
\(\mathbb R\)-valued BF SymTFT.

\section{Toward JT Gravity and 3D Gravity}
\label{sec:JT-Virasoro}
In this paper we have studied low-dimensional ensemble averaging holography from the perspective of
SymTFTs.  The central idea is that the
physical boundary condition is held fixed, while the topological boundary condition is varied.
In the closed Marolf--Maxfield model this becomes a groupoid sum over finite-set
caps, whereas in the Narain example it becomes an integral over (compact)
topological boundary conditions of an \(\mathbb{R}\)-valued BF SymTFT.  These two examples
suggest that some familiar holographic ensemble averages can be reinterpreted as averages
over topological completions of a fixed relative theory.  It would be interesting
to understand how far this interpretation extends beyond the topological and
abelian examples discussed above.

\subsection{Toward JT Gravity and Random Matrix Theory}

A natural next example is JT gravity and its relation to random matrix theory \cite{1903.11115,1907.03363}.
Although JT gravity is usually presented as a metric-dilaton theory, it also has
a first-order BF-theory formulation \cite{1812.00918,1905.02726}.  In its simplest bosonic form, one may write
schematically\footnote{We thank Patrick Jefferson for valuable discussions inspiring the content of this subsection.}
\begin{equation}
  S_{\mathrm{BF}}
  =
  i \int_Y {\mathrm{Tr}}\, \Phi F_A
\end{equation}
together with the Euler-characteristic term \(S_0 \chi(Y)\) and the appropriate
asymptotic boundary term.  Here \(A\) is an \(SL(2,\mathbb R)\) connection, and
\(\Phi\) is an adjoint-valued scalar field whose components include the JT
dilaton.  Integrating over \(\Phi\) imposes the flatness of \(A\), so the bulk
path integral reduces, up to the boundary degrees of freedom, to an integral over
a moduli space of flat \(SL(2,\mathbb R)\) connections.  The asymptotic
\(AdS_2\) boundary condition is not topological; it gives the Schwarzian boundary
mode and hence the usual one-dimensional quantum-mechanical observable
\(\mathrm{Tr}\,e^{-\beta H}\).

From the BF-theory point of view, the scalar field \(\Phi\) is a local zero-form
field in the 2D topological theory.  Therefore given gauge-invariant
functions of \(\Phi\), for example
\begin{equation}
  {\cal O}_f(x) = f\!\left(C(x)\right),
  \qquad
  C(x)=\frac12 {\mathrm{Tr}}\,\Phi(x)^2 ,
\end{equation}
one defines local operators of the underlying 2D TFT.  In the JT
boundary problem, the quadratic Casimir \(C\) is naturally related, up to
normalization conventions, to the Hamiltonian of the Schwarzian quantum
mechanics.  Thus insertions of functions of the BF scalar should be viewed as
topological representatives of spectral observables of the boundary quantum
mechanics.

This makes JT gravity structurally close to the Marolf--Maxfield example
discussed in Section~\ref{sec:MM-from-SymTFT}.  In that example, a fixed topological parent SymTFT was
capped by a topological boundary condition labelled by a finite set \(S\), and
the resulting absolute theory had
\begin{equation}
  Z_S(S^1)=|S| .
\end{equation}
Averaging over the finite-set caps then produced the Marolf--Maxfield moments.
For JT gravity, the analogous SymTFT would be the \(SL(2,\mathbb R)\)
BF theory, refined by the local algebra generated by the dilaton/BF scalar and
by the non-topological asymptotic JT boundary condition.  Given there are nontrivial local operators in the 2D bulk, the physical boundary
should prepare a class of relative one-dimensional quantum mechanics \cite{2411.14997} rather than a single quantum mechanics or
number.  A topological boundary condition \(L\) would then specify an absolute
completion of this relative boundary theory, namely a Hilbert space and a
Hamiltonian on which the Schwarzian/spectral observables act.

In this language, one expects a fixed topological boundary condition \(L\) to
produce a factorizing one-dimensional quantum mechanics,
\begin{equation}
  Z_L(\beta)
  =
  \big\langle L;S^1 \,\big|\, \Psi_{\mathrm{phys}}(\beta)\big\rangle
  =
  {\mathrm{Tr}}_{\mathcal H_L} e^{-\beta H_L}.
\end{equation}
For several asymptotic boundaries, the same topological cap should give the
corresponding product of traces in the fixed absolute theory.  Averaging over
topological boundary conditions would therefore take the schematic form
\begin{equation}
  \left\langle \prod_{i=1}^n Z(\beta_i) \right\rangle_{\mathrm{top}}
  =
  \int_{\mathcal L_{\mathrm{JT}}} d\mu(L)\,
  \big\langle L; \bigsqcup_{i=1}^n S^1_i
  \,\big|\,
  \Psi_{\mathrm{phys}}(\beta_1,\ldots,\beta_n)
  \big\rangle .
\end{equation}
The desired statement is that, for an appropriate space
\(\mathcal L_{\mathrm{JT}}\) of admissible topological boundary conditions and an
appropriate measure \(d\mu(L)\), this average reproduces the double-scaled random
matrix ensemble average
\begin{equation}
  \left\langle \prod_{i=1}^n {\mathrm{Tr}}\,e^{-\beta_i H}
  \right\rangle_{\mathrm{RMT}}.
\end{equation}
Different choices of additional topological data, such as orientation reversal,
spin or pin structures, and supersymmetry, should then correspond to different
random matrix universality classes \cite{1907.03363}.

We emphasize that the above perspective should be regarded as a very sloppy proposal.  The
main missing ingredients are a precise classification of topological boundary
conditions for the noncompact \(SL(2,\mathbb R)\) BF theory relevant to JT
gravity, and a derivation of the measure on this space that reproduces the
matrix-model eigenvalue measure and spectral curve.  Nevertheless, the analogy
with Section~\ref{sec:MM-from-SymTFT} suggests a useful organizing principle.

\subsection{Toward Virasoro TFT and 3D Gravity}

Another natural direction is pure 3D gravity with negative cosmological
constant.  Recent work suggests that the fixed-topology path integral of pure
\(\mathrm{AdS}_3\) gravity can be reformulated in terms of a topological theory built from
the quantization of Teichm\"{u}ller space, usually called Virasoro T(Q)FT
\cite{2304.13650,2401.13900}.  In particular, for suitable quasi-local
boundary conditions, fixed-length amplitudes can be written as Virasoro T(Q)FT
amplitude-squared, while fixed-angle amplitudes are naturally related to Conformal
Turaev--Viro theory \cite{2507.12696}. More precisely,
Virasoro T(Q)FT should be regarded as a chiral building block.  The gravitational answer
is obtained by combining a left-moving and a right-moving copy, schematically
\begin{equation}
  \mathcal{T}_{\mathrm{grav}}
  \sim
  \mathcal{T}_{\mathrm{Vir}}
  \boxtimes
  \overline{\mathcal{T}}_{\mathrm{Vir}} ,
\end{equation}
or, at the level of fixed-topology amplitudes, by an absolute-square type pairing of
Virasoro-T(Q)FT wavefunctions.  This is the non-abelian analogue of
the abelian Chern--Simons/BF structures discussed in Section~\ref{sec:narain-symtft}.

It is useful to phrase this point in the language of SymTFT.  An ordinary
rational-CFT SymTFT based on a modular tensor category (MTC) usually starts from a chosen
extended chiral algebra \(\mathcal{A}\).  The corresponding MTC
captures topological defect lines that commute with, or preserve, this full chiral
algebra.  By contrast, Virasoro T(Q)FT should be thought of as the universal version in
which only the Virasoro symmetry is fixed.  Its topological line operators are expected
to encode defects that commute with the left and right Virasoro algebras, without
requiring preservation of any further extended chiral algebra.  This is why it is the
natural candidate SymTFT for an ensemble in which the central charge and Virasoro
kinematics are fixed, while the extra chiral algebra, the spectrum of Virasoro primaries,
and the OPE data are supplied by the topological boundary.

Let \(\Sigma\) be the conformal boundary.  The Virasoro T(Q)FT assigns to \(\Sigma\) a
Hilbert space of Virasoro conformal blocks, which we denote schematically by
\begin{equation}
  \mathcal{H}_{\mathrm{Vir}}(\Sigma).
\end{equation}
For gravity one should use the left-right doubled space
\begin{equation}
  \mathcal{H}_{\mathrm{grav}}(\Sigma)
  =
  \mathcal{H}_{\mathrm{Vir}}(\Sigma)
  \otimes
  \overline{\mathcal{H}}_{\mathrm{Vir}}(\Sigma).
\end{equation}
This is the direct analogue of the Hilbert space of the \(\mathbb{R}\)-valued BF SymTFT in the
Narain example.  A physical conformal boundary condition, with complex structure,
operator insertions, and possible moduli collectively denoted by \(\Omega\), should prepare a
relative state
\begin{equation}
  \big|\Psi_{\mathrm{phys}}(\Sigma;\Omega)\big\rangle
  \in
  \mathcal{H}_{\mathrm{grav}}(\Sigma).
\end{equation}
This boundary is not topological.  It contains the Virasoro Hamiltonian.  For example,
on the torus one expects the physical boundary to carry
\begin{equation}
  H_{\mathrm{phys}}
  =
  L_0+\overline{L}_0-\frac{c}{12},
\end{equation}
so that, after choosing an absolute completion via a topological boundary, the torus observable has the usual form
\begin{equation}
  Z_L(\tau,\overline{\tau})
  =
  \mathrm{Tr}_{\mathcal{H}_L}
  \left(
    q^{L_0-c/24}\,
    \overline{q}^{\,\overline{L}_0-c/24}
  \right).
\end{equation}
Here \(L\) denotes a topological boundary condition of the doubled Virasoro T(Q)FT.
As in the previous examples, the topological boundary is required to carry zero
Hamiltonian.  It does not change the local Virasoro time evolution; rather, it specifies
the absolute Hilbert space \(\mathcal{H}_L\) on which this Hamiltonian acts.

This is very close in spirit to the Narain construction.  In Section~\ref{sec:narain-symtft},
the physical boundary supplied the oscillator sector, the current algebra descendants,
and the Hamiltonian, while the (compact) topological boundary condition supplied the
even self-dual charge lattice and hence the current-algebra primary spectrum.  In the
Virasoro case, the physical boundary should supply the universal Virasoro kinematics:
the conformal blocks, the descendant contributions, and the conformal Hamiltonian.
The topological boundary condition should supply the dynamical CFT data: the
spectrum of Virasoro primaries, their multiplicities, and the OPE coefficients obeying
crossing and modular invariance.

For example, on a torus one may write schematically
\begin{equation}
  Z_L(\tau,\overline{\tau})
  =
  \sum_{(h,\overline{h})\in \mathcal{S}_L}
  N_L(h,\overline{h})\,
  \chi_h(\tau)\,
  \overline{\chi}_{\overline{h}}(\overline{\tau}),
\end{equation}
where the characters are universal Virasoro objects, while the spectrum
\(\mathcal{S}_L\) and multiplicities \(N_L(h,\overline{h})\) are supplied by the Lagrangian boundary
\(L\).  Similarly, for a four-point function on the sphere, one expects an expansion of
the schematic form
\begin{equation}
  \big\langle
    \mathcal{O}_1\mathcal{O}_2\mathcal{O}_3\mathcal{O}_4
  \big\rangle_L
  =
  \sum_{p,\overline{p}}
  C^{(L)}_{12p}\,
  C^{(L)}_{34p}\,
  \mathcal{F}_{p}
  \left[
    \begin{matrix}
      h_1 & h_2\\
      h_3 & h_4
    \end{matrix}
  \right](z)
  \,
  \overline{\mathcal{F}}_{\overline{p}}
  \left[
    \begin{matrix}
      \overline{h}_1 & \overline{h}_2\\
      \overline{h}_3 & \overline{h}_4
    \end{matrix}
  \right](\overline{z}) .
\end{equation}
Again, the conformal blocks are universal and belong to the physical boundary data, while the coefficients \(C^{(L)}_{ijk}\) and the spectrum of exchanged
primaries are part of the topological completion.

Thus a possible SymTFT interpretation of the ensemble dual to pure
\(\mathrm{AdS}_3\) gravity is the following.  We conjecture that there should exist a space
\begin{equation}
  \mathcal{L}_{\mathrm{Vir}}
\end{equation}
of admissible topological boundary conditions of the doubled Virasoro T(Q)FT.  For
each \(L\in \mathcal{L}_{\mathrm{Vir}}\), the pairing
\begin{equation}
  Z_L(\Sigma;m)
  =
  \big\langle L;\Sigma \,\big|\,
  \Psi_{\mathrm{phys}}(\Sigma;m)
  \big\rangle
\end{equation}
defines an absolute 2D CFT-like object with fixed central charge and fixed
Virasoro symmetry.  The proposed ensemble average is then
\begin{equation}
  \big\langle Z(\Sigma;m)\big\rangle_{\mathrm{top}}
  =
  \int_{\mathcal{L}_{\mathrm{Vir}}}
  d\mu(L)\,
  \big\langle L;\Sigma \,\big|\,
  \Psi_{\mathrm{phys}}(\Sigma;m)
  \big\rangle .
\end{equation}
For several disconnected conformal boundaries, the same topological boundary condition
should be used on all components:
\begin{equation}
  \left\langle
    \prod_{a=1}^n Z(\Sigma_a;m_a)
  \right\rangle_{\mathrm{top}}
  =
  \int_{\mathcal{L}_{\mathrm{Vir}}}
  d\mu(L)\,
  \prod_{a=1}^n
  \big\langle L;\Sigma_a \,\big|\,
  \Psi_{\mathrm{phys}}(\Sigma_a;m_a)
  \big\rangle .
\end{equation}
At fixed \(L\), the answer factorizes, as it should for a single absolute CFT.  The
non-factorization of the gravitational answer is then produced by the integral over
topological caps.

This perspective also gives a natural way to interpret the wormhole computations in
Virasoro T(Q)FT.  A multi-boundary Euclidean wormhole computes a moment of CFT
data in the putative ensemble.  In the present language, this moment should arise because
the same topological boundary condition \(L\) is used to cap all physical boundaries
before being averaged over.  Thus the Virasoro-T(Q)FT wormhole is not changing the
local boundary Hamiltonian; it is correlating the absolute completion data, namely the
spectrum and OPE coefficients, across different copies of the physical boundary.

Here the main missing ingredient, which is also the most nontrivial step, is the correct definition of
\(\mathcal{L}_{\mathrm{Vir}}\) and of the measure \(d\mu(L)\).  For finite rational T(Q)FTs,
topological boundary conditions are described by Lagrangian algebras in a modular
tensor category.  For the \(R\)-valued BF theories used in the Narain example, this
becomes a continuous space of Lagrangian subgroups equipped with a Haar-type measure.
The Virasoro case should be simultaneously non-rational and non-compact analogue of these two
situations.  A topological boundary condition should be an object that pairs left and
right Virasoro conformal blocks into a single-valued, crossing-symmetric, modular
invariant full CFT.  In categorical language, it should be a kind of Lagrangian algebra
or full-center object in the non-rational braided category of Virasoro representations,
but this statement requires functional-analytic input beyond the finite semisimple
framework. 

The relation to quantum groups may be useful precisely here.  Quantum
Teichm\"uller theory is closely related to the modular double of
\(U_q(\mathfrak{sl}(2,\mathbb{R}))\) \cite{1109.6295,1302.3454}, with
\begin{equation}
  q=e^{i\pi b^2},
  \qquad
  \widetilde{q}=e^{i\pi b^{-2}},
  \qquad
  c=1+6(b+b^{-1})^2 .
\end{equation}
The fusion and braiding kernels of Virasoro conformal blocks can be interpreted in
terms of Clebsch--Gordan coefficients and \(6j\)-symbols of this modular-double
quantum group \cite{math/0007097,1202.4698,2309.11540}.  Therefore the algebra of topological line operators and the
Moore--Seiberg moves of Virasoro T(Q)FT should admit a quantum-group description.
From this point of view, the desired topological boundary conditions may be formulated
as suitable module categories, or commutative algebra objects, for the representation
category of the modular double.  The measure \(d\mu(L)\) should then be related to
the Plancherel measure \cite{math/0007097,1202.4698} or Haar-type measure naturally associated with this non-compact
quantum group.

Again, we emphasize that this should be taken only as a very sloppy proposal, not as a promising construction.  At generic
\(q\), the relevant representation theory is continuous and non-compact, so one should
not expect a finite groupoid sum over Lagrangian algebras.  At special rational or
root-of-unity points, finite quantum-group categories may give controlled toy models,
closer to ordinary Reshetikhin--Turaev \cite{Reshetikhin:1991tc} or Turaev--Viro theory \cite{Turaev:1992hq}.
The generic Virasoro
case, however, should involve a genuinely continuous topological-boundary average,
more like the Narain integral than like the finite Marolf--Maxfield groupoid sum.

The conceptual payoff would be significant.  Pure \(\mathrm{AdS}_3\) gravity would then
fit the same pattern as the examples studied in this paper:
For Narain theories, the topological boundary condition is equivalent to a charge lattice.  For the
Marolf--Maxfield model, it is a finite set.  For pure 3D gravity, it should
be the full Virasoro CFT data selected by a topological boundary condition of the
doubled Virasoro T(Q)FT.  Making this statement precise would require a classification
of admissible Virasoro topological boundaries, a derivation of the corresponding measure,
and a proof that this topological-boundary average reproduces the known Virasoro-T(Q)FT
gravity amplitudes.  These are open problems, but the analogy suggests a concrete route
for organizing them.

\addsec{Acknowledgement}
The author thanks Scott Collier, Max H\"{u}bner, Patrick Jefferson, Ho Tat Lam,  Alex Maloney, and Ling-Yan Hung for valuable discussions, with special thanks to Yikun Jiang for insightful discussions and encouragement at various stages of this project. The authors thanks Jonathan Heckman, Ethan Torres, and Andrew Turner for collaboration on related projects. This work is partially supported by the NSF
grant PHY-2310588.

\printbibliography[heading=bibliography]

@article{Heckman:2021vzx,
    author = "Heckman, Jonathan J. and Turner, Andrew P. and Yu, Xingyang",
    title = "{Disorder averaging and its UV discontents}",
    eprint = "2111.06404",
    archivePrefix = "arXiv",
    primaryClass = "hep-th",
    doi = "10.1103/PhysRevD.105.086021",
    journal = "Phys. Rev. D",
    volume = "105",
    number = "8",
    pages = "086021",
    year = "2022"
}

@article{2306.11783,
    author = "Lawrie, Craig and Yu, Xingyang and Zhang, Hao Y.",
    title = "{Intermediate defect groups, polarization pairs, and noninvertible duality defects}",
    eprint = "2306.11783",
    archivePrefix = "arXiv",
    primaryClass = "hep-th",
    reportNumber = "DESY-23-079; UPR-1323-T",
    doi = "10.1103/PhysRevD.109.026005",
    journal = "Phys. Rev. D",
    volume = "109",
    number = "2",
    pages = "026005",
    year = "2024"
}

@article{2311.16230,
    author = "Perez-Lona, A. and Robbins, D. and Sharpe, E. and Vandermeulen, T. and Yu, X.",
    title = "{Notes on gauging noninvertible symmetries. Part I. Multiplicity-free cases}",
    eprint = "2311.16230",
    archivePrefix = "arXiv",
    primaryClass = "hep-th",
    doi = "10.1007/JHEP02(2024)154",
    journal = "JHEP",
    volume = "02",
    pages = "154",
    year = "2024"
}

@article{2411.14997,
    author = "Yu, Xingyang",
    title = "{Gauging in parameter space: A top-down perspective}",
    eprint = "2411.14997",
    archivePrefix = "arXiv",
    primaryClass = "hep-th",
    doi = "10.1103/638n-qwnm",
    journal = "Phys. Rev. D",
    volume = "112",
    number = "2",
    pages = "025020",
    year = "2025"
}

@article{2401.10165,
    author = "Antinucci, Andrea and Benini, Francesco",
    title = "{Anomalies and gauging of U(1) symmetries}",
    eprint = "2401.10165",
    archivePrefix = "arXiv",
    primaryClass = "hep-th",
    reportNumber = "SISSA 01/2024/FISI",
    doi = "10.1103/PhysRevB.111.024110",
    journal = "Phys. Rev. B",
    volume = "111",
    number = "2",
    pages = "024110",
    year = "2025"
}

@article{Kong:2014qka,
    author = "Kong, Liang and Wen, Xiao-Gang",
    title = "{Braided fusion categories, gravitational anomalies, and the mathematical framework for topological orders in any dimensions}",
    eprint = "1405.5858",
    archivePrefix = "arXiv",
    primaryClass = "cond-mat.str-el",
    month = "5",
    year = "2014"
}

@article{Kong:2017hcw,
    author = "Kong, Liang and Wen, Xiao-Gang and Zheng, Hao",
    title = "{Boundary-bulk relation in topological orders}",
    eprint = "1702.00673",
    archivePrefix = "arXiv",
    primaryClass = "cond-mat.str-el",
    doi = "10.1016/j.nuclphysb.2017.06.023",
    journal = "Nucl. Phys. B",
    volume = "922",
    pages = "62--76",
    year = "2017"
}

@article{Kong:2020cie,
    author = "Kong, Liang and Lan, Tian and Wen, Xiao-Gang and Zhang, Zhi-Hao and Zheng, Hao",
    title = "{Algebraic higher symmetry and categorical symmetry -- a holographic and entanglement view of symmetry}",
    eprint = "2005.14178",
    archivePrefix = "arXiv",
    primaryClass = "cond-mat.str-el",
    doi = "10.1103/PhysRevResearch.2.043086",
    journal = "Phys. Rev. Res.",
    volume = "2",
    number = "4",
    pages = "043086",
    year = "2020"
}

@article{Kitaev:2011dxc,
    author = "Kitaev, Alexei and Kong, Liang",
    title = "{Models for Gapped Boundaries and Domain Walls}",
    eprint = "1104.5047",
    archivePrefix = "arXiv",
    primaryClass = "cond-mat.str-el",
    doi = "10.1007/s00220-012-1500-5",
    journal = "Commun. Math. Phys.",
    volume = "313",
    number = "2",
    pages = "351--373",
    year = "2012"
}

@article{2510.06319,
    author = "Torres, Ethan and Yu, Xingyang",
    title = "{SymTFT Entanglement and Holographic (Non)-Factorization}",
    eprint = "2510.06319",
    archivePrefix = "arXiv",
    primaryClass = "hep-th",
    reportNumber = "CERN-TH-2025-194",
    month = "10",
    year = "2025"
}

@article{Franco:2024mxa,
    author = "Franco, Sebastian and Yu, Xingyang",
    title = "{Generalized symmetries in 2D from string theory: SymTFTs, intrinsic relativeness, and anomalies of non-invertible symmetries}",
    eprint = "2404.19761",
    archivePrefix = "arXiv",
    primaryClass = "hep-th",
    doi = "10.1007/JHEP11(2024)004",
    journal = "JHEP",
    volume = "11",
    pages = "004",
    year = "2024"
}

@article{Jia:2025jmn,
    author = "Jia, Qiang and Luo, Ran and Tian, Jiahua and Wang, Yi-Nan and Zhang, Yi",
    title = "{Symmetry Topological Field Theory for Flavor Symmetry}",
    eprint = "2503.04546",
    archivePrefix = "arXiv",
    primaryClass = "hep-th",
    month = "3",
    year = "2025"
}

@article{Bhardwaj:2024igy,
    author = "Bhardwaj, Lakshya and Copetti, Christian and Pajer, Daniel and Schafer-Nameki, Sakura",
    title = "{Boundary SymTFT}",
    eprint = "2409.02166",
    archivePrefix = "arXiv",
    primaryClass = "hep-th",
    doi = "10.21468/SciPostPhys.19.2.061",
    journal = "SciPost Phys.",
    volume = "19",
    number = "2",
    pages = "061",
    year = "2025"
}

@article{Gagliano:2024off,
    author = "Gagliano, Finn and Garc{\'\i}a Etxebarria, I{\~n}aki",
    title = "{SymTFTs for $U(1)$ symmetries from descent}",
    eprint = "2411.15126",
    archivePrefix = "arXiv",
    primaryClass = "hep-th",
    month = "11",
    year = "2024"
}

@article{Cordova:2024iti,
    author = "Cordova, Clay and Holfester, Nicholas and Ohmori, Kantaro",
    title = "{Representation theory of solitons}",
    eprint = "2408.11045",
    archivePrefix = "arXiv",
    primaryClass = "hep-th",
    doi = "10.1007/JHEP06(2025)001",
    journal = "JHEP",
    volume = "06",
    pages = "001",
    year = "2025"
}

@article{2401.06128,
    author = "Brennan, T. Daniel and Sun, Zhengdi",
    title = "{A SymTFT for continuous symmetries}",
    eprint = "2401.06128",
    archivePrefix = "arXiv",
    primaryClass = "hep-th",
    doi = "10.1007/JHEP12(2024)100",
    journal = "JHEP",
    volume = "12",
    pages = "100",
    year = "2024"
}

@article{Luo:2025phx,
    author = "Luo, Ran and Wang, Yi-Nan and Bi, Zhen",
    title = "{Topological Holography for Mixed-State Phases and Phase Transitions}",
    eprint = "2507.06218",
    archivePrefix = "arXiv",
    primaryClass = "cond-mat.str-el",
    month = "7",
    year = "2025"
}

@article{Bonetti:2024cjk,
    author = "Bonetti, Federico and Del Zotto, Michele and Minasian, Ruben",
    title = "{SymTFTs for Continuous non-Abelian Symmetries}",
    eprint = "2402.12347",
    archivePrefix = "arXiv",
    primaryClass = "hep-th",
    month = "2",
    year = "2024"
}

@article{Argurio:2024oym,
    author = "Argurio, Riccardo and Benini, Francesco and Bertolini, Matteo and Galati, Giovanni and Niro, Pierluigi",
    title = "{On the symmetry TFT of Yang-Mills-Chern-Simons theory}",
    eprint = "2404.06601",
    archivePrefix = "arXiv",
    primaryClass = "hep-th",
    reportNumber = "SISSA 05/2024/FISI",
    doi = "10.1007/JHEP07(2024)130",
    journal = "JHEP",
    volume = "07",
    pages = "130",
    year = "2024"
}

@article{Kirillov2010TuraevViroIA,
      title={Turaev-Viro invariants as an extended TQFT},
      author={Alexander Kirillov Jr. and Benjamin Balsam},
      year={2010},
      eprint={1004.1533},
      archivePrefix={arXiv},
      primaryClass={math.GT}
}

@article{Apruzzi:2024htg,
    author = "Apruzzi, Fabio and Bedogna, Francesco and Dondi, Nicola",
    title = "{SymTh for non-finite symmetries}",
    eprint = "2402.14813",
    archivePrefix = "arXiv",
    primaryClass = "hep-th",
    month = "2",
    year = "2024"
}

@article{Cvetic:2024dzu,
    author = {Cveti{\v{c}}, Mirjam and Donagi, Ron and Heckman, Jonathan J. and H{\"u}bner, Max and Torres, Ethan},
    title = "{Cornering relative symmetry theories}",
    eprint = "2408.12600",
    archivePrefix = "arXiv",
    primaryClass = "hep-th",
    reportNumber = "UPR-1331-TH, CERN-TH-2024-139",
    doi = "10.1103/PhysRevD.111.085026",
    journal = "Phys. Rev. D",
    volume = "111",
    number = "8",
    pages = "085026",
    year = "2025"
}

@article{Heckman:2024zdo,
    author = {Heckman, Jonathan J. and H{\"u}bner, Max},
    title = "{Celestial Topology, Symmetry Theories, and Evidence for a NonSUSY D3-Brane CFT}",
    eprint = "2406.08485",
    archivePrefix = "arXiv",
    primaryClass = "hep-th",
    doi = "10.1002/prop.202400270",
    journal = "Fortsch. Phys.",
    volume = "73",
    number = "4",
    pages = "2400270",
    year = "2025"
}

@article{Heckman:2025lmw,
    author = {Heckman, Jonathan J. and H{\"u}bner, Max and Murdia, Chitraang},
    title = "{Symmetry Theories, Wigner's Function, Compactification, and Holography}",
    eprint = "2505.23887",
    archivePrefix = "arXiv",
    primaryClass = "hep-th",
    month = "5",
    year = "2025"
}

@article{Apruzzi:2025mdl,
    author = "Apruzzi, Fabio and Bedogna, Francesco and Mancani, Salvo",
    title = "{SymTFT construction of gapless exotic-foliated dual models}",
    eprint = "2504.11449",
    archivePrefix = "arXiv",
    primaryClass = "cond-mat.str-el",
    month = "4",
    year = "2025"
}

@article{Yu:2023nyn,
    author = "Yu, Xingyang",
    title = "{Non-invertible Symmetries in 2D from Type IIB String Theory}",
    eprint = "2310.15339",
    archivePrefix = "arXiv",
    primaryClass = "hep-th",
    month = "10",
    year = "2023"
}

@article{Antinucci:2022vyk,
    author = "Antinucci, Andrea and Benini, Francesco and Copetti, Christian and Galati, Giovanni and Rizi, Giovanni",
    title = "{The holography of non-invertible self-duality symmetries}",
    eprint = "2210.09146",
    archivePrefix = "arXiv",
    primaryClass = "hep-th",
    reportNumber = "SISSA 16/2022/FISI",
    month = "10",
    year = "2022"
}

@article{Kaidi:2022cpf,
    author = "Kaidi, Justin and Ohmori, Kantaro and Zheng, Yunqin",
    title = "{Symmetry TFTs for Non-Invertible Defects}",
    eprint = "2209.11062",
    archivePrefix = "arXiv",
    primaryClass = "hep-th",
    month = "9",
    year = "2022"
}

@article{2010.15890,
    author = "Gukov, Sergei and Hsin, Po-Shen and Pei, Du",
    title = "{Generalized global symmetries of $T[M]$ theories. Part I}",
    eprint = "2010.15890",
    archivePrefix = "arXiv",
    primaryClass = "hep-th",
    reportNumber = "CALT-TH-2020-045",
    doi = "10.1007/JHEP04(2021)232",
    journal = "JHEP",
    volume = "04",
    pages = "232",
    year = "2021"
}

@article{1412.5148,
    author = "Gaiotto, Davide and Kapustin, Anton and Seiberg, Nathan and Willett, Brian",
    title = "{Generalized Global Symmetries}",
    eprint = "1412.5148",
    archivePrefix = "arXiv",
    primaryClass = "hep-th",
    doi = "10.1007/JHEP02(2015)172",
    journal = "JHEP",
    volume = "02",
    pages = "172",
    year = "2015"
}

@article{1212.1692,
    author = "Freed, Daniel S. and Teleman, Constantin",
    title = "{Relative quantum field theory}",
    eprint = "1212.1692",
    archivePrefix = "arXiv",
    primaryClass = "hep-th",
    doi = "10.1007/s00220-013-1880-1",
    journal = "Commun. Math. Phys.",
    volume = "326",
    pages = "459--476",
    year = "2014"
}

@article{hep-th/0204148,
    author = "Fuchs, Jurgen and Runkel, Ingo and Schweigert, Christoph",
    title = "{TFT construction of RCFT correlators 1. Partition functions}",
    eprint = "hep-th/0204148",
    archivePrefix = "arXiv",
    reportNumber = "PAR-LPTHE-02-25",
    doi = "10.1016/S0550-3213(02)00744-7",
    journal = "Nucl. Phys. B",
    volume = "646",
    pages = "353--497",
    year = "2002"
}

@article{1012.0911,
    author = "Kapustin, Anton and Saulina, Natalia",
    editor = "Sati, Hisham and Schreiber, Urs",
    title = "{Surface operators in 3d Topological Field Theory and 2d Rational Conformal Field Theory}",
    eprint = "1012.0911",
    archivePrefix = "arXiv",
    primaryClass = "hep-th",
    reportNumber = "PI-STRINGS-195",
    pages = "175--198",
    month = "12",
    year = "2010"
}

@article{hep-th/9812012,
    author = "Witten, Edward",
    title = "{AdS / CFT correspondence and topological field theory}",
    eprint = "hep-th/9812012",
    archivePrefix = "arXiv",
    reportNumber = "IASSNS-HEP-98-96",
    doi = "10.1088/1126-6708/1998/12/012",
    journal = "JHEP",
    volume = "12",
    pages = "012",
    year = "1998"
}

@article{Apruzzi:2021nmk,
    author = "Apruzzi, Fabio and Bonetti, Federico and Garc\'\i{}a Etxebarria, I. and Hosseini, Saghar S. and Schafer-Nameki, Sakura",
    title = "{Symmetry TFTs from String Theory}",
    eprint = "2112.02092",
    archivePrefix = "arXiv",
    primaryClass = "hep-th",
    month = "12",
    year = "2021"
}

@article{1008.0654,
    author = "Kapustin, Anton and Saulina, Natalia",
    title = "{Topological boundary conditions in abelian Chern-Simons theory}",
    eprint = "1008.0654",
    archivePrefix = "arXiv",
    primaryClass = "hep-th",
    doi = "10.1016/j.nuclphysb.2010.12.017",
    journal = "Nucl. Phys. B",
    volume = "845",
    pages = "393--435",
    year = "2011"
}

@article{Heckman:2017uxe,
    author = "Heckman, Jonathan J. and Tizzano, Luigi",
    title = "{6D Fractional Quantum Hall Effect}",
    eprint = "1708.02250",
    archivePrefix = "arXiv",
    primaryClass = "hep-th",
    doi = "10.1007/JHEP05(2018)120",
    journal = "JHEP",
    volume = "05",
    pages = "120",
    year = "2018"
}

@article{Gaiotto:2020iye,
    author = "Gaiotto, Davide and Kulp, Justin",
    title = "{Orbifold groupoids}",
    eprint = "2008.05960",
    archivePrefix = "arXiv",
    primaryClass = "hep-th",
    doi = "10.1007/JHEP02(2021)132",
    journal = "JHEP",
    volume = "02",
    pages = "132",
    year = "2021"
}

@article{Baume:2023kkf,
    author = {Baume, Florent and Heckman, Jonathan J. and H\"ubner, Max and Torres, Ethan and Turner, Andrew P. and Yu, Xingyang},
    title = "{SymTrees and Multi-Sector QFTs}",
    eprint = "2310.12980",
    archivePrefix = "arXiv",
    primaryClass = "hep-th",
    reportNumber = "ZMP-HH/23-13, CERN-TH-2023-183",
    month = "10",
    year = "2023"
}

@article{DelZotto:2024tae,
    author = "Del Zotto, Michele and Meynet, Shani Nadir and Moscrop, Robert",
    title = "{Remarks on geometric engineering, symmetry TFTs and anomalies}",
    eprint = "2402.18646",
    archivePrefix = "arXiv",
    primaryClass = "hep-th",
    doi = "10.1007/JHEP07(2024)220",
    journal = "JHEP",
    volume = "07",
    pages = "220",
    year = "2024"
}

@article{Narain:1985jj,
    author = "Narain, K. S.",
    title = "{New Heterotic String Theories in Uncompactified Dimensions \ensuremath{<} 10}",
    reportNumber = "RAL-85-097",
    doi = "10.1016/0370-2693(86)90682-9",
    journal = "Phys. Lett. B",
    volume = "169",
    pages = "41--46",
    year = "1986"
}

@article{Narain:1986am,
    author = "Narain, K. S. and Sarmadi, M. H. and Witten, Edward",
    title = "{A Note on Toroidal Compactification of Heterotic String Theory}",
    reportNumber = "Print-86-0870 (RUTHERFORD), RAL-86-022",
    doi = "10.1016/0550-3213(87)90001-0",
    journal = "Nucl. Phys. B",
    volume = "279",
    pages = "369--379",
    year = "1987"
}

@article{Freed:2022qnc,
    author = "Freed, Daniel S. and Moore, Gregory W. and Teleman, Constantin",
    title = "{Topological symmetry in quantum field theory}",
    eprint = "2209.07471",
    archivePrefix = "arXiv",
    primaryClass = "hep-th",
    month = "9",
    year = "2022"
}

@article{Pace:2025hpb,
    author = {Pace, Salvatore D. and Aksoy, {\"O}mer M. and Lam, Ho Tat},
    title = "{Spacetime symmetry-enriched SymTFT: From LSM anomalies to modulated symmetries and beyond}",
    eprint = "2507.02036",
    archivePrefix = "arXiv",
    primaryClass = "cond-mat.str-el",
    reportNumber = "MIT-CTP/5884",
    doi = "10.21468/SciPostPhys.20.1.007",
    journal = "SciPost Phys.",
    volume = "20",
    number = "1",
    pages = "007",
    year = "2026"
}

@article{Apruzzi:2025hvs,
    author = "Apruzzi, Fabio and Dondi, Nicola and Garc{\'\i}a Etxebarria, I{\~n}aki and Lam, Ho Tat and Schafer-Nameki, Sakura",
    title = "{Symmetry TFTs for Continuous Spacetime Symmetries}",
    eprint = "2509.07965",
    archivePrefix = "arXiv",
    primaryClass = "hep-th",
    reportNumber = "MIT-CTP/5921",
    month = "9",
    year = "2025"
}

@article{Reshetikhin:1991tc,
    author = "Reshetikhin, N. and Turaev, V. G.",
    title = "{Invariants of three manifolds via link polynomials and quantum groups}",
    doi = "10.1007/BF01239527",
    journal = "Invent. Math.",
    volume = "103",
    pages = "547--597",
    year = "1991"
}

@article{Turaev:1992hq,
    author = "Turaev, V. G. and Viro, O. Y.",
    title = "{State sum invariants of 3 manifolds and quantum 6j symbols}",
    doi = "10.1016/0040-9383(92)90015-A",
    journal = "Topology",
    volume = "31",
    pages = "865--902",
    year = "1992"
}

@article{Barrett:1993ab,
    author = "Barrett, John W. and Westbury, Bruce W.",
    title = "{Invariants of piecewise linear three manifolds}",
    eprint = "hep-th/9311155",
    archivePrefix = "arXiv",
    doi = "10.1090/S0002-9947-96-01660-1",
    journal = "Trans. Am. Math. Soc.",
    volume = "348",
    pages = "3997--4022",
    year = "1996"
}

@article{Fuchs:2012dt,
    author = "Fuchs, Jurgen and Schweigert, Christoph and Valentino, Alessandro",
    title = "{Bicategories for boundary conditions and for surface defects in 3-d TFT}",
    eprint = "1203.4568",
    archivePrefix = "arXiv",
    primaryClass = "hep-th",
    reportNumber = "HAMBURGER-BEITR-ZUR-MATHEMATIK-NR-433, ZMP-HH-12-5, HAMBURGER-BEITR.-ZUR-MATHEMATIK-NR.-433",
    doi = "10.1007/s00220-013-1723-0",
    journal = "Commun. Math. Phys.",
    volume = "321",
    pages = "543--575",
    year = "2013"
}

@article{Freed:2018cec,
	archiveprefix = {arXiv},
	author = {Freed, Daniel S. and Teleman, Constantin},
	doi = {10.2140/gt.2022.26.1907},
	eprint = {1806.00008},
	journal = {Geom. Topol.},
	pages = {1907--1984},
	primaryclass = {math.AT},
	title = {{Topological dualities in the Ising model}},
	volume = {26},
	year = {2022}}

@article{Abrams:1996ty,
    author = "Abrams, Leonard S.",
    title = "{Two-dimensional topological quantum field theories and Frobenius algebras}",
    doi = "10.1142/S0218216596000333",
    journal = "J. Knot Theor. Ramifications",
    volume = "5",
    pages = "569--587",
    year = "1996"
}

@book{Kock_2003, place={Cambridge}, series={London Mathematical Society Student Texts}, title={Frobenius Algebras and 2-D Topological Quantum Field Theories}, publisher={Cambridge University Press}, author={Kock, Joachim}, year={2003}, collection={London Mathematical Society Student Texts}}

@article{9006cc9e-2dcc-3fd8-aada-e4af19b6e225,
 ISSN = {0003486X, 19398980},
 URL = {http://www.jstor.org/stable/1968346},
 author = {Alfred Haar},
 journal = {Annals of Mathematics},
 number = {1},
 pages = {147--169},
 publisher = {[Annals of Mathematics, Trustees of Princeton University on Behalf of the Annals of Mathematics, Mathematics Department, Princeton University]},
 title = {Der Massbegriff in der Theorie der Kontinuierlichen Gruppen},
 urldate = {2026-05-07},
 volume = {34},
 year = {1933}
}

@article{Coleman:1988cy,
    author = "Coleman, Sidney R.",
    title = "{Black holes as red herrings: Topological fluctuations and the loss of quantum coherence}",
    reportNumber = "HUTP-88/A008",
    doi = "10.1016/0550-3213(88)90110-1",
    journal = "Nucl. Phys. B",
    volume = "307",
    pages = "867--882",
    year = "1988"
}

@article{1903.11115,
    author = "Saad, Phil and Shenker, Stephen H. and Stanford, Douglas",
    title = "{JT gravity as a matrix integral}",
    eprint = "1903.11115",
    archivePrefix = "arXiv",
    primaryClass = "hep-th",
    month = "3",
    year = "2019"
}

@article{1907.03363,
    author = "Stanford, Douglas and Witten, Edward",
    title = "{JT gravity and the ensembles of random matrix theory}",
    eprint = "1907.03363",
    archivePrefix = "arXiv",
    primaryClass = "hep-th",
    doi = "10.4310/ATMP.2020.v24.n6.a4",
    journal = "Adv. Theor. Math. Phys.",
    volume = "24",
    number = "6",
    pages = "1475--1680",
    year = "2020"
}

@article{2002.08950,
    author = "Marolf, Donald and Maxfield, Henry",
    title = "{Transcending the ensemble: baby universes, spacetime wormholes, and the order and disorder of black hole information}",
    eprint = "2002.08950",
    archivePrefix = "arXiv",
    primaryClass = "hep-th",
    doi = "10.1007/JHEP08(2020)044",
    journal = "JHEP",
    volume = "08",
    pages = "044",
    year = "2020"
}

@article{2006.04855,
    author = "Maloney, Alexander and Witten, Edward",
    title = "{Averaging over Narain moduli space}",
    eprint = "2006.04855",
    archivePrefix = "arXiv",
    primaryClass = "hep-th",
    doi = "10.1007/JHEP10(2020)187",
    journal = "JHEP",
    volume = "10",
    pages = "187",
    year = "2020"
}

@article{2006.04839,
    author = "Afkhami-Jeddi, Nima and Cohn, Henry and Hartman, Thomas and Tajdini, Amirhossein",
    title = "{Free partition functions and an averaged holographic duality}",
    eprint = "2006.04839",
    archivePrefix = "arXiv",
    primaryClass = "hep-th",
    doi = "10.1007/JHEP01(2021)130",
    journal = "JHEP",
    volume = "01",
    pages = "130",
    year = "2021"
}

@article{2405.20366,
    author = "Dymarsky, Anatoly and Shapere, Alfred",
    title = "{TQFT gravity and ensemble holography}",
    eprint = "2405.20366",
    archivePrefix = "arXiv",
    primaryClass = "hep-th",
    doi = "10.1007/JHEP02(2025)091",
    journal = "JHEP",
    volume = "02",
    pages = "091",
    year = "2025"
}

@article{2212.00195,
    author = "Freed, Daniel S.",
    title = "{Introduction to topological symmetry in QFT.}",
    eprint = "2212.00195",
    archivePrefix = "arXiv",
    primaryClass = "hep-th",
    doi = "10.1090/pspum/107/01946",
    journal = "Proc. Symp. Pure Math.",
    volume = "107",
    pages = "93--106",
    year = "2024"
}

@article{0712.0155,
    author = "Maloney, Alexander and Witten, Edward",
    title = "{Quantum Gravity Partition Functions in Three Dimensions}",
    eprint = "0712.0155",
    archivePrefix = "arXiv",
    primaryClass = "hep-th",
    doi = "10.1007/JHEP02(2010)029",
    journal = "JHEP",
    volume = "02",
    pages = "029",
    year = "2010"
}

@article{1407.6008,
    author = "Keller, Christoph A. and Maloney, Alexander",
    title = "{Poincare Series, 3D Gravity and CFT Spectroscopy}",
    eprint = "1407.6008",
    archivePrefix = "arXiv",
    primaryClass = "hep-th",
    reportNumber = "RUNHETC-2014-13",
    doi = "10.1007/JHEP02(2015)080",
    journal = "JHEP",
    volume = "02",
    pages = "080",
    year = "2015"
}

@article{2201.00903,
    author = "Banerjee, Anindya and Moore, Gregory W.",
    title = "{Comments on summing over bordisms in TQFT}",
    eprint = "2201.00903",
    archivePrefix = "arXiv",
    primaryClass = "hep-th",
    doi = "10.1007/JHEP09(2022)171",
    journal = "JHEP",
    volume = "09",
    pages = "171",
    year = "2022"
}

@article{2310.13044,
    author = "Barbar, Ahmed and Dymarsky, Anatoly and Shapere, Alfred D.",
    title = "{Global Symmetries, Code Ensembles, and Sums over Geometries}",
    eprint = "2310.13044",
    archivePrefix = "arXiv",
    primaryClass = "hep-th",
    doi = "10.1103/PhysRevLett.134.151603",
    journal = "Phys. Rev. Lett.",
    volume = "134",
    number = "15",
    pages = "151603",
    year = "2025"
}

@article{2511.04311,
    author = "Barbar, Ahmed",
    title = "{Automorphism-weighted ensembles from TQFT gravity}",
    eprint = "2511.04311",
    archivePrefix = "arXiv",
    primaryClass = "hep-th",
    month = "11",
    year = "2025"
}

@article{Barbar:2025krh,
    author = "Barbar, Ahmed and Dymarsky, Anatoly and Shapere, Alfred",
    title = "{Holographic description of 4d Maxwell theories and their code-based ensembles}",
    eprint = "2510.03392",
    archivePrefix = "arXiv",
    primaryClass = "hep-th",
    doi = "10.1007/JHEP06(2026)153",
    journal = "JHEP",
    volume = "06",
    pages = "153",
    year = "2026"
}

@article{Dymarsky:2026asf,
    author = "Dymarsky, Anatoly and Shapere, Alfred",
    title = "{Mass formula for topological boundary conditions from TQFT gravity}",
    eprint = "2602.00224",
    archivePrefix = "arXiv",
    primaryClass = "hep-th",
    month = "1",
    year = "2026"
}

@article{Jia:2026tfh,
    author = "Jia, Qiang and Zhang, Yi",
    title = "{Flat Gauging of Continuous (Non-invertible) Symmetries and Non-compact BF SymTFT for Compact Boson}",
    eprint = "2606.15732",
    archivePrefix = "arXiv",
    primaryClass = "hep-th",
    month = "6",
    year = "2026"
}

@article{2412.11486,
    author = "Takahashi, Shunta",
    title = "{Anyon Condensation in Virasoro TQFT: Wormhole Factorization}",
    eprint = "2412.11486",
    archivePrefix = "arXiv",
    primaryClass = "hep-th",
    doi = "10.1007/JHEP06(2025)243",
    journal = "JHEP",
    volume = "06",
    pages = "243",
    year = "2025"
}

@misc{Lam:KITP2025,
    author = "Lam, Ho Tat",
    title = "{Talk at KITP conference ``Generalized Symmetries: High-Energy, Condensed Matter and Mathematics''}",
    howpublished = "KITP, Santa Barbara, April 2025",
    year = "2025"
}

@misc{Lam:Harvard2025,
    author = "Lam, Ho Tat",
    title = "{Talk at the Harvard workshop ``Workshop on Symmetries and Gravity''}",
    howpublished = "Harvard University, 2025",
    year = "2025"
}

@article{1812.00918,
    author = "Blommaert, Andreas and Mertens, Thomas G. and Verschelde, Henri",
    title = "{Fine Structure of Jackiw-Teitelboim Quantum Gravity}",
    eprint = "1812.00918",
    archivePrefix = "arXiv",
    primaryClass = "hep-th",
    doi = "10.1007/JHEP09(2019)066",
    journal = "JHEP",
    volume = "09",
    pages = "066",
    year = "2019"
}

@article{1905.02726,
    author = "Iliesiu, Luca V. and Pufu, Silviu S. and Verlinde, Herman and Wang, Yifan",
    title = "{An exact quantization of Jackiw-Teitelboim gravity}",
    eprint = "1905.02726",
    archivePrefix = "arXiv",
    primaryClass = "hep-th",
    reportNumber = "PUPT-2584",
    doi = "10.1007/JHEP11(2019)091",
    journal = "JHEP",
    volume = "11",
    pages = "091",
    year = "2019"
}

@article{2304.13650,
    author = "Collier, Scott and Eberhardt, Lorenz and Zhang, Mengyang",
    title = "{Solving 3d gravity with Virasoro TQFT}",
    eprint = "2304.13650",
    archivePrefix = "arXiv",
    primaryClass = "hep-th",
    doi = "10.21468/SciPostPhys.15.4.151",
    journal = "SciPost Phys.",
    volume = "15",
    number = "4",
    pages = "151",
    year = "2023"
}

@article{2401.13900,
    author = "Collier, Scott and Eberhardt, Lorenz and Zhang, Mengyang",
    title = "{3d gravity from Virasoro TQFT: Holography, wormholes and knots}",
    eprint = "2401.13900",
    archivePrefix = "arXiv",
    primaryClass = "hep-th",
    doi = "10.21468/SciPostPhys.17.5.134",
    journal = "SciPost Phys.",
    volume = "17",
    pages = "134",
    year = "2024"
}

@article{2210.12127,
    author = "Chen, Lin and Ji, Kaixin and Zhang, Haochen and Shen, Ce and Wang, Ruoshui and Zeng, Xiangdong and Hung, Ling-Yan",
    title = "{CFTD from TQFTD+1 via Holographic Tensor Network, and Precision Discretization of CFT2 }",
    eprint = "2210.12127",
    archivePrefix = "arXiv",
    primaryClass = "hep-th",
    doi = "10.1103/PhysRevX.14.041033",
    journal = "Phys. Rev. X",
    volume = "14",
    number = "4",
    pages = "041033",
    year = "2024"
}

@article{2311.18005,
    author = "Cheng, Gong and Chen, Lin and Gu, Zheng-Cheng and Hung, Ling-Yan",
    title = "{Precision Reconstruction of Rational Conformal Field Theory from Exact Fixed-Point Tensor Network}",
    eprint = "2311.18005",
    archivePrefix = "arXiv",
    primaryClass = "cond-mat.str-el",
    doi = "10.1103/PhysRevX.15.011073",
    journal = "Phys. Rev. X",
    volume = "15",
    number = "1",
    pages = "011073",
    year = "2025"
}

@article{2403.03179,
    author = "Chen, Lin and Hung, Ling-Yan and Jiang, Yikun and Lao, Bing-Xin",
    title = "{Deriving the non-perturbative gravitational dual of quantum Liouville theory from BCFT operator algebra}",
    eprint = "2403.03179",
    archivePrefix = "arXiv",
    primaryClass = "hep-th",
    doi = "10.21468/SciPostPhys.19.6.163",
    journal = "SciPost Phys.",
    volume = "19",
    number = "6",
    pages = "163",
    year = "2025"
}

@article{2412.12045,
    author = "Bao, Ning and Hung, Ling-Yan and Jiang, Yikun and Liu, Zhihan",
    title = "{QG from SymQRG: AdS$_3$/CFT$_2$ Correspondence as Topological Symmetry-Preserving Quantum RG Flow}",
    eprint = "2412.12045",
    archivePrefix = "arXiv",
    primaryClass = "hep-th",
    month = "12",
    year = "2024"
}

@article{2504.21660,
    author = "Hung, Ling-Yan and Jiang, Yikun and Lao, Bing-Xin",
    title = "{Universal Structures and Emergent Geometry from Large-$c$ BCFT Ensemble}",
    eprint = "2504.21660",
    archivePrefix = "arXiv",
    primaryClass = "hep-th",
    month = "4",
    year = "2025"
}

@article{2507.12696,
    author = "Hartman, Thomas",
    title = "{Triangulating quantum gravity in AdS$_3$}",
    eprint = "2507.12696",
    archivePrefix = "arXiv",
    primaryClass = "hep-th",
    month = "7",
    year = "2025"
}

@article{2203.09537,
    author = "Benini, Francesco and Copetti, Christian and Di Pietro, Lorenzo",
    title = "{Factorization and global symmetries in holography}",
    eprint = "2203.09537",
    archivePrefix = "arXiv",
    primaryClass = "hep-th",
    reportNumber = "SISSA 05/2022/FISI, SISSA 05/2022/FISI",
    doi = "10.21468/SciPostPhys.14.2.019",
    journal = "SciPost Phys.",
    volume = "14",
    number = "2",
    pages = "019",
    year = "2023"
}

@article{1511.00295,
    author = "Sharma, Amit and Voronov, Alexander A.",
    title = "{Categorification of Dijkgraaf{\textendash}Witten theory}",
    eprint = "1511.00295",
    archivePrefix = "arXiv",
    primaryClass = "math.AT",
    reportNumber = "IPMU15-0216",
    doi = "10.4310/ATMP.2017.v21.n4.a5",
    journal = "Adv. Theor. Math. Phys.",
    volume = "21",
    pages = "1023--1061",
    year = "2017"
}

@article{2206.12448,
    author = "Menger, Leon",
    title = "{Introduction to 2-dimensional Topological Quantum Field Theory}",
    eprint = "2206.12448",
    archivePrefix = "arXiv",
    primaryClass = "math.QA",
    month = "6",
    year = "2022"
}

@article{1109.6295,
    author = "Ellegaard Andersen, J{\o}rgen and Kashaev, Rinat",
    title = {{A TQFT from Quantum Teichm{\"u}ller Theory}},
    eprint = "1109.6295",
    archivePrefix = "arXiv",
    primaryClass = "math.QA",
    doi = "10.1007/s00220-014-2073-2",
    journal = "Commun. Math. Phys.",
    volume = "330",
    pages = "887--934",
    year = "2014"
}

@article{1302.3454,
    author = "Nidaiev, I. and Teschner, J.",
    title = "{On the relation between the modular double of $U_q(sl(2,R))$ and the quantum Teichmueller theory}",
    eprint = "1302.3454",
    archivePrefix = "arXiv",
    primaryClass = "math-ph",
    reportNumber = "DESY-13-026",
    month = "2",
    year = "2013"
}

@article{math/0007097,
    author = "Ponsot, B. and Teschner, J.",
    title = "{Clebsch-Gordan and Racah-Wigner coefficients for a continuous series of representations of U(q)(sl(2,R))}",
    eprint = "math/0007097",
    archivePrefix = "arXiv",
    reportNumber = "DIAS-STP-00-15, BERLIN-SFB-288, LPM-00-21",
    doi = "10.1007/PL00005590",
    journal = "Commun. Math. Phys.",
    volume = "224",
    pages = "613--655",
    year = "2001"
}

@article{1202.4698,
    author = {Teschner, J{\"o}rg and Vartanov, Grigory},
    title = "{6j symbols for the modular double, quantum hyperbolic geometry, and supersymmetric gauge theories}",
    eprint = "1202.4698",
    archivePrefix = "arXiv",
    primaryClass = "hep-th",
    reportNumber = "DESY-12-035",
    doi = "10.1007/s11005-014-0684-3",
    journal = "Lett. Math. Phys.",
    volume = "104",
    pages = "527--551",
    year = "2014"
}

@article{2309.11540,
    author = "Eberhardt, Lorenz",
    title = "{Notes on crossing transformations of Virasoro conformal blocks}",
    eprint = "2309.11540",
    archivePrefix = "arXiv",
    primaryClass = "hep-th",
    month = "9",
    year = "2023"
}

@article{2006.05499,
    author = "Belin, Alexandre and de Boer, Jan",
    title = "{Random statistics of OPE coefficients and Euclidean wormholes}",
    eprint = "2006.05499",
    archivePrefix = "arXiv",
    primaryClass = "hep-th",
    reportNumber = "CERN-TH-2020-096",
    doi = "10.1088/1361-6382/ac1082",
    journal = "Class. Quant. Grav.",
    volume = "38",
    number = "16",
    pages = "164001",
    year = "2021"
}

@article{2006.08648,
    author = "Cotler, Jordan and Jensen, Kristan",
    title = "{AdS$_{3}$ gravity and random CFT}",
    eprint = "2006.08648",
    archivePrefix = "arXiv",
    primaryClass = "hep-th",
    doi = "10.1007/JHEP04(2021)033",
    journal = "JHEP",
    volume = "04",
    pages = "033",
    year = "2021"
}

@article{2006.11317,
    author = "Maxfield, Henry and Turiaci, Gustavo J.",
    title = "{The path integral of 3D gravity near extremality; or, JT gravity with defects as a matrix integral}",
    eprint = "2006.11317",
    archivePrefix = "arXiv",
    primaryClass = "hep-th",
    doi = "10.1007/JHEP01(2021)118",
    journal = "JHEP",
    volume = "01",
    pages = "118",
    year = "2021"
}

@article{2006.13414,
    author = "Witten, Edward",
    title = "{Matrix Models and Deformations of JT Gravity}",
    eprint = "2006.13414",
    archivePrefix = "arXiv",
    primaryClass = "hep-th",
    doi = "10.1098/rspa.2020.0582",
    journal = "Proc. Roy. Soc. Lond. A",
    volume = "476",
    number = "2244",
    pages = "20200582",
    year = "2020"
}

@article{2006.13971,
    author = "Blommaert, Andreas",
    title = "{Dissecting the ensemble in JT gravity}",
    eprint = "2006.13971",
    archivePrefix = "arXiv",
    primaryClass = "hep-th",
    doi = "10.1007/JHEP09(2022)075",
    journal = "JHEP",
    volume = "09",
    pages = "075",
    year = "2022"
}

@article{2006.16289,
    author = "Bousso, Raphael and Wildenhain, Elizabeth",
    title = "{Gravity/ensemble duality}",
    eprint = "2006.16289",
    archivePrefix = "arXiv",
    primaryClass = "hep-th",
    doi = "10.1103/PhysRevD.102.066005",
    journal = "Phys. Rev. D",
    volume = "102",
    number = "6",
    pages = "066005",
    year = "2020"
}

@article{2007.15653,
    author = "Cotler, Jordan and Jensen, Kristan",
    title = "{AdS$_3$ wormholes from a modular bootstrap}",
    eprint = "2007.15653",
    archivePrefix = "arXiv",
    primaryClass = "hep-th",
    doi = "10.1007/JHEP11(2020)058",
    journal = "JHEP",
    volume = "11",
    pages = "058",
    year = "2020"
}

@article{2012.15830,
    author = "Dymarsky, Anatoly and Shapere, Alfred",
    title = "{Comments on the holographic description of Narain theories}",
    eprint = "2012.15830",
    archivePrefix = "arXiv",
    primaryClass = "hep-th",
    doi = "10.1007/JHEP10(2021)197",
    journal = "JHEP",
    volume = "10",
    pages = "197",
    year = "2021"
}

@article{2102.03136,
    author = "Meruliya, Viraj and Mukhi, Sunil and Singh, Palash",
    title = "{Poincar{\'e} Series, 3d Gravity and Averages of Rational CFT}",
    eprint = "2102.03136",
    archivePrefix = "arXiv",
    primaryClass = "hep-th",
    doi = "10.1007/JHEP04(2021)267",
    journal = "JHEP",
    volume = "04",
    pages = "267",
    year = "2021"
}

@article{2102.12509,
    author = "Datta, Shouvik and Duary, Sarthak and Kraus, Per and Maity, Pronobesh and Maloney, Alexander",
    title = "{Adding flavor to the Narain ensemble}",
    eprint = "2102.12509",
    archivePrefix = "arXiv",
    primaryClass = "hep-th",
    reportNumber = "CERN-TH-2021-020",
    doi = "10.1007/JHEP05(2022)090",
    journal = "JHEP",
    volume = "05",
    pages = "090",
    year = "2022"
}

@article{2103.15826,
    author = "Benjamin, Nathan and Keller, Christoph A. and Ooguri, Hirosi and Zadeh, Ida G.",
    title = "{Narain to Narnia}",
    eprint = "2103.15826",
    archivePrefix = "arXiv",
    primaryClass = "hep-th",
    doi = "10.1007/s00220-021-04211-x",
    journal = "Commun. Math. Phys.",
    volume = "390",
    number = "1",
    pages = "425--470",
    year = "2022"
}

@article{2103.16754,
    author = "Saad, Phil and Shenker, Stephen H. and Stanford, Douglas and Yao, Shunyu",
    title = "{Wormholes without averaging}",
    eprint = "2103.16754",
    archivePrefix = "arXiv",
    primaryClass = "hep-th",
    doi = "10.1007/JHEP09(2024)133",
    journal = "JHEP",
    volume = "09",
    pages = "133",
    year = "2024"
}

@article{2104.01184,
    author = "Gao, Ping and Jafferis, Daniel Louis and Kolchmeyer, David K.",
    title = "{An effective matrix model for dynamical end of the world branes in Jackiw-Teitelboim gravity}",
    eprint = "2104.01184",
    archivePrefix = "arXiv",
    primaryClass = "hep-th",
    doi = "10.1007/JHEP01(2022)038",
    journal = "JHEP",
    volume = "01",
    pages = "038",
    year = "2022"
}

@article{2104.10178,
    author = "Meruliya, Viraj and Mukhi, Sunil",
    title = "{AdS$_{3}$ gravity and RCFT ensembles with multiple invariants}",
    eprint = "2104.10178",
    archivePrefix = "arXiv",
    primaryClass = "hep-th",
    doi = "10.1007/JHEP08(2021)098",
    journal = "JHEP",
    volume = "08",
    pages = "098",
    year = "2021"
}

@article{2104.14710,
    author = "Ashwinkumar, Meer and Dodelson, Matthew and Kidambi, Abhiram and Leedom, Jacob M. and Yamazaki, Masahito",
    title = "{Chern-Simons invariants from ensemble averages}",
    eprint = "2104.14710",
    archivePrefix = "arXiv",
    primaryClass = "hep-th",
    doi = "10.1007/JHEP08(2021)044",
    journal = "JHEP",
    volume = "08",
    pages = "044",
    year = "2021"
}

@article{2105.02142,
    author = "Verlinde, Herman",
    title = "{Deconstructing the Wormhole: Factorization, Entanglement and Decoherence}",
    eprint = "2105.02142",
    archivePrefix = "arXiv",
    primaryClass = "hep-th",
    month = "5",
    year = "2021"
}

@article{2105.08207,
    author = "Mukhametzhanov, Baur",
    title = "{Half-wormholes in SYK with one time point}",
    eprint = "2105.08207",
    archivePrefix = "arXiv",
    primaryClass = "hep-th",
    doi = "10.21468/SciPostPhys.12.1.029",
    journal = "SciPost Phys.",
    volume = "12",
    number = "1",
    pages = "029",
    year = "2022"
}

@article{2106.09048,
    author = "Johnson, Clifford V.",
    title = "{Quantum Gravity Microstates from Fredholm Determinants}",
    eprint = "2106.09048",
    archivePrefix = "arXiv",
    primaryClass = "hep-th",
    doi = "10.1103/PhysRevLett.127.181602",
    journal = "Phys. Rev. Lett.",
    volume = "127",
    number = "18",
    pages = "181602",
    year = "2021"
}

@article{2106.12760,
    author = "Collier, Scott and Maloney, Alexander",
    title = "{Wormholes and spectral statistics in the Narain ensemble}",
    eprint = "2106.12760",
    archivePrefix = "arXiv",
    primaryClass = "hep-th",
    doi = "10.1007/JHEP03(2022)004",
    journal = "JHEP",
    volume = "03",
    pages = "004",
    year = "2022"
}

@article{2107.02178,
    author = "Blommaert, Andreas and Kruthoff, Jorrit",
    title = "{Gravity without averaging}",
    eprint = "2107.02178",
    archivePrefix = "arXiv",
    primaryClass = "hep-th",
    doi = "10.21468/SciPostPhys.12.2.073",
    journal = "SciPost Phys.",
    volume = "12",
    number = "2",
    pages = "073",
    year = "2022"
}

@article{2107.13130,
    author = "Saad, Phil and Shenker, Stephen H. and Yao, Shunyu",
    title = "{Comments on wormholes and factorization}",
    eprint = "2107.13130",
    archivePrefix = "arXiv",
    primaryClass = "hep-th",
    doi = "10.1007/JHEP10(2024)076",
    journal = "JHEP",
    volume = "10",
    pages = "076",
    year = "2024"
}

@article{2110.14649,
    author = "Belin, Alexandre and de Boer, Jan and Liska, Diego",
    title = "{Non-Gaussianities in the statistical distribution of heavy OPE coefficients and wormholes}",
    eprint = "2110.14649",
    archivePrefix = "arXiv",
    primaryClass = "hep-th",
    reportNumber = "CERN-TH-2021-166",
    doi = "10.1007/JHEP06(2022)116",
    journal = "JHEP",
    volume = "06",
    pages = "116",
    year = "2022"
}

@article{2111.07863,
    author = "Blommaert, Andreas and Iliesiu, Luca V. and Kruthoff, Jorrit",
    title = "{Gravity factorized}",
    eprint = "2111.07863",
    archivePrefix = "arXiv",
    primaryClass = "hep-th",
    doi = "10.1007/JHEP09(2022)080",
    journal = "JHEP",
    volume = "09",
    pages = "080",
    year = "2022"
}

@article{2111.14856,
    author = "Peng, Cheng and Tian, Jia and Yu, Jianghui",
    title = "{Baby universes, ensemble averages and factorizations with matters}",
    eprint = "2111.14856",
    archivePrefix = "arXiv",
    primaryClass = "hep-th",
    month = "11",
    year = "2021"
}

@article{2112.09143,
    author = "Anous, Tarek and Belin, Alexandre and de Boer, Jan and Liska, Diego",
    title = "{OPE statistics from higher-point crossing}",
    eprint = "2112.09143",
    archivePrefix = "arXiv",
    primaryClass = "hep-th",
    reportNumber = "CERN-TH-2021-201",
    doi = "10.1007/JHEP06(2022)102",
    journal = "JHEP",
    volume = "06",
    pages = "102",
    year = "2022"
}

@article{2203.06511,
    author = "Chandra, Jeevan and Collier, Scott and Hartman, Thomas and Maloney, Alexander",
    title = "{Semiclassical 3D gravity as an average of large-c CFTs}",
    eprint = "2203.06511",
    archivePrefix = "arXiv",
    primaryClass = "hep-th",
    doi = "10.1007/JHEP12(2022)069",
    journal = "JHEP",
    volume = "12",
    pages = "069",
    year = "2022"
}

@article{2208.14457,
    author = "Henriksson, Johan and McPeak, Brian",
    title = "{Averaging over codes and an SU(2) modular bootstrap}",
    eprint = "2208.14457",
    archivePrefix = "arXiv",
    primaryClass = "hep-th",
    doi = "10.1007/JHEP11(2023)035",
    journal = "JHEP",
    volume = "11",
    pages = "035",
    year = "2023"
}

@article{2209.02131,
    author = "Jafferis, Daniel Louis and Kolchmeyer, David K. and Mukhametzhanov, Baur and Sonner, Julian",
    title = "{Jackiw-Teitelboim gravity with matter, generalized eigenstate thermalization hypothesis, and random matrices}",
    eprint = "2209.02131",
    archivePrefix = "arXiv",
    primaryClass = "hep-th",
    doi = "10.1103/PhysRevD.108.066015",
    journal = "Phys. Rev. D",
    volume = "108",
    number = "6",
    pages = "066015",
    year = "2023"
}

@article{2306.07321,
    author = "Kames-King, Joshua and Kanargias, Alexandros and Knighton, Bob and Usatyuk, Mykhaylo",
    title = "{The lion, the witch, and the wormhole: ensemble averaging the symmetric product orbifold}",
    eprint = "2306.07321",
    archivePrefix = "arXiv",
    primaryClass = "hep-th",
    doi = "10.1007/JHEP07(2024)236",
    journal = "JHEP",
    volume = "07",
    pages = "236",
    year = "2024"
}

@article{2307.03707,
    author = "Di Ubaldo, Gabriele and Perlmutter, Eric",
    title = "{AdS$_{3}$/RMT$_{2}$ duality}",
    eprint = "2307.03707",
    archivePrefix = "arXiv",
    primaryClass = "hep-th",
    doi = "10.1007/JHEP12(2023)179",
    journal = "JHEP",
    volume = "12",
    pages = "179",
    year = "2023"
}

@article{2308.01787,
    author = "Di Ubaldo, Gabriele and Perlmutter, Eric",
    title = "{AdS3 Pure Gravity and Stringy Unitarity}",
    eprint = "2308.01787",
    archivePrefix = "arXiv",
    primaryClass = "hep-th",
    doi = "10.1103/PhysRevLett.132.041602",
    journal = "Phys. Rev. Lett.",
    volume = "132",
    number = "4",
    pages = "041602",
    year = "2024"
}

@article{2308.03829,
    author = "Belin, Alexandre and de Boer, Jan and Jafferis, Daniel Louis and Nayak, Pranjal and Sonner, Julian",
    title = "{Approximate CFTs and random tensor models}",
    eprint = "2308.03829",
    archivePrefix = "arXiv",
    primaryClass = "hep-th",
    reportNumber = "CERN-TH-2023-068",
    doi = "10.1007/JHEP09(2024)163",
    journal = "JHEP",
    volume = "09",
    pages = "163",
    year = "2024"
}

@article{2309.10846,
    author = {Collier, Scott and Eberhardt, Lorenz and M{\"u}hlmann, Beatrix and Rodriguez, Victor A.},
    title = "{The Virasoro minimal string}",
    eprint = "2309.10846",
    archivePrefix = "arXiv",
    primaryClass = "hep-th",
    doi = "10.21468/SciPostPhys.16.2.057",
    journal = "SciPost Phys.",
    volume = "16",
    number = "2",
    pages = "057",
    year = "2024"
}

@article{2310.06012,
    author = "Aharony, Ofer and Dymarsky, Anatoly and Shapere, Alfred D.",
    title = "{Holographic description of Narain CFTs and their code-based ensembles}",
    eprint = "2310.06012",
    archivePrefix = "arXiv",
    primaryClass = "hep-th",
    doi = "10.1007/JHEP05(2024)343",
    journal = "JHEP",
    volume = "05",
    pages = "343",
    year = "2024"
}

@article{2311.00699,
    author = "Ashwinkumar, Meer and Kidambi, Abhiram and Leedom, Jacob M. and Yamazaki, Masahito",
    title = "{Generalized Narain theories $\mathfrak{decoded}$: $\mathfrak{d}$iscussions on $\mathfrak{E}$isenstein series, $\mathfrak{c}$haracteristics, $\mathfrak{o}$rbifolds, $\mathfrak{d}$iscriminants {\&} $\mathfrak{e}$nsembles in any $\mathfrak{d}$imension}",
    eprint = "2311.00699",
    archivePrefix = "arXiv",
    primaryClass = "hep-th",
    reportNumber = "DESY-23-170",
    doi = "10.4310/atmp.250524025221",
    journal = "Adv. Theor. Math. Phys.",
    volume = "29",
    number = "1",
    pages = "1--55",
    year = "2025"
}

@article{2311.08132,
    author = "de Boer, Jan and Liska, Diego and Post, Boris and Sasieta, Martin",
    title = "{A principle of maximum ignorance for semiclassical gravity}",
    eprint = "2311.08132",
    archivePrefix = "arXiv",
    primaryClass = "hep-th",
    doi = "10.1007/JHEP02(2024)003",
    journal = "JHEP",
    volume = "2024",
    pages = "003",
    year = "2024"
}

@article{2312.02276,
    author = "Raeymaekers, Joris and Rossi, Paolo",
    title = "{Wormholes and surface defects in rational ensemble holography}",
    eprint = "2312.02276",
    archivePrefix = "arXiv",
    primaryClass = "hep-th",
    doi = "10.1007/JHEP01(2024)104",
    journal = "JHEP",
    volume = "01",
    pages = "104",
    year = "2024"
}

@article{2403.02976,
    author = "Forste, Stefan and Jockers, Hans and Kames-King, Joshua and Kanargias, Alexandros and Zadeh, Ida G.",
    title = "{Ensemble averages of {\ensuremath{\mathbb{Z}}}$_{2}$ orbifold classes of Narain CFTs}",
    eprint = "2403.02976",
    archivePrefix = "arXiv",
    primaryClass = "hep-th",
    reportNumber = "BONN-TH-2024-07, MITP/24-034",
    doi = "10.1007/JHEP05(2024)240",
    journal = "JHEP",
    volume = "05",
    pages = "240",
    year = "2024"
}

@article{2404.10035,
    author = "Hern{\'a}ndez-Cuenca, Sergio",
    title = "{Wormholes and factorization in exact effective theory}",
    eprint = "2404.10035",
    archivePrefix = "arXiv",
    primaryClass = "hep-th",
    reportNumber = "MIT-CTP/5708",
    doi = "10.1007/JHEP05(2025)024",
    journal = "JHEP",
    volume = "05",
    pages = "024",
    year = "2025"
}

@article{2405.13111,
    author = "de Boer, Jan and Liska, Diego and Post, Boris",
    title = "{Multiboundary wormholes and OPE statistics}",
    eprint = "2405.13111",
    archivePrefix = "arXiv",
    primaryClass = "hep-th",
    doi = "10.1007/JHEP10(2024)207",
    journal = "JHEP",
    volume = "10",
    pages = "207",
    year = "2024"
}

@article{2407.02649,
    author = "Jafferis, Daniel L. and Rozenberg, Liza and Wong, Gabriel",
    title = "{3d gravity as a random ensemble}",
    eprint = "2407.02649",
    archivePrefix = "arXiv",
    primaryClass = "hep-th",
    doi = "10.1007/JHEP02(2025)208",
    journal = "JHEP",
    volume = "02",
    pages = "208",
    year = "2025"
}

@article{2503.00101,
    author = "Boruch, Jan and Di Ubaldo, Gabriele and Haehl, Felix M. and Perlmutter, Eric and Rozali, Moshe",
    title = "{Modular-Invariant Random Matrix Theory and AdS3 Wormholes}",
    eprint = "2503.00101",
    archivePrefix = "arXiv",
    primaryClass = "hep-th",
    reportNumber = "RIKEN-iTHEMS-Report-25",
    doi = "10.1103/4hhn-c6mp",
    journal = "Phys. Rev. Lett.",
    volume = "135",
    number = "12",
    pages = "121602",
    year = "2025"
}

@article{2504.08724,
    author = "Dymarsky, Anatoly and Henriksson, Johan and McPeak, Brian",
    title = "{Holographic duality from Howe duality: Chern-Simons gravity as an ensemble of code CFTs}",
    eprint = "2504.08724",
    archivePrefix = "arXiv",
    primaryClass = "hep-th",
    reportNumber = "CERN-TH-2025-060",
    doi = "10.1007/JHEP02(2026)257",
    journal = "JHEP",
    volume = "02",
    pages = "257",
    year = "2026"
}

@article{2506.19817,
    author = "Jafferis, Daniel L. and Rozenberg, Liza and Wang, Diandian",
    title = "{Open-closed 3d gravity as a random ensemble}",
    eprint = "2506.19817",
    archivePrefix = "arXiv",
    primaryClass = "hep-th",
    doi = "10.1007/JHEP10(2025)228",
    journal = "JHEP",
    volume = "10",
    pages = "228",
    year = "2025"
}

@article{Zamolodchikov:1986gt,
    author = "Zamolodchikov, A. B.",
    title = "{Irreversibility of the Flux of the Renormalization Group in a 2D Field Theory}",
    journal = "JETP Lett.",
    volume = "43",
    pages = "730--732",
    year = "1986"
}

@article{Giddings:1988cx,
    author = "Giddings, Steven B. and Strominger, Andrew",
    title = "{Loss of incoherence and determination of coupling constants in quantum gravity}",
    reportNumber = "HUTP-88/A006",
    doi = "10.1016/0550-3213(88)90109-5",
    journal = "Nucl. Phys. B",
    volume = "307",
    pages = "854--866",
    year = "1988"
}

@article{Giddings:1988wv,
    author = "Giddings, Steven B. and Strominger, Andrew",
    title = "{Baby Universes, Third Quantization and the Cosmological Constant}",
    reportNumber = "HUTP-88/A036",
    doi = "10.1016/0550-3213(89)90353-2",
    journal = "Nucl. Phys. B",
    volume = "321",
    pages = "481--508",
    year = "1989"
}

@article{Atiyah:1989vu,
    author = "Atiyah, M.",
    title = "{Topological quantum field theories}",
    doi = "10.1007/BF02698547",
    journal = "Inst. Hautes Etudes Sci. Publ. Math.",
    volume = "68",
    pages = "175--186",
    year = "1989"
}

@article{2603.12323,
    author = {Bergman, Oren and Heckman, Jonathan J. and H{\"u}bner, Max and Migliorati, Daniele and Yu, Xingyang and Zhang, Hao Y.},
    title = "{On the SymTFTs of Finite Non-Abelian Symmetries}",
    eprint = "2603.12323",
    archivePrefix = "arXiv",
    primaryClass = "hep-th",
    month = "3",
    year = "2026"
}
\vspace{0.5\baselineskip}

\end{document}